\documentclass[apj,numberedappendix]{emulateapj}

\usepackage[]{natbib}
\bibpunct{(}{)}{;}{a}{}{,}
\usepackage{graphicx}
\usepackage{rotating}
\usepackage{afterpage}
\usepackage{xcolor}
\usepackage{txfonts}
\newcommand{\um}{$\mu$m}
\newcommand{\brgamma}{Br$\gamma$}
\newcommand{\kms}{km\thinspace s$^{-1}$}

\def\degr{\hbox{$^\circ$}}
\def\arcmin{\hbox{$^\prime$}}
\def\arcsec{\hbox{$^{\prime\prime}$}}
\def\utw{\smash{\rlap{\lower5pt\hbox{$\sim$}}}}
\def\udtw{\smash{\rlap{\lower6pt\hbox{$\approx$}}}}

\def\Lsun{\hbox{\it L$_\odot$}}
\def\Lstar{\hbox{\it L$_*$}}
\def\Teff{\hbox{\it T$_{\rm eff}$}}

\def\Msun{\hbox{\it M$_\odot$}}

\def\Mbol{\hbox{\it M$_{bol}$}}
\def\Mbolone{\hbox{\it M$_{bol1}$}}

\def\Teff{\hbox{\it T$_{\rm eff}$}}

\def\K{\hbox{\it K}}

\def\Mk{\hbox{\it M$_{\rm K}$}}
\newcommand{\Ks}{{\it K$_{\rm s}$}}

\newcommand{\Aks}{{\it A$_{\rm K_{\rm s}}$}}

\newcommand{\Av}{{\it A$_{\rm V}$}}

\def\BCK{\hbox{\it BC$_{\rm K}$}}
\def\BCKs{\hbox{\it BC$_{\rm K_S}$}}

\def\simgr{\mathrel{\hbox{\rlap{\hbox{\lower4pt\hbox{$\sim$}}}\hbox{$>$}}}}
\def\HH{H{\sc ii}}	% HII region
\def\ew{\hbox{\it EW$_{\rm 2.29~\micron}$}}

\def\brg{\hbox{\it Br$_\gamma$}}

\def\Vlsr{\hbox{\it V$_{\rm lsr}$}}

\def\nodata{\hbox{ $..$}}

\DeclareRobustCommand{\ion}[2]{%
\relax\ifmmode
\ifx\testbx\f@series
{\mathbf{#1\,\mathsc{#2}}}\else
{\mathrm{#1\,\mathsc{#2}}}\fi
\else\textup{#1\,{\mdseries\textsc{#2}}}%
\fi}
\shorttitle{Massive stars in the disk of the Milky Way.}
   
\shortauthors{Messineo et al.}

\begin{document}

%% LaTeX will automatically break titles if they run longer than
%% one line. However, you may use \\ to force a line break if
%% you desire.

\title{Massive Stars  in Molecular Clouds Rich in High-energy Sources: \\
The Bridge of G332.809$-$0.132 and CS 78 in NGC 6334.
       }\thanks{Based on observations collected at the European Southern Observatory 
   (ESO Programmes 087.D-09609).} \thanks{MM is currently employed by USTC. 
   This works was designed  during her  ESA fellowship (2010), and
   partially carried out at the MPIfR.
   }

%% Use \author, \affil, and the \and command to format
%% author and affiliation information.
%% Note that \email has replaced the old \authoremail command
%% from AASTeX v4.0. You can use \email to mark an email address
%% anywhere in the paper, not just in the front matter.
%% As in the title, use \\ to force line breaks.

\author{Maria Messineo\altaffilmark{1}, 
        Karl M. Menten\altaffilmark{2},
        Donald F. Figer\altaffilmark{3},
        J. Simon Clark\altaffilmark{4}
        }
\altaffiltext{1}{Key Laboratory for Researches in Galaxies and Cosmology, University of 
Science and Technology of China, Chinese Academy of Sciences, Hefei, Anhui, 230026, China
\email{messineo@ustc.edu.cn}}
\altaffiltext{2}{Max-Planck-Institut f\"ur Radioastronomie, Auf dem H\"ugel 69, D-53121 Bonn, Germany}
\altaffiltext{3}{Center for Detectors, Rochester Institute of Technology, 54 Memorial Drive, Rochester, NY 14623, USA}
\altaffiltext{4}{Department of Physics and Astronomy, The Open University, Walton Hall, Milton Keynes, MK7 6AA, UK}

\begin{abstract} 

Detections of massive stars in the direction of the \HH\ region CS 78 
in NGC 6334 and of G332.809$-$0.132 are here presented. 
The region covered by the G332.809$-$0.132 complex coincides with the
RCW 103 stellar association.
In its core (40\arcmin\ in radius),  approximately 110  OB  candidate stars 
(\Ks $< 10$ mag and $0.4<$\Aks$<1.6$ mag) 
were identified using 2MASS, DENIS, and GLIMPSE data.
This  number of OB stars   accounts 
for more than 50\% of the observed  number of Lyman continuum 
photons from this region.
Medium-resolution  $K$-band spectra were obtained for seven early types, 
including one WN 8 star and one Ofpe/WN 9 star; 
the latter is located near the RCW 103 remnant and its 
luminosity is consistent with a distance of $\approx 3$ kpc. 
The area analyzed encloses 9 of the 34 OB stars 
previously known   in RCW 103, as well as IRAS  16115$-$5044 which we
reclassify as a candidate luminous blue variable.
The line of sight is particularly interesting,  crossing three
spiral arms;  a molecular cloud at $-50$ (with  RCW 103  in the Scutum-Crux arm) 
and another at $-90$ \kms\ (in the Norma arm)
are detected, both rich in massive stars and supernova remnants.

We also report the detection of a B supergiant as the main ionizing
source of CS 78, 2MASS J17213513$-$3532415. 
Medium-resolution  $H$ and $K$-band 
spectra display H I and He I lines, as well as Fe II lines.
By assuming a distance of 1.35 kpc, 
we estimate a bolometric magnitude of  $ -6.16$,
which is typical of supergiants.

\end{abstract}

%% Keywords should appear after the \end{abstract} command. The uncommented
%% example has been keyed in ApJ style. See the instructions to authors
%% for the journal to which you are submitting your paper to determine
%% what keyword punctuation is appropriate.


%% From the front matter, we move on to the body of the paper.
%% In the first two sections, notice the use of the natbib \citep
%% and \citet commands to identify citations.  The citations are
%% tied to the reference list via symbolic KEYs. The KEY corresponds
%% to the KEY in the \bibitem in the reference list below. We have
%% chosen the first three characters of the first author's name plus
%% the last two numeral of the year of publication as our KEY for
%% each reference.

%% Authors who wish to have the most important objects in their paper
%% linked in the electronic edition to a data center may do so by tagging
%% their objects with \objectname{} or \object{}.  Each macro takes the
%% object name as its required argument. The optional, square-bracket 
%% argument should be used in cases where the data center identification
%% differs from what is to be printed in the paper.  The text appearing 
%% in curly braces is what will appear in print in the published paper. 
%% If the object name is recognized by the data centers, it will be linked
%% in the electronic edition to the object data available at the data centers  
%%
%% Note that for sources with brackets in their names, e.g. [WEG2004] 14h-090,
%% the brackets must be escaped with backslashes when used in the first
%% square-bracket argument, for instance, \object[\[WEG2004\] 14h-090]{90}).
%%  Otherwise, LaTeX will issue an error. 

\section{Introduction} 
Massive stars ( $<\approx 8$ \Msun)  are key contributors to the
chemical enrichment of the Milky Way with their high mass-loss rates.
They explode as supernovae creating neutron stars and black holes, 
and are believed to be sources of $\gamma$-ray bursts, which are the most
energetic explosions in the Universe.
Much of what we know about the masses of supernova progenitors and the 
fates of massive stars come from theoretical work 
\citep{heger03}. Ideally, progenitor masses could be measured by 
studying young pulsars 
and supernova remnants associated with young stellar clusters/massive stars. 
However, because of their short
life spans, massive stars are rare and predominantly observed in young
(a few megayear) massive ($\sim 10^{4}$ \Msun) clusters.  
Stellar clusters are important building 
blocks of a Galaxy, and the best natural laboratories to test theories of 
star formation and stellar evolution.
However, in recent years,  analyses of open clusters 
and of their surroundings have revealed the  enigmatic existence of 
several massive stars in isolation \citep{gvaramadze12}.
Massive stars seen in isolation could be either  stars born in 
isolation,   or in  an already dissolved cluster, or  
ejected from a nearby cluster. Theoretical simulations show that 
a cluster may lose  up to 25\% of massive stars via ejection, 
including its most massive members \citep[e.g.,][]{oh14}.
A  fraction of stars may be born  by hierarchical 
star formation and not by monolithic collapses, 
such as stars in the Cyg OB1 association \citep[e.g.,][]{wright15}.

%%%%%%%%%%%%%%%%%%%%%%%%%%%
 
Ideally, one should  analyze 
the content of evolved massive stars per  
 region, where  ``region''
is the environmental unit in which the cluster was formed, i.e., the 
parental molecular cloud. Eventually, the comparison  between 
various units will yield new insights into the
distribution of various types of massive stars 
and  the remnants of core-collapse supernovae (SNRs). 
Currently, this topic is  widely discussed, and several
works have  been published \citep[e.g.,][]{humphreys17,smith15}.

%%%%%%%%%%%%%%%%%%%%%%%%%%%

A total of 295 SNRs  \citep{green19,green91}
are known  in the Milky Way.
Almost  80 very-high-energy $\gamma$--ray emitters have been 
discovered with the High Energy Stereoscopic System (HESS) \citep{hess18}; 
they are mostly found associated with young pulsar wind nebulae and SNRs. 
Several new high-mass X-ray binaries have been discovered in Chandra 
and XMM data.  However, very few collapsed objects (pulsars and black holes) and
remnants  are currently  confirmed to be associated with young stellar clusters
 rich in massive stars. 
For example, the soft gamma repeater 
SGR 1806$-$20 is associated with the  massive cluster Cl 1806$-$20 \citep{figer05};
an X-ray pulsar was detected in  Westerlund 1 by \citet{muno06}; 
the TeV source HESS J1837-069 was likely originated from a member of
the Red Supergiant cluster 1 (RSGC 1) \citep{figer06,gotthelf08}; 
the magnetar SGR 1900+14 is found to be associated with 
a cluster of two red supergiants (RSGs)  \citep{davies09};
the TeV gamma-ray source HESS 1640$-$465
is located a few arcminutes away
from the stellar cluster Mercer 81   \citep{davies12}, and
SGR J1745-29 is located  near the Galactic center (GC) cluster \citep{mori13}.

%%%%%%%%%%%%%%%%%%%%%%%%%%%
%%%% snr
 
With regard to SNRs, a letter from \citet{pauls77} reports  the serendipitous discovery of
the positional coincidence of SNR G127.1+00.5 with the open cluster NGC 559. A longer
discussion of positional coincidences of SNRs with stellar clusters  
is given by \citet{kumar78}. The author concluded that most coincidences are
random, while he suggested that SNR G291.0-0.1 is located in the open cluster 
Trumpler 18. While the first-mentioned association 
(NGC 559 with an SNR) still holds, 
that of Trumpler 18  could not be confirmed \citep{acero16,kharchenko16}.
More recently, \citet{safi07} investigated the
likelihood of an association between  the  NGC 5281 cluster, which
contains  the enigmatic X-ray emitter HD 119682, 
and the SNR G309.2$-$00.6, with the 
result that it could not be established. \citet{messineo08}
reported that the SNR G12.72$-$0.00 positionally coincided
with the massive  stellar cluster Cl 1713$-$178.
However, we estimated that 
almost 90\% of the celestial
superpositions of clusters and SNRs are chance coincidences 
(see the Appendix).

Certainly, it  is important to carefully examine stellar associations and 
giant molecular cloud complexes with care  for  their content 
of massive stars, pulsars, and SNRs.
Examples include the Cygnus region that contains about 10 SNRs and 14 pulsars,
along with some of the most massive stellar associations of the Milky Way
(9 OB associations) \citep{verytas19}.
Sparse massive stars and  one young stellar cluster (GLIMPSE9)
are found to populate the G23.3$-$0.3 (”[23,78]”) complex 
\citep{messineo14a}, which contains four SNRs:
W41 (which hosts the TeV emitter HESS J$1834-087$), 
G$22.7-0.2$ (SNR2), G$22.7583-0.4917$ (SNR3), and G$22.9717-0.3583$ (SNR4).

%%%%%%%%%%%%%%%%%%%%%%%%%%%

Comparisons of the spatial distribution
of different types of evolved massive stars (e.g, Wolf-Rayet stars (WRs),
B supergiants, RSGs) 
have recently been reported \citep[e.g.,][]{smith15}.
The rarity of known massive stars and
the mounting evidence that a large fraction of them is found
in isolation (loose associations or ejected stars)
motivate us to  look for them also outside  stellar clusters,
by using other tracers, such as molecular clouds rich in  
high-energy sources and SNRs.
Inspired by this idea and by the availability of large
surveys of the Galactic plane at near- and mid-infrared wavelengths,
we searched for massive stars in the direction of a few regions 
rich in high-energy sources.

%________________________________________________________________

In Sections \ref{sample} and   \ref{obs}, we describe the targets,  
the spectroscopic observations, and available photometric data. 
In Section \ref{analysis}, we analyze the infrared spectra.
In Sections \ref{region1} and   \ref{region2}, we briefly describe the 
core of the giant molecular complex G332.809$-$0.132 
and the \HH\ region CS 78 in NGC 6334,
and we discuss the likely association of the massive  stars detected
with these regions.

\begin{table*}
\caption{\label{obs.table} List of observed regions.}
\begin{tabular}{llllllrr}
\hline
\hline
ID &Molecular Cloud         & Radio/X-rays   &  Ra[J2000]   &  Dec[J2000]  & Radius \\
   &                        &                & (hh mm ss)   &  (dd mm ss)  & (arcmin)\\
\hline
1 & G332.809$-$0.132-Bridge     & 3 SNRs + 1 PWN +  6 PSRs &   16 16 08  &  $-$51 01 04  &  40 \\   
2 & NGC 6334/CS 78$^a$            & PSR B1718-35   &   17 21 35  &  $-$35 32 41  &  2  \\   
\hline
\end{tabular}
\begin{list}{}{}
\item[{\bf Notes.}] 
CS 78 (with a radius of 2\arcmin) is a bubble identified with 
GLIMPSE data \citep{churchwell07}. 
It is detached from the bulk of emission of the molecular complex NGC 6334 
(which has a radius about 20\arcmin), and it is located in its northeastern periphery
\citep[e.g., see figures in][]{willis13}.
\end{list}
\end{table*}

\section{The Sample}
\label{sample}

In this work, we present  bright infrared stars in the direction of 
two prominent Galactic star-forming regions 
that have a remarkable number of high-energy sources associated with them:
the core  of the giant \HH\ region  G332.809$-$0.132
(hereafter called ``The Bridge'') \citep{rahman10}
and  CS 78 in NGC 6334 \citep{churchwell07}.
The two regions are listed in  Table \ref{obs.table}.

\begin{figure}
\resizebox{1.0\hsize}{!}{\includegraphics[angle=0]{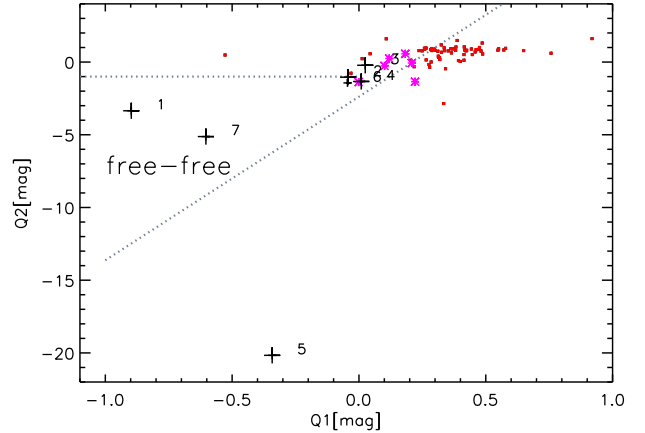}}
\caption{\label{ranges.fig} $Q1$ versus $Q2$ diagram of targets.
Plus signs  indicate the early-type stars from Table \ref{obs.early}; 
those with emission lines are marked with larger crosses.
Small red boxes mark the late-type stars in Table  \ref{table.latespectra}, 
and magenta asterisks those with water absorption (Mira-like stars).
The two  gray lines  are from \citet{messineo12} 
and mark the region where free-free emitters resides.
} 
\end{figure}

The list of  observed targets is  provided in 
 Tables \ref{obs.early} (8 early-type stars), {\bf 3, and \ref{table.latespectra} }
(76 late-type stars).
\Ks\ magnitudes of the targets and serendipitous detections range from 3.9 mag
to 12.05 mag; $H-$\Ks\  colors are mostly from 0.5 mag to 1.2 mag.
The majority of the targets were selected as infrared-bright and 
obscured stars in the directions of the chosen regions.
In Fig.\  \ref{ranges.fig}, we plot their $Q2$ versus the $Q1$ values, 
as defined in \citet{messineo12}\footnote{$Q1~=~(J-H)-1.8 \times (H-K_{\rm s}); $
$Q2~=~(J- K_{\rm s})-2.69\times (K_{\rm s} -[8.0]).$}.
In the  $J-H$ versus $H-$\Ks\ plane, $Q1$ is equivalent to the distance 
on the $y$-axis of a datapoint to a vector that passes through the origin 
and follows the vector of interstellar extinction 
\citep[e.g.,][]{messineo12,messineo14a}. 
 In the  $J-$\Ks\ versus \Ks$-[8.0]$ plane, $Q2$ is equivalent 
to the distance on the $y$-axis of a datapoint to a vector that passes 
through the origin and follows the vector of interstellar extinction. 
$Q2$ is a measure of the 8.0 \um\ excess.
The area below $Q2= -1$ mag and to the left of the oblique line
is rich in free-free emitters \citep[e.g.,][]{messineo12,mauerhan11}.
This diagram is a powerful diagnostic  for locating free-free emitters 
(84\% of the Wolf-Rayet stars, WRs, and 67\% of known luminous blue variables,
LBVs, are located in this area).
Normal OB stars are located to the right of the oblique line and 
 late-type stars to its left side. 
Due to stellar variability and nonsimultaneity of the measurements, 
additional criteria are  required to identify OB stars, for example,
the classical $J-H$ versus $J-$\Ks\ diagram or the $I-J$ versus  $J-$\Ks\ diagram.   
Details on the number of candidate OB stars  are given in Section \ref{earlyselection}.

\section{Data}
\label{obs}

\subsection{Spectroscopic Observations and Data Reduction}

Observations were made  with the Son of Isaac (SofI) instrument on the 
ESO New Technology Telescope (NTT) on La Silla on 2011 June 10--12 
(under ESO Program 087.D-09609, see also \citet{messineo14a}).
Spectra with the medium-resolution grism, a slit width of 1\arcsec, and the \Ks\ filter 
that covers from 2.0 \um\ to 2.3 \um\ at a resolving power $R \sim 2200$, 
were obtained for all targets. Additionally, low-resolution spectra covering 
from 1.53 \um\ to 2.52 \um\  with $ R \sim 980$  were taken for a few targets. 
Medium-resolution spectra in the $H-$band with a  slit width of 1\arcsec, 
covering from 1.5 \um\ to 1.8 \um\ with a resolving power of $\sim 1250$, 
were taken for one  interesting target (star E1). 

Pairs of frames were subtracted and flat-fielded.  One to three stellar traces were
identified in each frame and extracted. Wavelength calibration was performed with Xe 
and Ne arc spectra.
Correction for atmospheric and instrumental responses was performed by dividing 
the target spectrum by spectra  of  B-type stars  taken in the same manner and 
at a similar air mass to the target.
Telluric lines were removed with  linear interpolation.  

Observed targets are listed in Tables \ref{obs.early},
{\bf  3, and \ref{table.latespectra}.}

\begin{table*}
\caption{ \label{obs.early} List of detected early-type stars. }
\begin{tabular} {llllllllll}
\hline 
\hline
ID  &   $alpha$[J2000] & $\delta$[J2000]    &     Sp. type   &   \Teff\        &           Field           &   2MASS ID/Alias            \\
    &   &{\rm (hh mm ss)}        & {\rm (deg mm ss)} & {\rm (K)}    &              &                &                \\
\hline
 
   E1 & 17 21 35.13 & $-$35~32~41.60 &           B3-5  & 14600$\pm$  900 &            CS 78             &      J17213513$-$3532415 \\
   E2 & 16 17 42.92 & $-$50~54~25.14 &        Ofpe/WN 9 &26500$\pm$ 6500 &            G332.809$-$0.132 &     J16174291$-$5054251 \\
   E3 & 16 17 42.04 & $-$50~56~37.26 &            B0-5 &20600$\pm$ 6900 &            G332.809$-$0.132 &     J16174204$-$5056372 \\
   E4 & 16 20 16.07 & $-$50~58~07.68 &            WN 8 &45000$\pm$ 5000 &            G332.809$-$0.132 &     J16201606$-$5058076 \\
   E5 & 16 17 02.20 & $-$50~47~03.18 &              Oe &37700$\pm$ 6700 &            G332.809$-$0.132 &     J16170220$-$5047031/$$[RA2004a]$$~IRS~1 \\
   E6 & 16 20 09.03 & $-$50~48~58.84 &            B0-5 &20600$\pm$ 6900 &            G332.809$-$0.132 &     J16200902$-$5048588/TYC~8324-906-1\\
   E7 & 16 17 09.23 & $-$50~47~14.70 &              Oe &37700$\pm$ 6700 &            G332.809$-$0.132 &     J16170923$-$5047147/$$[RA2004a]$$~IRS~2\\
   E8 & 16 17 03.01 & $-$50~47~30.84 &              Oe &37700$\pm$ 6700 &            G332.809$-$0.132 &     J16170300$-$5047308 \\

\hline
\end{tabular}
\begin{list}{}{}
\item[{\bf Notes.}] Identification numbers are
followed by  coordinates,  spectral types, effective temperatures, \Teff, and field name.~
Sources are listed per field and  are sorted by \Ks\ magnitude within each field.
\end{list}
\end{table*}

\begin{table}
\caption{ \label{spberta} List of  Lines Detected in the Spectrum of Star E1.}
\begin{tabular}{lllllll}
\hline
\hline
Line & Vacuum $\lambda$  & Obs. $\lambda$  & EW & FWHM \\
     & ($\mu$m)          &($\mu$m) & (\AA) & (\AA) \\
\hline 
H I 18-4 &  1.53460$^b$ & 1.53409 & $2.1\pm0.8$ & $30\pm2$\\
H I 17-4 &  1.54432$^b$ & 1.54389 & $2.6\pm0.7$ & $27\pm2$\\
H I 16-4 &  1.55605$^a$ & 1.55600 & $5.0\pm1.5$ & $37\pm2$\\
H I 15-4 &  1.57050$^b$ & 1.57051 & $1.6\pm0.3$ & $23\pm2$\\
FeII  $z^2$ I$_{11/2}-$3 d$^5$ 4 s$^2$ I$_{11/2}$ & 1.5776$^c$& 1.57624 & $1.1\pm0.1$ & $18\pm3$ \\
H I 14-4 &  1.58849$^b$ & 1.58865 & $2.9\pm0.3$ & $32\pm2$\\
H I 13-4 &  1.61138$^b$ & 1.61145 & $3.2\pm0.6$ & $28\pm2$\\
H I 12-4 &  1.64114$^a$ & 1.64089 & $3.8\pm0.6$ & $27\pm1$\\
H I 11-4 &  1.68111$^a$ & 1.68108 & $3.1\pm0.6$ & $22\pm2$\\
FeII  $z^4$ F$_{9/2}-$ c$^4$ F$_{9/2}$ & 1.68778$^c$  & 1.68840 & $3.6\pm0.8$ & $34\pm2$\\
H I 10-4 &  1.73669$^a$ & 1.73644 & $4.9\pm0.5$ & $32\pm1$ \\
He I     &  2.05869$^a$ & 2.06190 &   $..$      & $25\pm8$\\ 
Fe II $z^4$ F$^0_{3/2}-$ c$^4$ F$_{3/2}$ ?  & 2.091$^c$& 2.08911 & $0.6\pm0.4$ & $18\pm3$ \\
H I  7-4 &  2.16612$^a$ & 2.165871 & $8.2\pm1.4$ & $27\pm3$ \\ 
\hline
\end{tabular}
\begin{list}{}{}
\item[{\bf Notes.}] $^a$ from the NIST line list.~
$^b$ \citet{storey95}. ~$^c$ \citet{morris96}, \citet{clark99}.~
\end{list}
\end{table}

\begin{table*}
\caption{\label{table.latespectra} Spectra of Detected Late-type Stars.
}
{\footnotesize
\begin{tabular}{@{\extracolsep{-.03in}}rrrrrrrrll}
\hline 
\hline 
 {\rm ID}   & {\rm RA(J2000)} & {\rm DEC(J2000)}  &  {\rm EW(CO)-high}$^*$  & {\rm EW(CO)-low} &{\rm Sp[rgb]}  & {\rm Sp[rsg]} & Type & Field & 2MASS-ID/Alias\\ 
 &{\rm [hh mm ss]}            & {\rm (deg mm ss)}    & {\rm (K)}   &    {\rm (K)} &	              &               &      &  &\\ 
\hline 

     L1 &  16 17 29.76 & $-$50 55 39.64 &    41$\pm$ 2&   $..$&    M7    &   K5    &      $..$    &           G332.809$-$0.132 & J16172975$-$5055396/IRAS~16137$-$5048\\
     L2 &  16 20 20.86 & $-$50 53 37.21 &    51$\pm$ 1&   51$\pm$ 4&  $..$    &   M2    &      $..$    &           G332.809$-$0.132 & J16202085$-$5053372                     \\
     L3 &  16 16 58.67 & $-$51 04 01.69 &    49$\pm$ 2&   $..$&  $..$    &   M2    &      $..$    &           G332.809$-$0.132 & J16165866$-$5104016/IRAS~16132$-$5056\\
     L4 &  16 18 14.49 & $-$50 56 13.45 &    41$\pm$ 2&   $..$&    M7    &   K5    &      Mira    &           G332.809$-$0.132 & J16181449$-$5056134                  \\
     L5 &  16 18 25.88 & $-$51 02 05.56 &    48$\pm$ 3&   $..$&  $..$    &   M1    &      Mira    &           G332.809$-$0.132 & J16182588$-$5102055/IRAS~16145$-$5054\\
     L6 &  16 17 30.93 & $-$50 58 35.12 &    30$\pm$ 3&   $..$&    M2    &   K2    &      $..$    &           G332.809$-$0.132 & J16173092$-$5058351/TYC~8323$-$726$-$1\\
     L7 &  16 17 08.76 & $-$51 02 17.22 &    28$\pm$ 1&   $..$&    M1    &   K2    &      Mira    &           G332.809$-$0.132 & J16170875$-$5102172/TYC~8323$-$928$-$1\\
     L8 &  16 16 52.03 & $-$51 05 33.59 &    46$\pm$ 2&   $..$&  $..$    &   M1    &      $..$    &           G332.809$-$0.132 & J16165202$-$5105335                     \\
     L9 &  16 17 18.77 & $-$51 01 36.43 &    27$\pm$ 3&   $..$&    M1    &   K2    &      $..$    &           G332.809$-$0.132 & J16171876$-$5101364                     \\
    L10 &  16 16 58.27 & $-$50 50 47.40 &    30$\pm$ 3&   $..$&    M2    &   K2    &      $..$    &           G332.809$-$0.132 & J16165827$-$5050474                     \\
    L11 &  16 17 18.06 & $-$50 57 29.24 &    40$\pm$ 6&   $..$&    M7    &   K5    &      $..$    &           G332.809$-$0.132 & J16171805$-$5057292                     \\
    L12 &  16 16 49.59 & $-$50 51 17.24 &    44$\pm$ 2&   $..$&    M7    &   M0    &      $..$    &           G332.809$-$0.132 & J16164958$-$5051172                     \\
    L13 &  16 17 04.73 & $-$50 59 41.71 &    34$\pm$ 2&   $..$&    M4    &   K3    &      $..$    &           G332.809$-$0.132 & J16170473$-$5059417                     \\
    L14 &  16 16 16.69 & $-$51 14 59.81 &    37$\pm$ 3&   $..$&    M6    &   K4    &      $..$    &           G332.809$-$0.132 & J16161669$-$5114598                     \\
    L15 &  16 19 08.56 & $-$50 45 09.09 &    40$\pm$ 4&   $..$&    M7    &   K5    &      Mira    &           G332.809$-$0.132 & J16190855$-$5045090                     \\
    L16 &  16 22 28.57 & $-$50 49 28.07 &    44$\pm$ 3&   $..$&    M7    &   M0    &      $..$    &           G332.809$-$0.132 & J16222857$-$5049280                     \\
    L17 &  16 17 24.47 & $-$51 03 32.02 &    37$\pm$ 2&   $..$&    M6    &   K4    &      $..$    &           G332.809$-$0.132 & J16172446$-$5103320/GSC~8323$-$1211\\
    L18 &  16 19 37.96 & $-$50 57 29.86 &    38$\pm$ 5&   $..$&    M6    &   K5    &      $..$    &           G332.809$-$0.132 & J16193796$-$5057298                     \\
    L19 &  16 19 16.22 & $-$51 02 17.65 &    41$\pm$ 4&   $..$&    M7    &   K5    &      $..$    &           G332.809$-$0.132 & J16191621$-$5102176                     \\
    L20 &  16 21 58.62 & $-$51 05 14.86 &    46$\pm$ 1&   $..$&  $..$    &   M1    &      Mira    &           G332.809$-$0.132 & J16215861$-$5105148                     \\
    L21 &  16 17 38.73 & $-$50 59 38.32 &    49$\pm$ 2&   $..$&  $..$    &   M2    &      $..$    &           G332.809$-$0.132 & J16173873$-$5059383                     \\
    L22 &  16 18 43.30 & $-$50 56 40.10 &    43$\pm$ 3&   $..$&    M7    &   M0    &      $..$    &           G332.809$-$0.132 & J16184330$-$5056401                     \\
    L23 &  16 18 13.16 & $-$51 02 10.65 &    31$\pm$ 3&   $..$&    M3    &   K3    &      Mira    &           G332.809$-$0.132 & J16181315$-$5102106                     \\
    L24 &  16 21 56.79 & $-$51 01 14.53 &    40$\pm$ 2&   41$\pm$ 1&    M7    &   K5    &      $..$    &           G332.809$-$0.132 & J16215678$-$5101145                    \\
    L25 &  16 17 03.52 & $-$50 55 02.68 &    39$\pm$ 3&   $..$&    M7    &   K5    &      $..$    &           G332.809$-$0.132 & J16170351$-$5055026                     \\
    L26 &  16 22 25.10 & $-$50 48 35.90 &    44$\pm$ 4&   $..$&    M7    &   M0    &      $..$    &           G332.809$-$0.132 & J16222509$-$5048358                     \\
    L27 &  16 18 01.46 & $-$50 55 58.85 &    37$\pm$ 2&   $..$&    M6    &   K4    &      $..$    &           G332.809$-$0.132 & J16180145$-$5055588                     \\
    L28 &  16 17 36.17 & $-$50 55 23.64 &    39$\pm$ 3&   $..$&    M7    &   K5    &      $..$    &           G332.809$-$0.132 & J16173617$-$5055236                     \\
    L29 &  16 16 19.70 & $-$51 14 01.78 &    35$\pm$ 2&   $..$&    M4    &   K4    &      $..$    &           G332.809$-$0.132 & J16161970$-$5114017                     \\
    L30 &  16 17 45.40 & $-$50 58 02.02 &    36$\pm$ 4&   $..$&    M5    &   K4    &      $..$    &           G332.809$-$0.132 & J16174539$-$5058020                     \\
    L31 &  16 17 35.34 & $-$50 57 09.18 &    19$\pm$ 3&   $..$&    K2    &   K0    &      $..$    &           G332.809$-$0.132 & J16173534$-$5057091                     \\
    L32 &  16 17 06.52 & $-$51 00 05.84 &    47$\pm$ 2&   $..$&  $..$    &   M1    &      $..$    &           G332.809$-$0.132 & J16170651$-$5100058                     \\
    L33 &  16 19 24.79 & $-$51 08 20.98 &    38$\pm$ 1&   40$\pm$ 1&    M7    &   K5    &      $..$    &           G332.809$-$0.132 & J16192478$-$5108209                \\
    L34 &  16 16 41.97 & $-$51 06 05.57 &    36$\pm$ 1&   $..$&    M5    &   K4    &      $..$    &           G332.809$-$0.132 & J16164196$-$5106055                     \\
    L35 &  16 20 22.94 & $-$50 53 29.08 &    46$\pm$ 2&   47$\pm$ 4&  $..$    &   M1    &      $..$    &           G332.809$-$0.132 & J16202293$-$5053290                 \\
    L36 &  16 16 50.14 & $-$51 12 23.61 &    39$\pm$ 3&   $..$&    M7    &   K5    &      $..$    &           G332.809$-$0.132 & J16165014$-$5112236                     \\
    L37 &  16 18 57.81 & $-$50 52 29.36 &    39$\pm$ 1&   $..$&    M7    &   K5    &      $..$    &           G332.809$-$0.132 & J16185780$-$5052293                     \\
    L38 &  16 16 43.75 & $-$51 11 13.24 &    43$\pm$ 2&   $..$&    M7    &   M0    &      $..$    &           G332.809$-$0.132 & J16164375$-$5111132                     \\

\hline
\end{tabular}
}
\begin{list}{}{}
\item[{\bf Notes.}] Sources are  sorted by \Ks\ magnitude.~
Identification numbers are followed by
celestial coordinates,  EW(CO) from the high-resolution mode, EW(CO) from the low-resolution mode, 
inferred spectral types (for giants and supergiants), comment on the type, and field name.~
$(^*)$ EW(CO) from the high-resolution mode (slit of 1\arcsec) have been 
multiplied by a factor 1.85 to match those
from the low-resolution mode (slit of 0\farcs6).
\end{list}
\end{table*}

\addtocounter{table}{-1}
\begin{table*}
\caption{  Continuation  of Table \ref{table.latespectra}.}
{\footnotesize
\begin{tabular}{@{\extracolsep{-.03in}}rrrrrrrrll}
\hline 
\hline 
 {\rm ID}   & {\rm RA(J2000)} & {\rm DEC(J2000)}  &  {\rm EW(CO)-high}  & {\rm EW(CO)-low} &{\rm Sp[rgb]}  & {\rm Sp[rsg]} & Type & Field & 2MASS-ID/Alias\\ 
 &{\rm (hh mm ss)}            & {\rm (deg mm ss)}    & {\rm (K)}   &    {\rm (K)} &	              &               &      &  &\\ 
\hline 
    L39 &  16 16 46.98 & $-$51 08 43.46 &    36$\pm$ 3&   $..$&    M5    &   K4    &      $..$    &           G332.809$-$0.132 & J16164698$-$5108434                    \\
    L40 &  16 19 29.03 & $-$50 42 25.71 &    38$\pm$ 2&   $..$&    M6    &   K5    &      $..$    &           G332.809$-$0.132 & J16192903$-$5042257                    \\
    L41 &  16 16 56.37 & $-$50 54 26.34 &    40$\pm$ 2&   $..$&    M7    &   K5    &      $..$    &           G332.809$-$0.132 & J16165636$-$5054263                     \\
    L42 &  16 21 56.89 & $-$51 03 56.01 &    33$\pm$ 0&   36$\pm$ 0&    M3    &   K3    &      $..$    &           G332.809$-$0.132 & J16215688$-$5103560               \\
    L43 &  16 18 05.17 & $-$50 54 35.53 &    37$\pm$ 3&   $..$&    M6    &   K4    &      $..$    &           G332.809$-$0.132 & J16180517$-$5054355                     \\
    L44 &  16 17 29.04 & $-$51 00 23.50 &    43$\pm$ 1&   $..$&    M7    &   M0    &      $..$    &           G332.809$-$0.132 & J16172904$-$5100234                    \\
    L45 &  16 16 40.40 & $-$51 07 32.75 &    17$\pm$ 2&   $..$&    K1    &   K0    &      $..$    &           G332.809$-$0.132 & J16164039$-$5107327/TYC~8323$-$2680$-$1\\
    L46 &  16 17 26.06 & $-$51 02 56.05 &    43$\pm$ 1&   $..$&    M7    &   M0    &      $..$    &           G332.809$-$0.132 & J16172605$-$5102560                     \\
    L47 &  16 16 44.17 & $-$51 04 08.21 &    35$\pm$ 2&   $..$&    M5    &   K4    &      $..$    &           G332.809$-$0.132 & J16164417$-$5104082                     \\
    L48 &  16 20 27.39 & $-$50 54 42.00 &    35$\pm$ 1&   34$\pm$ 1&    M5    &   K4    &      $..$    &           G332.809$-$0.132 & J16202739$-$5054420/$$[BRA2016]$$$-$DBS100$-$Obj4\\
    L49 &  16 18 43.49 & $-$50 56 11.19 &    35$\pm$ 2&   $..$&    M4    &   K4    &      $..$    &           G332.809$-$0.132 & J16184348$-$5056111                     \\
    L50 &  16 18 10.35 & $-$51 01 57.36 &    33$\pm$ 2&   $..$&    M4    &   K3    &      $..$    &           G332.809$-$0.132 & J16181035$-$5101573                     \\
    L51 &  16 17 51.06 & $-$50 53 14.34 &    30$\pm$ 4&   $..$&    M2    &   K2    &      $..$    &           G332.809$-$0.132 & J16175105$-$5053143                     \\
    L52 &  16 19 37.74 & $-$50 58 13.68 &    41$\pm$ 4&   $..$&    M7    &   K5    &      $..$    &           G332.809$-$0.132 & J16193774$-$5058136                     \\
    L53 &  16 17 58.23 & $-$51 00 59.49 &    36$\pm$ 1&   $..$&    M5    &   K4    &      $..$    &           G332.809$-$0.132 & J16175823$-$5100594                     \\
    L54 &  16 16 48.47 & $-$51 08 39.38 &    30$\pm$ 1&   $..$&    M2    &   K2    &      $..$    &           G332.809$-$0.132 & J16164846$-$5108393                     \\
    L55 &  16 17 44.77 & $-$50 57 45.13 &    39$\pm$ 2&   $..$&    M7    &   K5    &      $..$    &           G332.809$-$0.132 & J16174477$-$5057451                     \\
    L56 &  16 20 16.21 & $-$50 58 52.28 &    36$\pm$ 3&   $..$&    M5    &   K4    &      $..$    &           G332.809$-$0.132 & J16201621$-$5058522                     \\
    L57 &  16 17 46.28 & $-$50 58 23.54 &    40$\pm$ 2&   $..$&    M7    &   K5    &      $..$    &           G332.809$-$0.132 & J16174628$-$5058235                     \\
    L58 &  16 17 01.80 & $-$50 47 30.49 &    21$\pm$ 2&   $..$&    K3    &   K0    &      $..$    &           G332.809$-$0.132 & J16170180$-$5047303                     \\
    L59 &  16 17 14.37 & $-$50 47 18.83 &    40$\pm$ 2&   $..$&    M7    &   K5    &      $..$    &           G332.809$-$0.132 & J16171437$-$5047188                     \\
    L60 &  16 20 07.49 & $-$50 47 12.40 &    30$\pm$ 0&   $..$&    M2    &   K2    &      $..$    &           G332.809$-$0.132 & J16200749$-$5047123                     \\
    L61 &  16 18 57.83 & $-$50 51 53.86 &    30$\pm$ 0&   $..$&    M2    &   K3    &      $..$    &           G332.809$-$0.132 & J16185783$-$5051538                     \\
    L62 &  16 20 09.46 & $-$50 49 25.61 &    16$\pm$ 0&   $..$&    K1    &   K0    &      $..$    &           G332.809$-$0.132 & J16200945$-$5049256                     \\
    L63 &  16 20 28.27 & $-$50 54 49.35 &    27$\pm$ 0&   $..$&    M0    &   K1    &      $..$    &           G332.809$-$0.132 & J16202826$-$5054493                     \\
    L64 &  16 17 34.73 & $-$50 55 10.61 &    27$\pm$ 1&   $..$&    M1    &   K2    &      $..$    &           G332.809$-$0.132 & J16173473$-$5055106                     \\
    L65 &  16 20 20.94 & $-$50 54 46.75 &    34$\pm$ 1&   $..$&    M4    &   K4    &      $..$    &           G332.809$-$0.132 & J16202094$-$5054467                     \\
    L66 &  16 20 18.66 & $-$50 55 11.53 &    40$\pm$ 1&   $..$&    M7    &   K5    &      $..$    &           G332.809$-$0.132 & J16201866$-$5055114                     \\
    L67 &  16 16 12.48 & $-$51 16 21.77 &    29$\pm$ 2&   $..$&    M1    &   K2    &      $..$    &           G332.809$-$0.132 & J16161247$-$5116217                     \\
    L68 &  16 20 23.06 & $-$50 52 32.90 &    32$\pm$ 0&   30$\pm$ 3&    M3    &   K3    &      $..$    &           G332.809$-$0.132 & J16202305$-$5052328                \\
    L69 &  16 20 26.32 & $-$50 55 10.07 &    37$\pm$ 1&   $..$&    M6    &   K4    &      $..$    &           G332.809$-$0.132 & J16202632$-$5055100                     \\
    L70 &  16 17 13.60 & $-$50 48 11.98 &    49$\pm$ 2&   $..$&  $..$    &   M1    &      $..$    &           G332.809$-$0.132 & J16171360$-$5048119                     \\
    L71 &  16 20 24.71 & $-$50 54 04.46 &    39$\pm$ 1&   $..$&    M7    &   K5    &      $..$    &           G332.809$-$0.132 & J16202470$-$5054044/$$[BRA2016]$$$-$DBS100$-$ysoc\\
    L72 &  16 17 07.84 & $-$50 46 45.47 &    25$\pm$ 0&   $..$&    K5    &   K1    &      $..$    &           G332.809$-$0.132 & J16170784$-$5046454                    \\
    L73 &  16 20 27.69 & $-$50 54 56.21 &    43$\pm$ 3&   $..$&    M7    &   M0    &      $..$    &           G332.809$-$0.132 & J16202772$-$5054564                    \\
    L74 &  16 17 04.37 & $-$50 47 16.44 &    29$\pm$ 2&   $..$&    M2    &   K2    &      $..$    &           G332.809$-$0.132 & J16170437$-$5047164/$$[RA2004a]$$~IRS~9\\
    L75 &  16 17 06.11 & $-$50 47 31.61 &    39$\pm$ 4&   $..$&    M7    &   K5    &      $..$    &           G332.809$-$0.132 & J16170610$-$5047314                     \\
    L76 &  16 20 08.90 & $-$50 48 51.17 &    40$\pm$ 0&   $..$&    M7    &   K5    &      $..$    &           G332.809$-$0.132 & J16200887$-$5048510                     \\

\hline
\end{tabular}
}
\end{table*}

\subsection{Photometric Data}
For every spectroscopically observed star, we searched for counterparts 
in the Two Micron All Sky Survey (2MASS)  Catalog of Point Sources 
\citep{cutri03, skrutskie06}, in the third release of Deep Near Infrared Survey of the Southern Sky (DENIS) data 
available at CDS (catalog  B/denis) \citep{epchtein94}, 
in the Galactic Legacy Infrared Midplane Survey Extraordinaire (GLIMPSE) catalog \citep{spitzer09}, 
and in the  Wide-field Infrared
Survey Explorer (WISE) catalog \citep{wright10b} using the closest match within a search radius of 2\arcsec.
The  II/293 GLIMPSE catalog  from CDS is a combination of the original GLIMPSE-I (v2.0),
GLIMPSE-II (v2.0), and GLIMPSE-3D catalogs. 
WISE counterparts were retained only if their signal-to-noise ratios were larger than 2.5.
We searched  for counterparts in the Version 2.3 of the The Midcourse Space Experiment (MSX) 
Point Source Catalog (PSC)  \citep{price01} using a search radius of 5\arcsec. 
MSX upper limits were removed. 

Due to their brightness, targets are mostly saturated in the VISTA Variables in the Via 
Lactea (VVV) dataset \citep{saito12}.
For a few fainter targets with 2MASS upper limits, 
VVV photometric measurements were adopted.

Using 2MASS positions, we searched for $BVR$ photometry in The Naval Observatory Merged 
Astrometric Dataset (NOMAD) \citep{zacharias05} and for 
Gaia data in the second release (DR2) \citep{gaia2}.
The photometric data are listed in the Appendix A.

In the following sections, Gaia DR2 parallactic distances will be
considered, but only for those objects with fractional 
error on the parallax ($\varpi$) smaller than 20\% and with  weighed errors (UWE)
within the limits advised by \citet{gaiaplx2}. 
A zero point of +0.029 mas was used.

\begin{figure}
\resizebox{0.9\hsize}{!}{\includegraphics[angle=0]{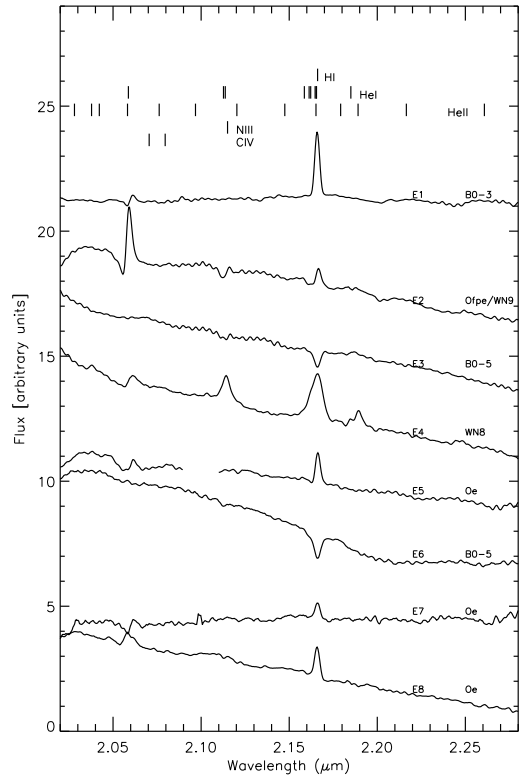}}
\caption{\label{early.fig} $K$-band spectra of detected early-type stars.} 
\end{figure}

\begin{figure}[h]
\begin{center}
\resizebox{1.0\hsize}{!}{\includegraphics[angle=0]{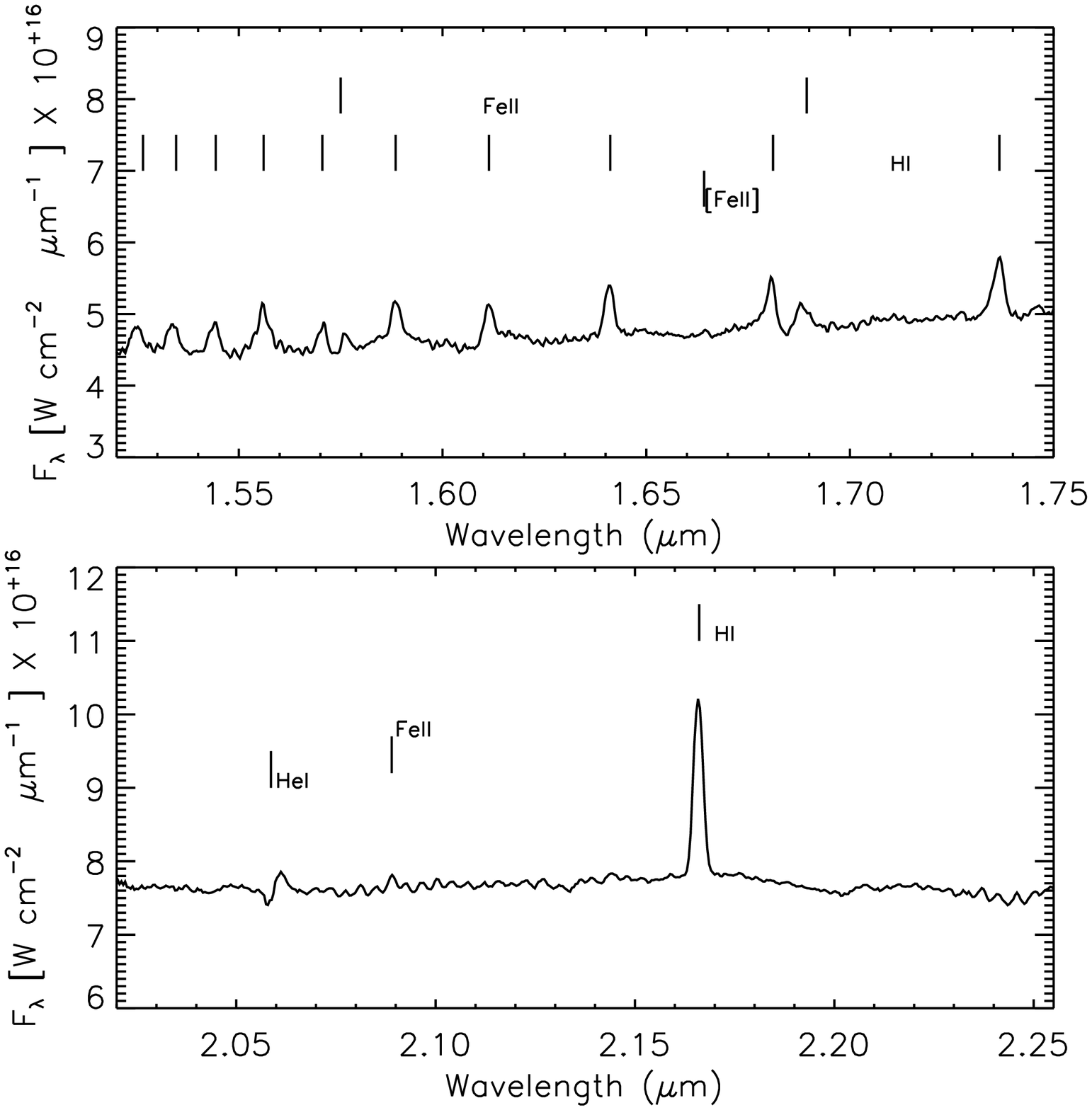}}
\begin{center}
\resizebox{1.0\hsize}{!}{\includegraphics[angle=0]{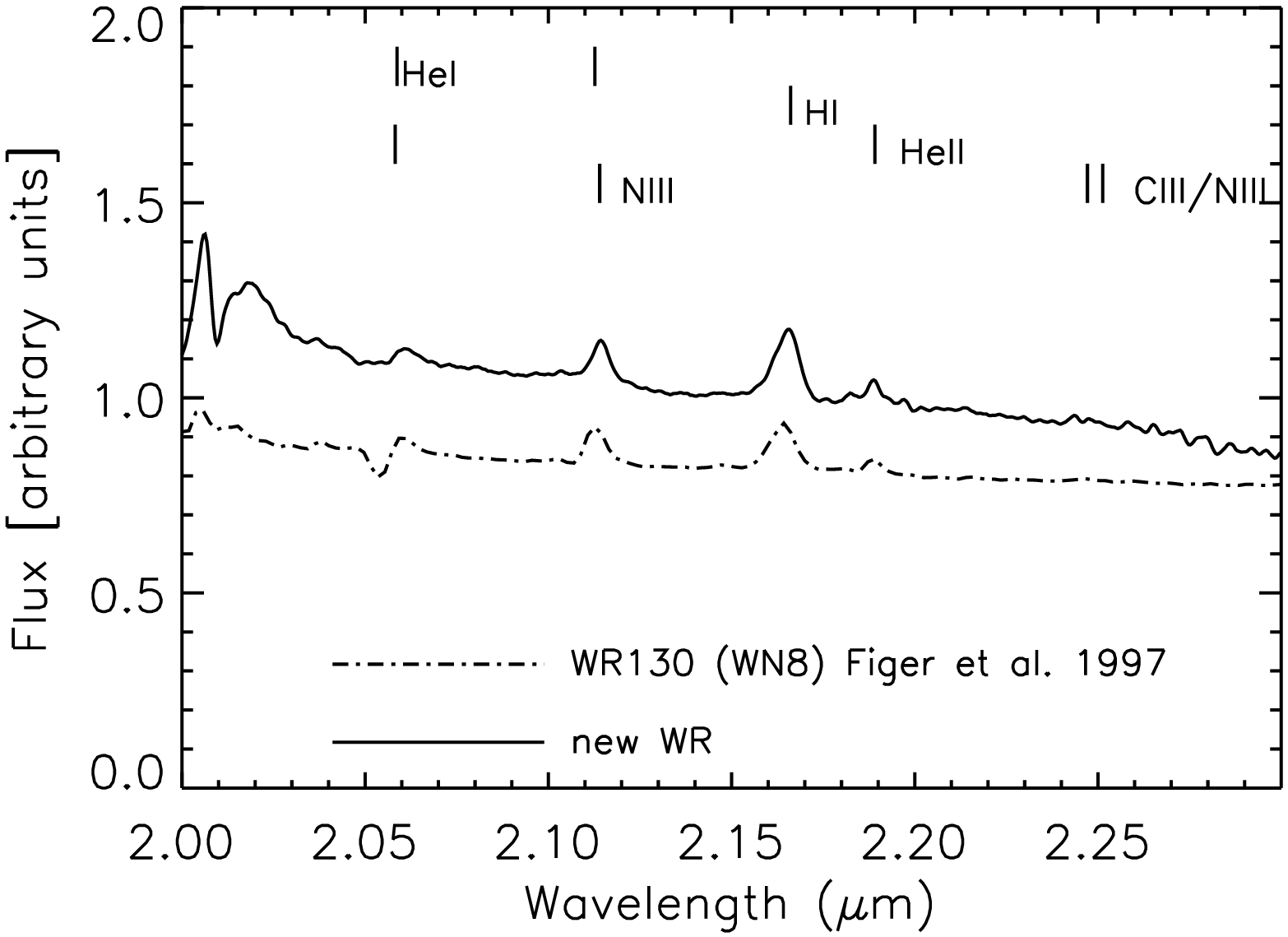}}
\end{center}
\caption{ {\it Upper and middle panels:} \label{berta.sp} $H$- and $K$-bands 
spectra of star E1. \label{wr.fig} {\it Bottom panel:} 
$K$-band spectrum of star E4, a newly discovered WR star, of WN 8 type.
For a comparison, a spectrum of WR 130 from the atlas of \citet{figer97} is  
shown with a dotted line.
} 
\end{center}
\end{figure}

\section{Analysis}
\label{analysis}
\subsection{Spectra of Early-type Stars}

We detected  eight early-type stars,
with associated spectra  shown in Figs.\ \ref{early.fig} and \ref{berta.sp}, 
and listed in Table \ref{obs.early}.

Star E1 was observed in both $H$ and $K$ bands, as shown in Fig.\ \ref{berta.sp} and Table \ref{spberta}.
The $K$ band shows the He I line at 2.058 \um\ in emission, the iron line at 
2.089 \um\ in emission, and the \brg\ in emission.
The $H$-band spectrum displays the  H I lines in emission, along with two \ion{Fe}{II} 
lines at 1.5778 and 1.6878 \um\ \citep{morris96,steele01}. There is also a hint of
[FeII] at 1.6642 \um. 
The absence of He I emission at 2.112 \um\  and the lack of
Mg II lines 2.138 \um\ and 2.144 \um\ suggest \Teff\
 lower than that of  [OMN2000]LS1  
\citep[$\approx 13000$ K][]{clark09} 
and [MDF2011]15 \citep[$\approx 16000$ K][]{messineo11}. 
On the other hand, star E1  has infrared spectral features 
similar to those seen in the low-resolution spectra of S Dor taken  in 1993 September
by \citet{morris96}.  Its spectrum also lacks He I emission.
 S Dor is an LBV and has a variable spectral type that ranges from B3 to 
F type \citep{massey00}. The spectrophotometric work of   \citet{wolf97} reports
a minimum of S Dor  between the  beginning of 1993 and the end of 1994.
\citet{vangenderen97}  provide  a spectral type of B3 (16000 K) 
for observations taken on 1993, April 22.
We conclude that the temperature of star E1 must be in the range of 13000--16000 K (B5-B3).

The spectrum of star E2   shows a strong  \ion{He}{I} line at 2.058 \um, 
a \brg\ line in emission, a \ion{He}{I} line at 2.112 \um\ in absorption, 
and \ion{N}{III} at 2.115 \um\ in emission. 
The ratio of the \ion{He}{I} and \brg\ lines is typical of  Ofpe/WN 9 stars, 
which are massive evolved stars 
in transition to the  WR phase \citep[e.g.,][]{najarro97,messineo09}.

The spectra of stars E3 and E6 have  \ion{He}{I} lines 
at 2.112 \um\ (in absorption), and \brg\ lines (in absorption). 
The lack of \ion{He}{II} lines suggests that  these are  B0-5 stars (B0-8I or B0-3V). 

The spectrum of star E4  has the \ion{He}{I} line  at 2.058 \um\ in emission, 
the \ion{N}{III} complex at  2.115 \um\ in emission, the \brg\ line 
in emission,  and the \ion{He}{II} line at  2.189 \um\ in emission. 
This broad-line spectrum resembles  that of  WR 130 \citep{figer97}, 
which is classified as a  WN 8 star by \citet{vanderhucht01}.

The spectra of  stars E5, E7, and E8  display  \brg\ lines in emission.
Because $H$-band spectra are not available, we will call them OBe stars. 
Star E5  appears as  a blend of two stars, 
which are surrounded by a  gaseous ring (see charts in Appendix C). 
Stars E5 ($[RA2004a]$~IRS~1), E7 ($[RA2004a]$~IRS~2), and 
L74 ($[RA2004a]$~IRS~9) have been analyzed with infrared photometry by
\citet{romanlopes04}. For stars E5 and  E7,
the authors have inferred spectral types O5 and mid-O, respectively.

\subsection{Spectra of Late Types}
\Ks-band spectra of late-type stars are characterized by CO bands in absorption.  
The equivalent widths of the CO bands, EW(CO), 
linearly increase with decreasing temperatures. EWs were determined 
with a line taken from 2.285 \um\ to 2.307 \um\  (high-resolution) or
from 2.285 \um\ to 2.315 \um\  (low-resolution) and  using two 
estimates of the continuum; one continuum  was taken as the average flux density
from 2.285 \um\ to 2.290 \um\ \citep[][]{figer06}  
and a second continuum was taken as a linear fit 
to the flux densities  from 2.250 \um\ to 2.257 \um, from 2.270 \um\ to 2.277 \um, and
from 2.285 \um\ to 2.290 \um\ \citep{ramirez00}. 
Spectral types were derived by  comparing the obtained EWs with those of stellar 
templates of K and M types  \citep{kleinmann86}. 
Spectral types derived from the seven low-resolution spectra
are typically accurate within two subclasses \citep{figer06,messineo17}.

For the 76 spectra with the high-resolution mode, 
the transmission quickly drops down longward of  2.3 \um, 
so EW(CO) from the high-resolution mode
are  more uncertain than those from the low mode.
We recalibrate the  EW from the high-resolution spectra 
with those  measurements in common with  the  low-resolution spectra.
Because the EW(CO) and temperatures of giants 
and supergiants follow two different linear relations
\citep{blum03,figer06}, spectral types were derived for giants and supergiants.
Detected late-type stars are listed in Table \ref{table.latespectra}. 
A few targets with spectra that show curved continua 
are marked in the Table as Mira stars.

In the catalog by \citet{skiff14}, GSC 8323-1211 (L17) is reported as an M2 star;
we estimated an M6 type by assuming a giant.
L48 (M5) coincides with   $\text{[BRA2016]-DBS100-Obj4}$ and L71 (M7) with 
the candidate young stellar object $\text{[BRA2016]-DBS100-ysoc}$ \citep{borissova16}.

\section{The Core of G332.809$-$0.132: ``The Bridge'' }

\label{region1}

\begin{table*}
\caption{\label{table.xray}  List of the Three SNRs and of One Candidate Pulsar Wind Nebula  projected over G332.809$-$0.132-Bridge. }
\begin{tabular}{llllllll}
\hline
\hline
Name                       & Ra[J2000] & Dec [J2000] & Diam      & Related pulsar           & PSR distance & Ref.\\
                           & [hh mm ss]& [hh mm ss]  &  [\arcmin]&                           &              &      \\
\hline
SNR RCW 103 (G332.4-0.4)   & 16 17 33   & $-$51 02 00 & 10        & 1E 1613.7-5055                           &       3.3$^{+1.3}_{-0.2}$   & 2,5,6,7,9,10,11, 13\\
                           & 16 16 29.89& $-$50 17 14.9 &  &           PSR J1616$-$5017     &       3.5 & 12 \\
\hline
G332.0+00.2                & 16 13 17  & $-$50 53 00 & 12  	     &                                          &              &2,5\\
Kes 32/ G332.4+00.1        & 16 15 20.0& $-$50 42 00 & 15        & PSR J1614-5048/  PSR B1610-50 (?)                       &        7.3,  $7.5^{+3.5}_{-0.9}$    &  2,5,8,11, 12, 13\\
HESS J1616$-$508             & 16 16 23  & $-$50 53 48 & 8.16      & CXOU J161729.3-505512/PSR J1617-5055  (?) &       6.5      & 1,2,3,4, 12\\
                           & 16 14 45.7& $-$51 44 49 &   &      PSR  J1614$-$5144      &       6.1  &  12\\
                           & 16 16 30.9& $-$51 09 17 &  &         PSR  J1616$-$5109      &       6.8  &  12\\
\hline
\end{tabular}
\begin{list}{}
  \item {\bf References.}
  1 = \citet{aharonian06};    2 = \citet{hare17};         3 = \citet{kargaltsev09};  
  4 = \citet{chang08}; 5 = \citet{green14};   6 = \citet{liu15};  7 = \citet{rea16};  
  8 = \citet{brinkmann99}; 9 = \citet{ho17}; 10 = \citet{borghese18};  11=\citet{acero16}; 
   12= \citet{manchester05};  13=\citet{stafford19}.
\end{list}
\end{table*}

\begin{table*}
\caption{\label{table.mbolearly} Parameters of Detected Early-type Stars.  }
\begin{tabular}{@{\extracolsep{-.08in}}rlrrrrrrrrrrllr}
\hline 
\hline
{\rm ID} &Sp. Type & \Ks$_o$ & $J-H$  &  $H-$\Ks & \Aks($JH$) &\Aks($J$\Ks) &\Aks($H$\Ks) & \BCKs  &\Mbolone  & $Q1$    & $Q2$    &MD & MD($\varpi$) &\\ 
         &         &  [mag]  &  [mag] &  [mag]   &   [mag]    &    [mag]    &  [mag]      &  [mag] &   [mag]  &  [mag]  &  [mag]  &  [mag] & [mag]  \\
\hline 
    E1 &       B3-5& 5.24 & 0.00 &-0.02 & 1.54 & 1.81 & 2.30 &  $-$0.75 & $-$6.16 $\pm$  0.57 &  $-$0.90 $\pm$    0.17 &  $-$3.35 $\pm$    0.00 & 10.65 $\pm$ $^{  0.18}_{ 0.26}$  & $..$    & \\
   E2 &  Ofpe/WN 9& 5.12 &$-$0.10 &$-$0.07 & 0.93 & 0.94 & 0.98 &  $-$2.91 &$-$10.45 $\pm$  0.71 &  $-$0.04 $\pm$    0.16 &  $-$1.04 $\pm$    0.01 & 12.66 $\pm$ $^{  0.41}_{  0.50}$  & $..$    & \\
   E3 &      B0-5& 5.13 &$-$0.03 &$-$0.02 & 0.95 & 0.94 & 0.93 &  $-$2.12 & $-$9.64 $\pm$  0.71 &     0.02 $\pm$    0.10 &  $-$0.22 $\pm$    0.01 & 12.66 $\pm$ $^{  0.41}_{  0.50}$  & $..$    & \\
   E4 &       WN 8& 6.90 &   0.02 &   0.11 & 0.53 & 0.47 & 0.37 &  $-$3.40 & $-$9.16 $\pm$  0.71 &     0.01 $\pm$    0.09 &  $-$1.33 $\pm$    0.01 & 12.66 $\pm$ $^{  0.41}_{  0.50}$  & $..$    & \\
   E5 &        Oe& 8.53 &$-$0.11 &$-$0.10 & 1.20 & 1.32 & 1.53 &  $-$4.38 & $-$8.51 $\pm$  0.71 &  $-$0.34 $\pm$    0.14 &                  $..$  & 12.66 $\pm$ $^{  0.41}_{  0.50}$  & $..$    & \\
   E6 &      B0-5& 9.91 &$-$0.03 &$-$0.02 & 0.19 & 0.20 & 0.23 &  $-$2.12 & $-$4.87 $\pm$  0.71 &  $-$0.05 $\pm$    0.07 &  $-$1.43 $\pm$    0.03 & 12.66 $\pm$ $^{  0.41}_{  0.50}$  & 12.51   & \\
   E7 &        Oe& 8.56 &$-$0.11 &$-$0.10 & 1.59 & 1.79 & 2.14 &  $-$4.38 & $-$8.48 $\pm$  0.71 &  $-$0.60 $\pm$    0.11 &  $-$5.12 $\pm$    0.04 & 12.66 $\pm$ $^{  0.41}_{  0.50}$  & $..$    & \\
   E8 &        Oe& 9.48 &$-$0.11 &$-$0.10 & 0.81 & 0.78 & 0.73 &  $-$4.38 & $-$7.56 $\pm$  0.71 &     0.16 $\pm$    0.16 &    $..$                & 12.66 $\pm$ $^{  0.41}_{  0.50}$  & $..$    & \\
\hline
P1$^a$ &       WC 8& 9.69 & 0.05 & 0.38 & 1.85 & 1.66 & 1.30 &  $-$3.60 & $-$6.57 $\pm$  0.71 &   0.01 $\pm$    0.21 &  $-$1.88 $\pm$    0.02 & 12.66 $\pm$ $^{  0.41}_{  0.50}$   & $..$  &\\
P2$^a$ &       WN 7& 8.40 & 0.02 & 0.11 & 0.40 & 0.43 & 0.47 &  $-$3.90 & $-$8.16 $\pm$  0.71 &  $-$0.26 $\pm$    0.10 &  $-$2.72 $\pm$    0.01 & 12.66 $\pm$ $^{  0.41}_{  0.50}$   & $..$  &\\
end{tabular}
\end{tabular}
\begin{list}{}{}
\item[{\bf Notes.}] Identification numbers, which are taken from Table  \ref{obs.early}, 
are followed by  spectral types,   de-reddened \Ks\ magnitudes,
three estimates of total  extinction (\Aks(JH), \Aks(JK), \Aks(HKs)), \BCKs, 
bolometric magnitudes (\Mbol~ = \Ks$_{\rm abs}$ + \BCKs),  $Q1$ and $Q2$ parameters.
Appended to the table, are the values for the two previously known WR stars.~\\
($^a$) P1 = $[$SMG2009$]$~1059$-$34,  2MASS J16143723$-$5126263; 
P2 = WR~74, 2MASS J16161381$-$5136417.
\end{list}
\end{table*}

\begin{table}
\caption{\label{wrrsgknown} List of Isolated Known Massive Stars in the Direction of ``The Bridge''.}
\begin{tabular}{@{\extracolsep{-.07in}}llllll}
\hline
\hline
ID & 2MASS-ID            &    Ra[J2000]   & Dec[J2000]   &  Sp. type   & References \\
   &                 &    (hh mm ss)  & (dd mm ss)   &             &   \\
\hline
 P1  &   J16143723-5126263        & 16:14:37.24   &$-$51:26:26.33   &      WC 8 & 1,2  \\
 P2  &   J16161381-5136417        & 16:16:13.81   &$-$51:36:41.77   &      WN 7 & 1,3  \\
 P3  &   J16125464-5044510        & 16:12:54.65   &$-$50:44:50.99   &  K3Ib-II & 1,4 \\
 P4  &   J16151992-5026463        & 16:15:19.94   &$-$50:26:46.12   &    M2Iab & 1,4 \\
 P5  &   J16164067-5014370        & 16:16:40.70   &$-$50:14:37.12   &      M3I & 1,4 \\
 P6  &   J16191454-5024243        & 16:19:14.53   &$-$50:24:24.45   &    M5.5I & 1,4  \\
 P7  &   J16191948-5131264        & 16:19:19.48   &$-$51:31:26.52   &     M3Ia & 1,4  \\
 P8  &   J16202307-5104580        & 16:20:23.06   &$-$51:04:58.12   &     M1Ib & 1,4  \\

\hline
\end{tabular}
\begin{list}{}
\item {\bf References.} 1 = \citet{skiff14}; 2 = \citet{shara09}; 3 = \citet{houk78}; 
4 = \citet{messineo19}.
\end{list}
\end{table}

\begin{figure}
\begin{center}
\resizebox{1.0\hsize}{!}{\includegraphics[angle=0]{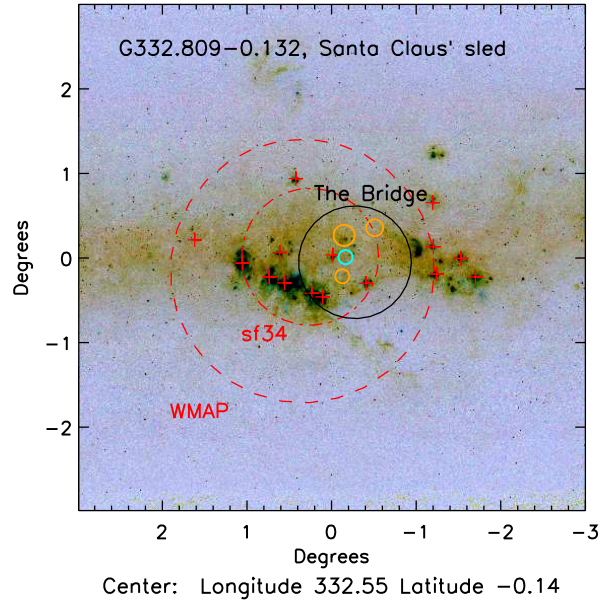}}
\end{center}
\caption{\label{g332mapMSX} Composite map of the G332.809$-$0.132 region with 
MSX data in Galactic coordinates. Longitude is on the $x$-axis and latitude on 
the $y$-axis. The blue is the $A$ band, the green is the $D$ band, and the red  
is the $E$ band. Orange circles indicate SNRs listed by \citet{green91,green14}. 
The  cyan circle shows the positional uncertainty of the $\gamma$-ray source
HESS $J1616-508$. The large black circle  encloses the central area (r = 40\arcmin), 
referred to as ``The Bridge''.
Red circles show the locations and semimajor axis
of the WMAP source from Table 1 of \citet{rahman10}
and the associated SF region n.34 from their Table 2.
\Vlsr\ from $-38$ to $-$60 \kms\ from \citet{caswell87} are marked with red plus sign symbols.}
\end{figure}

Because of the estimated number of Lyman continuum photons,
\citet{rahman10} reported  G332 as one of the 14 extended giant  
complexes of the Galaxy identified with WMAP data (with a semi-major axis of 93\farcm6),
 where one-third of Galactic free-free emission resides.
G332.809$-$0.132 is the associated  star forming complex at 3.4 kpc
(with a semi-major axis of 48\arcmin), object n. 34 in  Table 2 of \citet{rahman10}.
We estimated that the 14 extended regions (Table 1 of Rahman et al.) cover 
27\% of the disk ($360^\circ \times 1^\circ$)
and include 50\% of the Conti \&\ Crowther giant \HH\ regions, 
while the identified star-forming (SF) complexes (Table 2 of Rahman et al.)
cover 4\% of the disk and include 33\% of 
Conti \&\ Crowther giant \HH\ regions.\footnote{ 
The giant HII region complexes identified by Conti are strong radio and mid-infrared (MIR) 
emitters. The presence of MIR emission identifies a radio source as an HII region 
(with thermal emission). SNRs, which are associated with nonthermal (synchrotron) 
radiation, show no or very little MIR emission.}
Later, in Section \ref{associations and clusters?}, we will see that this complex
coincides with  the Nor OB4/RCW 103 association 
\citep[with 34 stellar members listed in][]{melnik17}.
At mid-infrared wavelengths, the complex  appears shaped like ``Santa Claus's sled''. 
In the MSX composite image, warm dust emission ($A$ band) traces
the right and left tracks of ``the sled'', while the central area 
(hereafter called ``the Bridge'') is particularly rich in high-energy sources 
(see Figs. \ref{g332mapMSX} and \ref{g332map}).
We defined ``the Bridge''  by visually inspecting of the MSX composite image
 and locations of the three SNRs, by wanting to center and delineate a good portion
of the complex  and to exclude the two strong ridges of 8 \um\ emission on the two sides. 
The  radius used (40\arcmin) would allow us to sample the evolved massive
stars along the line of sight and to avoid regions of  higher nebulosity, which could be
perhaps  younger -- triggered seeds of  star formation.
One can imagine that supernova explosions and  ionizing flux 
from  massive  stars  dominate the central area.

For this complex, \citet{rahman10} estimated a total number of Lyman continuum photons,
$Q_0$, $5.0 \times 10 ^ {51}$ s$^{-1}$. This number corresponds to an incredibly 
large number of OB stars; it is equivalent to 1000 O7 dwarfs \citep{martins05}
or   250 O7 supergiants \citep{martins05},
or   200 O7 supergiants  \citep{panagia73}. {\it  Where are they?} 
 Only 34 massive stars  have been 
reported in  RCW 103 \citep{melnik17}. 
Because the complex is  bubbling with supernova explosions 
(see Section \ref{highenergy}),  
we can assume that a large fraction is made up of evolved massive stars. 
The quest to detect the earlier generation of massive stars
that is undergoing final destruction via supernovae is an imperative
to understand the intricate mechanism of sequential star formation and 
the possible overlap along the same line of sight of multiple clouds.
With this aim in mind, we tried to identify candidate massive OB stars in this
complex and to count them. 
Indeed, a second smaller molecular cloud (a few arcminutes wide) was detected along 
the same line of sight by \citet{rahman10}; it is object number 35  
with $V_{\mathrm LSR} \approx -91$ \kms, i.e with a near-kinematic distance of about 5 kpc.
Furthermore, by looking at the original velocity measurements around 
$-52$ \kms, covering this complex and
listed in \citet{caswell87}, we notice two groupings of velocities  
along the  structure;
in most positions, velocities  are from $-$41.0 to $-$48.4 \kms, 
but in the southern central area there are four positions with velocities 
from $-$52.0 to $-$57.3 \kms\ with only one position with both  velocities. 
Toward IRAS 16132$-$5039, two  absorption lines appear 
in the \ion{H}{I} spectrum of \citet{corti16}, one at $\approx-52$ 
and  the second deeper one at $-42$ \kms, with  a map of the
$-42$ \kms\ component centered  on the  IRAS source.
 $\Vlsr = -46.7\pm2.6$ \kms\ for the high-density 
tracing CS molecule   was  measured  toward the IRAS source  \citep[][]{bronfman96}. 
\Vlsr $= -42$ or  $-47$ yields a near-kinematic 
distance of $2.9$ or $3.1$ kpc \citep{reid09}. This is
a relative shift in velocity of 5-10 \kms\ relative to that used by 
\citet[$-52\pm7.0$ \kms, $\approx 3.4$ kpc,][]{murray10} to identify 
the G332.809$-$0.132 complex.
It is unclear at this stage if the two velocities ($-46$ and $-52$) 
are tracing two different clouds 
along the same line of sight, or a peculiar velocity structure of 
the same complex like in W33 \citep{immer14}. 
Both velocity components are here regarded as belonging to the same complex in 
the Scutum-Crux arm (Fig. \ref{whereareyou}).

\subsection{High-energy Sources in ``The Bridge''}
\label{highenergy}

\begin{figure*}
\begin{center}
\resizebox{0.8\hsize}{!}{\includegraphics[angle=0]{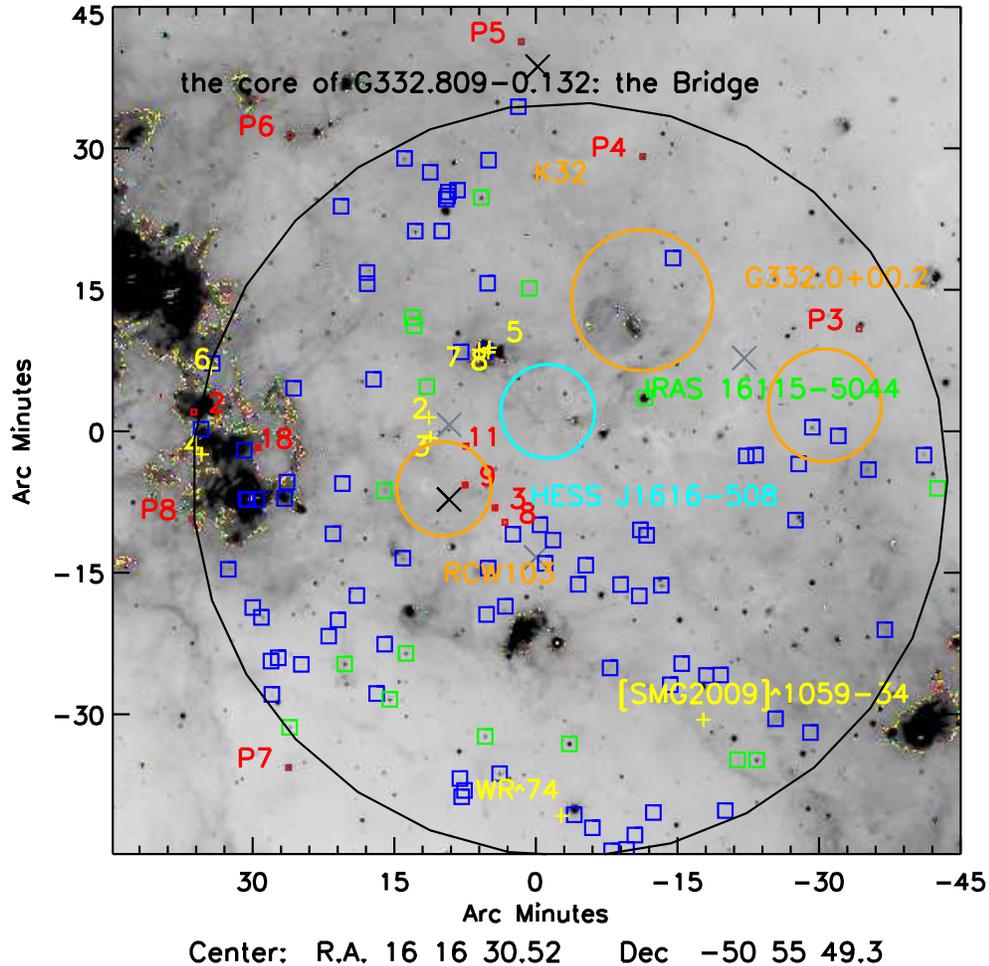}}
\end{center}
\caption{\label{g332map}  
A zoomed-in image of the central $1^\circ \times1^\circ$  of G332.809$-$0.132
(``The Bridge''). The gray image represents the  WISE $W3$ band. 
North is up and East to the left.
Detected early-type stars are marked with yellow crosses ( Table \ref{table.mbolearly}), 
late-type stars with red boxes (Table \ref{tablelatelum}). 
Green squares indicate    free-free emitters with \Ks $< 10.0$ mag 
and $J-$\Ks\ from  0.7 to 3.0 mag  (from Tables \ref{obknown} and \ref{table.wr}), 
while blue squares indicate   normal early-type stars with  \Ks $< 10.0$ mag and $J-$\Ks\ 
from  0.7 to 3.0 mag  (from Tables  \ref{obknown} and \ref{table.ob}).
Orange circles indicate SNRs listed by \citet{green91,green14}. 
The cyan circle shows the positional uncertainty of the $\gamma$-ray source
HESS $J1616-508$. 
Black crosses (at $\approx 3$ kpc) and gray crosses (at $\approx 6$ kpc) 
mark the location of X-ray point sources listed in Table \ref{table.xray}.
} 
\end{figure*}

``The Bridge'' region is rich in high-energy sources, with three
known SNRs and one TeV emitter, HESS J1616$-$508
\citep{green14,hare17},  as listed in {\bf Table \ref{table.xray}.}

The SNR RCW 103 has a diameter of 10\arcmin\
and is reported to be at  the likely distance of 3.3 kpc \citep[e.g.,][]{caswell75,xing14},
and therefore, associated with the star-forming complex G332.809$-$0.132.
RCW 103 is a young SNR ($\approx 2000$ yr) with a known 
central compact object, 1E 161348$-$5055. 
The recently detected X-ray-burst activity points to a magnetar 
\citep{ho17,borghese18} and therefore to a massive 
star progenitor of 18-20 \Msun\ \citep{frank15}.

Kes 32 (G332.4+00.1) is an SNR with a diameter of 15\arcmin, about 20\arcmin\ 
north of RCW 103,  
while SNR G332.0+00.2 is 15\arcmin\ west of Kes 32 and has a diameter of 12\arcmin.
The PSR J1614$-$5048 is located between Kes 32 and SNR G332.0+00.2.
The high column density indicates that PSR J1614$-$5048 is located behind ``The Bridge''
at  7.3 kpc \citep{brinkmann99}, or it could 
experience massive local extinction (in ``The Bridge'').

The HESS source J1616$-$508,  detected by \citet{aharonian06},
is one of the brightest HESS sources with a flux (1-10 TeV) of 
1.7 $\times 10^{-11}$ erg cm$^{-2}$ s$^{-1}$. It is  possible
that the TeV emission is associated with a pulsar wind nebula powered by  
the off-centered PSR, PSR J1617$-$5055, 
which coincides with CXOU J161729.3$-$505512 \citep{chang08,kargaltsev09,hare17}.
In any case (pulsar wind nebula or hadronic processes in SNR), 
the HESS J1616$-$508  is another tracer for a massive evolved star.
It is located only a few arcminutes away from RCW 103 (see Fig. \ref{g332map}), 
and both sources are likely to contribute to the $\gamma$-ray emission
detected by Fermi  \citep{xing14}.
For PSR J1617$-$5055, \citet{kargaltsev09} estimated a distance of 6.5 kpc from the 
dispersion measure and the Galactic electron density distribution
model by \citet{taylor93}. This confirms the superposition of two
molecular clouds along the line of sight ($\approx 3$ and $\approx 6$ kpc).
The ATNF pulsar catalog  provides  three additional pulsars \citep{manchester05}\footnote{
About 2500 pulsars are listed in the catalog of \citet{manchester05}.}.
PSR J1616$-$5017 is at an estimated distance of 
3.5 kpc, compatible with  that of RWC 103 in the Scutum-Crux arm.
PSRs J1614$-$5144 and J1616$-$5109 are at 6.1 and 6.8 kpc, which means
they may be in the Norma arm along with  HESS  J1616$-$508.

\begin{figure}[!]
\begin{center}
\resizebox{0.9\hsize}{!}{\includegraphics[angle=0]{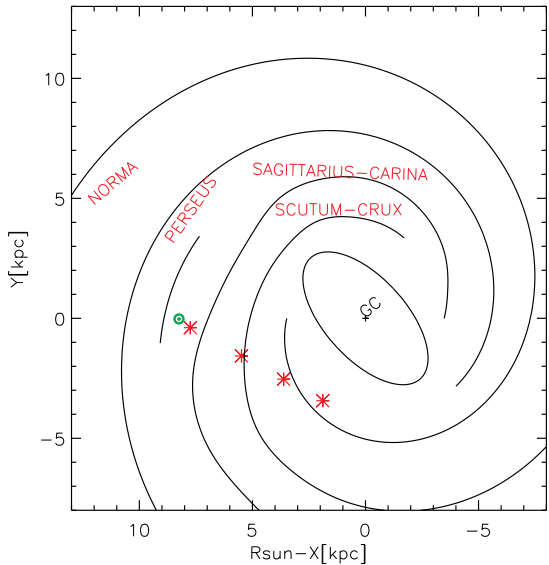}}
\end{center}
\caption{\label{whereareyou} Top view of the Galactic disk, with a sketch of the central Bar and 
spiral arms. Spiral arms are taken from \citet{lazio02}.
The Galactic center is marked with a cross and our location with a green dot.
Red asterisks mark distances of 850 pc ($\approx -8$ \kms), 
3400 pc ($-52$ \kms), 5500 pc ($-102$ \kms), and 7450 pc ($-140$ \kms) 
along the line of sight of  G332.809$-$0.132. 
}
\end{figure}

\begin{figure*}
\begin{center}
\resizebox{0.33\hsize}{!}{\includegraphics[angle=0]{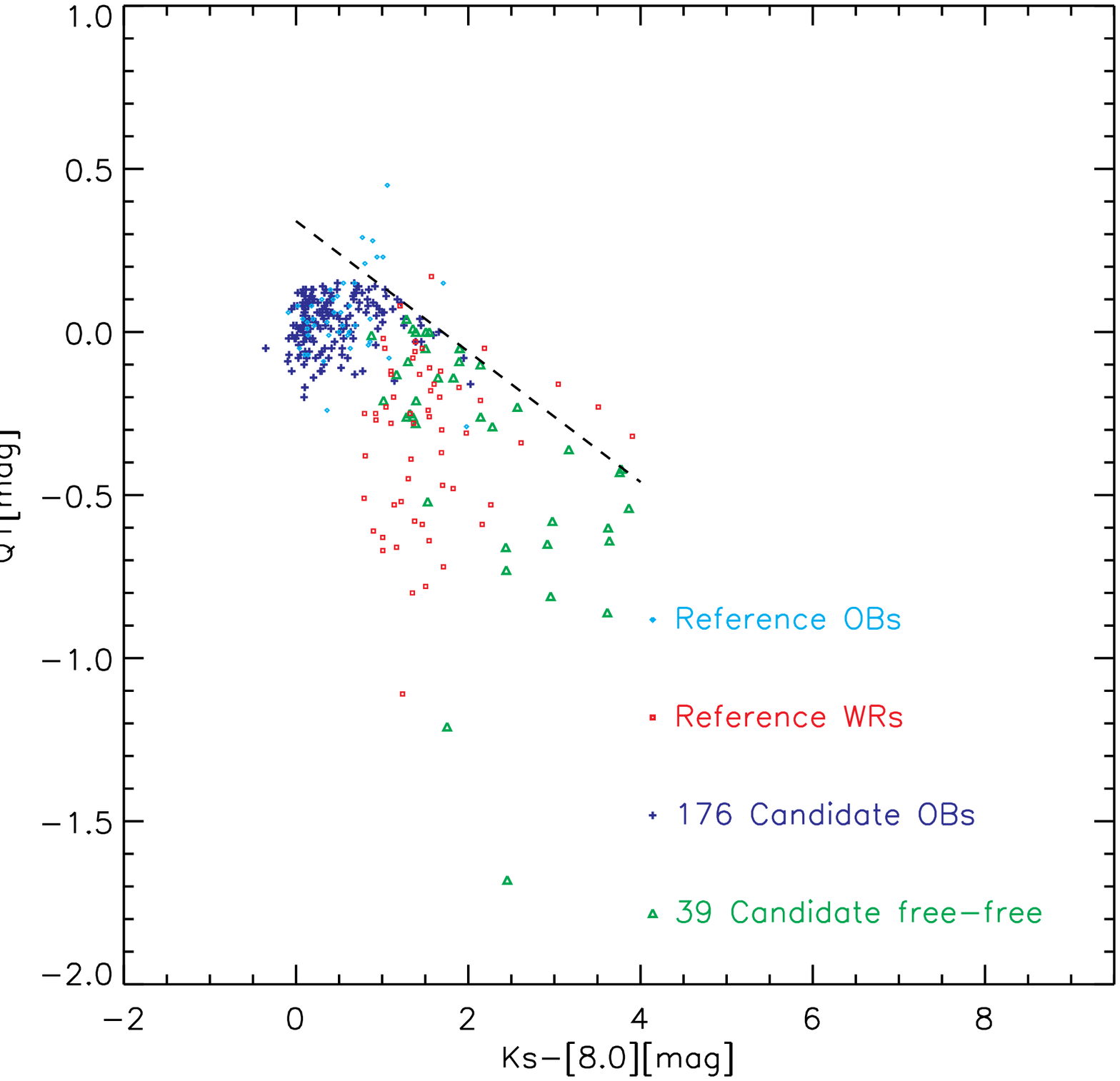}}
\resizebox{0.33\hsize}{!}{\includegraphics[angle=0]{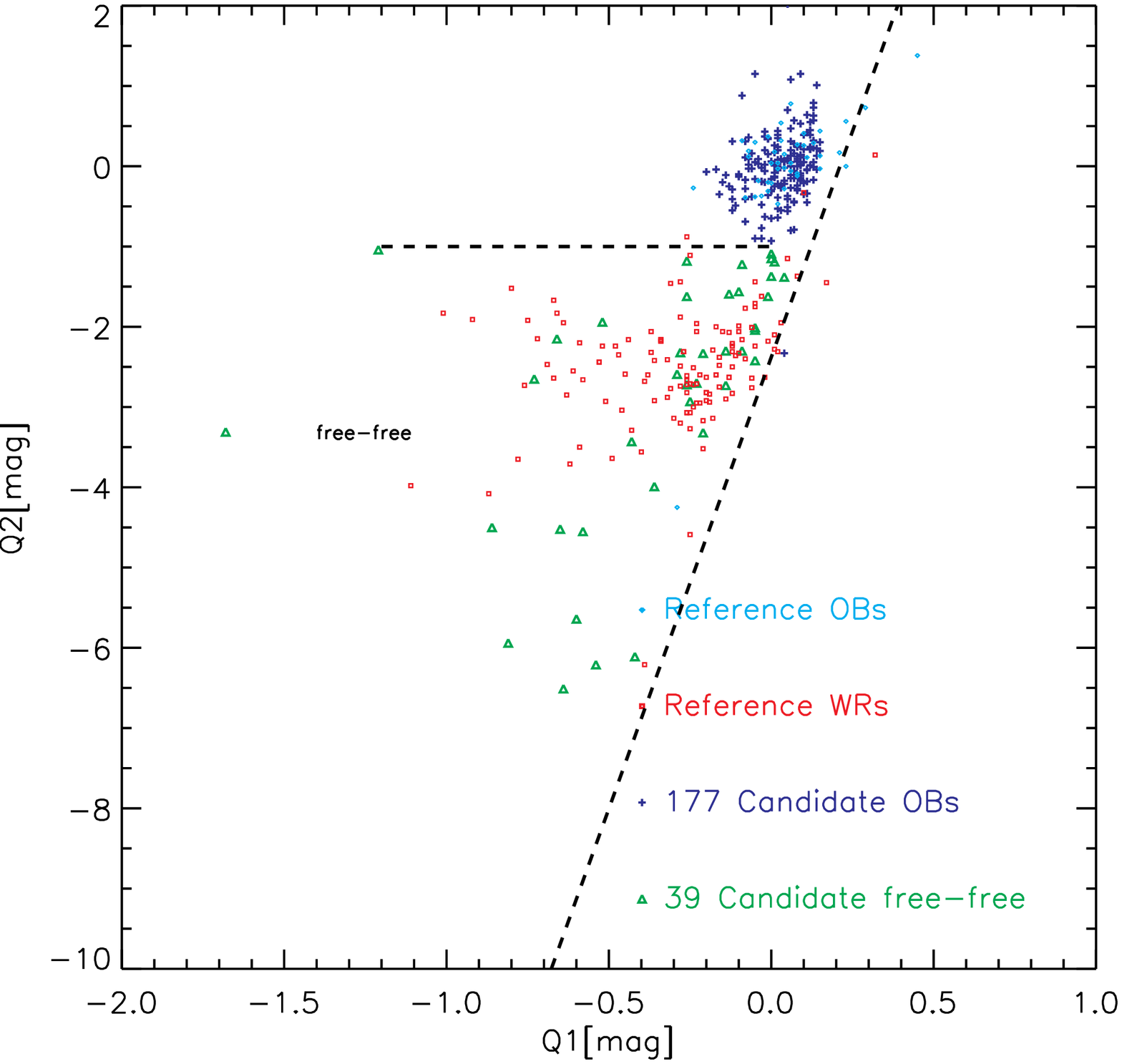}}
\resizebox{0.33\hsize}{!}{\includegraphics[angle=0]{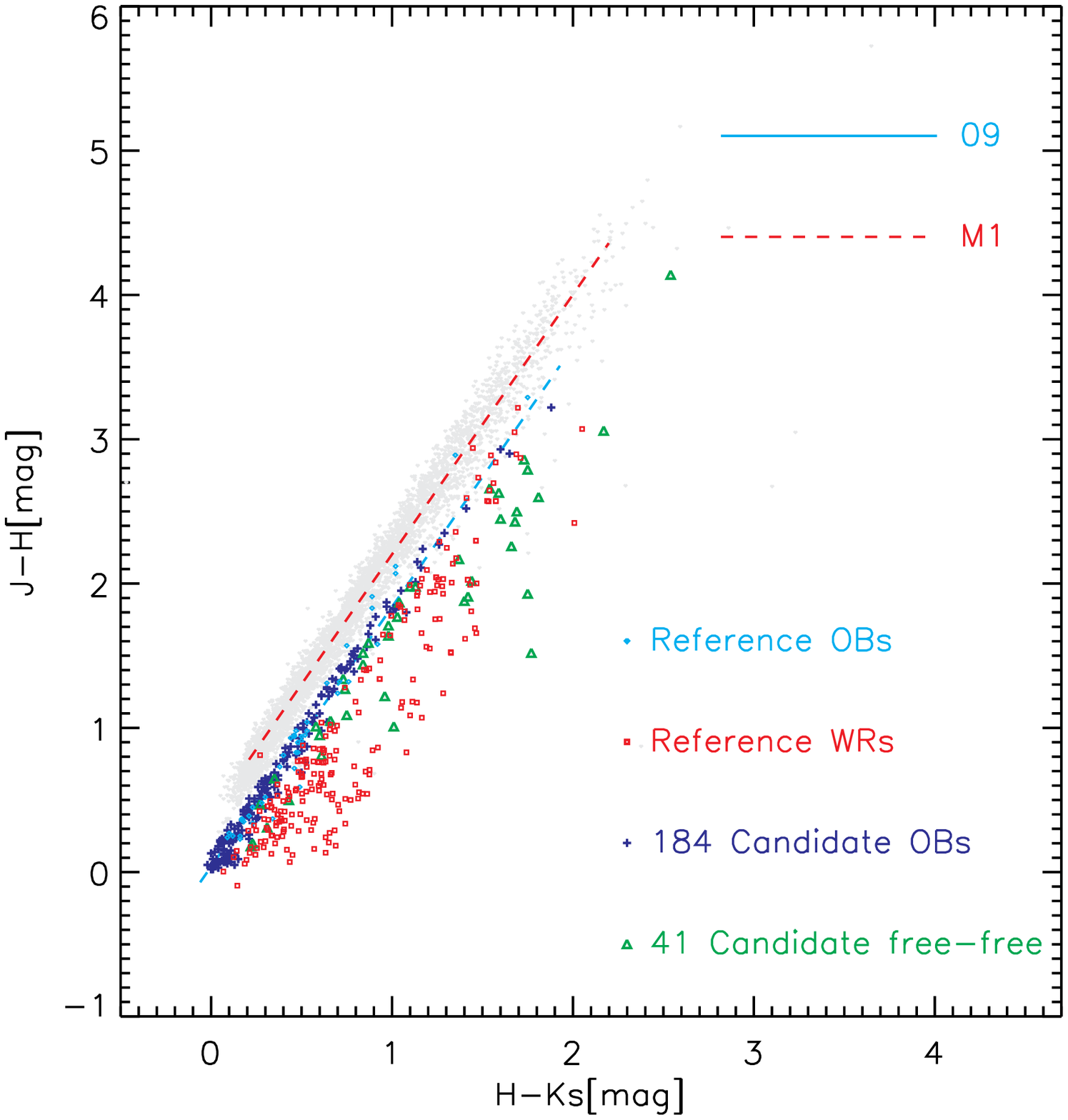}}
\caption{\label{free.fig}  
{\it  Left panel:} $Q1$ versus \Ks$-[8.0]$ diagram. 
The dashed curve marks the line below which free-free emitters are located. 
The region above it is mainly populated by late-type stars 
\citep[for more details, see][]{messineo12}.
{\it Middle panel:} $Q2$ versus $Q1$ diagram. Free-free emitters are located below 
the line $Q2~=~-1$ mag and to the left of the oblique curve 
\citep[for more details see][]{messineo12}. 
{\it  Right panel:}  $J-H$ versus $H-$\Ks\ diagram. The location of M1 stars 
with \Aks\ from 0 to 3 mag is marked with a long-dashed red line. The location of  O9 stars 
with \Aks\ from 0 to 3 mag is indicated with a  long-dashed cyan line.  
 Gray dots show 2MASS datapoints in ``The Bridge'' area with \Ks $<10$ mag.
\\
In all panels, red squares indicate the positions of reference WRs and 
cyan diamonds those of reference OB stars \citep{messineo12}. 
 Thirty-five new, very probable, free-free emitters  from Table \ref{table.wr} 
are located and marked with green triangles;
the other  122 stars that are likely  early-type stars  from Table \ref{table.ob} 
are indicated with  blue crosses.
}
\end{center} 
\end{figure*}

\begin{figure*}
\begin{center}
\resizebox{1.0\hsize}{!}{\includegraphics[angle=0]{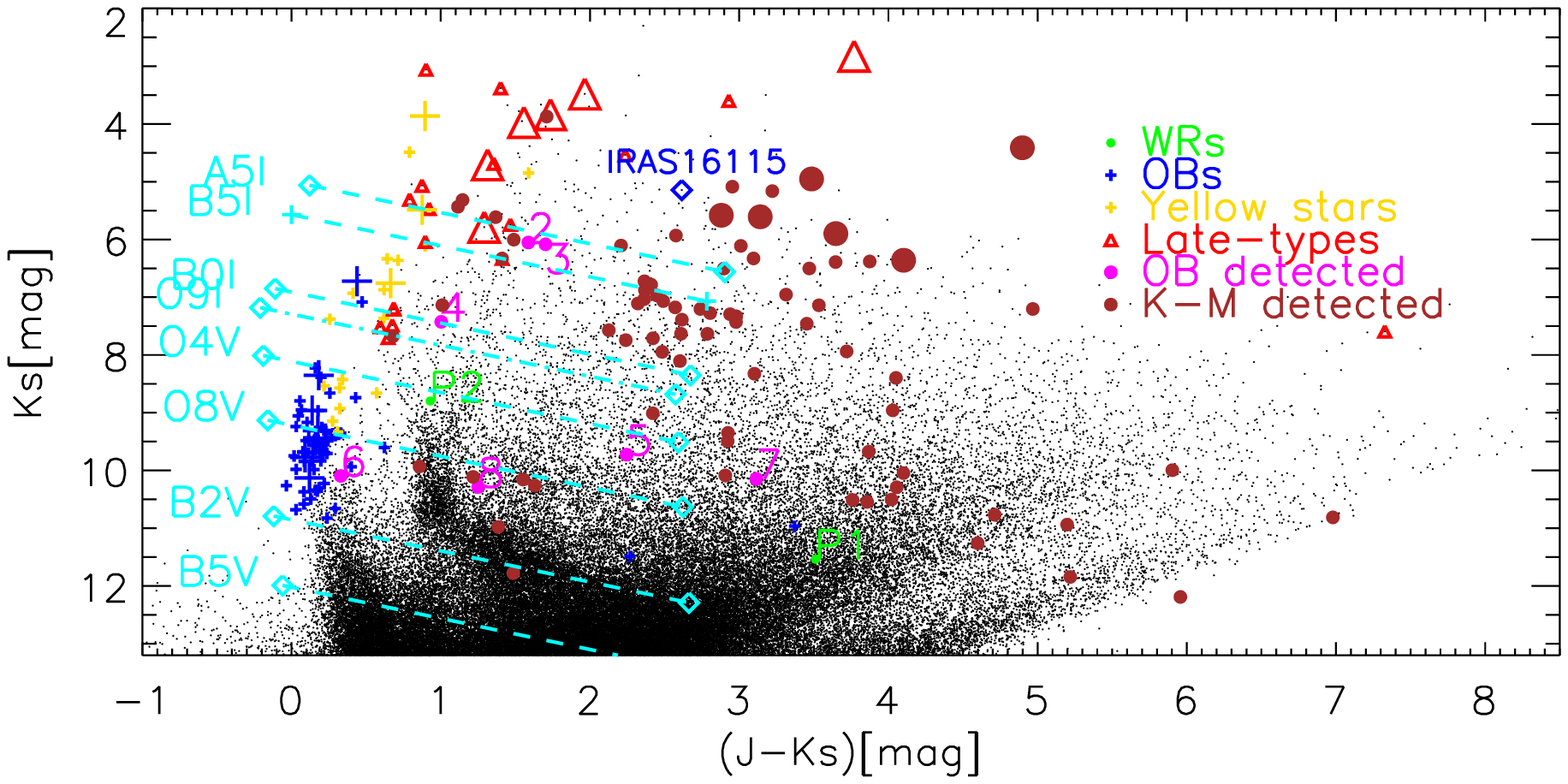}}
\caption{\label{cmdmassSKIFF.fig}  
\Ks\ versus $J-$\Ks\ diagram of 2MASS stars in the direction of 
``The Bridge''. The locations of O4, O8, B2, and B5 dwarfs for DM = 12.66 mag
and \Aks\ = 0.0 mag are marked with cyan diamonds 
\citep[][and references therein]{messineo11,messineo14a}. 
Dashed lines connect the zero-extinction 
locations to those at \Aks\ = 1.5 mag. 
Red triangles indicate the locations of previously known late-type stars, 
gold pluses those of yellow stars, and blue plus signs those of known early-type stars,  
(see Table \ref{obknown}).  
OB stars from Table \ref{obs.early} are marked and 
labeled in magenta and K-M stars from Table \ref{table.latespectra} 
are indicated with brown filled circles.
Bigger symbols are used for stars of class I.
 The position of the candidate free-free emitter IRAS  16115$-$5044 is indicated with a
blue diamond and labeled.
}
\end{center} 
\end{figure*}

\subsection{Newly Detected Evolved Massive Stars in ``The Bridge''}
\label{massivebridge}

We targeted stars  in ``The  Bridge'' area, 
mainly in isolation and in the direction of the HESS J1616$-$508 source. 
Only a few targets  are located   within 5\arcmin\  from  known clusters --
star E4 is near the  [DBS2003]-100 cluster and E5, E7, and E8 are 
in [DBS2003]-160/[DBS2003]-161.

\subsubsection{Detected  Early-type Stars}
Star E2 is a rare transitional object on its way to the WR stage, an Ofpe/WN 9.
Ofpe/WN 9 stars were first recognized  in the GC by their  
strong He I lines  \citep[e.g.,][]{najarro97,blum95,figer95}.
They have masses from 25 to 60 \Msun\ and may be the progenitors of WN 8 stars
\citep{martins07}.
E2 is located in isolation between SNR RCW 103 and the circle error of HESS J1616$-$508. 
By assuming the distance of G332.809$-$0.132 (DM = 12.66 mag),
\Aks\ = 0.94 mag, and \BCK\ = $-2.91\pm0.19$ \citep[][]{martins07},
we derived \Mk\ = $-7.54$ mag, \Mbol\ = $-10.45$ mag, i.e., $L~=~1.2 \times 10^6$ \Lsun.
This luminosity is consistent with that of the  Ofpe/WN 9 
stars in the Quartet cluster
\citep{messineo09}, in the Quintuplet \citep{figer99},
and in the GC \citep{martins07}. 
Star E2 is possibly associated with G332.809$-$0.132
and with the progenitor of SNR RCW 103.
For a distance of 6 kpc, \Mbol\ would be $-11.68$ mag, which would make E2
brighter than the Ofpe/WN 9 stars in the GC \citep[min = $-8.5$, max = $-11.3$ mag,][]{martins06}.
Its Gaia DR2 parallax, $\varpi=0.57\pm0.14$  mas (corresponding to a distance 
from 1.3 to 2.2 kpc), has a fractional error larger than 20\%. With the current Gaia DR2, 
deviations up to 50\% from the actual distance
are measured for OB stars   \citep[$>1.5$ kpc,][]{shull19}.

Star E3 (B0-5) is 2\farcm2   (about 2 pc)
away from star E2 and has a similar value of
\Ks\ and interstellar extinction (dereddened \Ks\ = 5.13 mag and \Aks\ = 0.94). 
Its colors are consistent with those of  normal early-type stars
and its magnitude suggests a supergiant for a distance between 1.5 and 3.5 kpc.
By assuming DM = 12.66 mag, for star E3  we estimate an absolute 
\Mk\  = $-7.53$ mag, which is a value typical for  B5 supergiants.
This indicates that star E3 is another evolved massive star (B0$-$5 I)  
  associated with the progenitor of SNR RCW 103 in 
G332.809$-$0.132.
For a distance of 6 kpc, \Mk\ would be $-8.76$ mag, a value that is  
too bright for a B0-5 star \citep{martins06}.  The Gaia DR2 parallax is of poor quality.

Star E4 is a WR  of WN 8 type. It is located in isolation 
on the southwest side of ``The Bridge'', about 4\arcmin\ away from the center of
the cluster [DBS2003]-100 \citep[see][]{borissova16},
and it was targeted because of its free-free excess. By assuming intrinsic 
$J-H$ = 0.02 mag and $H-$\Ks\ = 0.11 mag,  and a bolometric correction \BCK\ = $-3.4$ mag 
from the work by \citet{crowther06}, we derived \Aks\ = 0.46 mag, and a dereddened 
\Ks$~=~+$6.90 mag, and  apparent bolometric magnitude $m_{bol}~=~3.5$ mag.
The two WN 8 stars detected in Westerlund1 have \Mk$\approx-5.8$ mag, 
while WR66 (WN 8) has \Mk~= $-4.89$ mag \citep{crowther06}.
By assuming \Mk\ = $-5.8$ mag, we obtain a distance modulus of DM = 12.66 mag, i.e., a distance of
3.5 kpc. By assuming an average \Mk\ = $-5.34$ mag, we obtain DM = 12.25 mag (2.8 kpc).
  The Gaia DR2 parallax is of poor quality.
 It is plausible that star E4 is associated with the giant molecular complex 
G332.809$-$0.132 (and SNR RCW 103). Typical initial masses of WN 8 stars 
are larger than $40$ \Msun.

In the direction of the stellar cluster [DBS2003]-160/[DBS2003]-161 \citep{dutra03}, 
we detected three emission line stars (E5, E7, and E8 in Table \ref{obs.early}).
This cluster was analyzed by \citet{romanlopes04} and is
centered on IRAS 16132$-$5039. IRAS 16132$-$5039 harbors two  MSX components.
From near-infrared (NIR) and mid-infrared  photometry, by assuming that it is 
on the zero-age main sequence (ZAMS),
\citet[][]{romanlopes04}  have inferred an O5 type for the  star E5, which is the
brightest star in the MSX A component (E5  coincides with star IRS 1 in their Table 2).
The obtained low-resolution \K-band spectra do not allow a class refinement.
With \Aks\ = 1.27, 1.74, and 0.74  mag, stars E5, E7, and E8 
have dereddened \Ks\ = 8.57, 8.60, and 9.52 mag. 
By assuming DM = 12.66 mag, we derive \Mk\ = $-4.09, -4.06, -3.14$ mag,
typical values for  O5, O5, and O8 stars on the ZAMS 
\citep[or O6, O6, and O9.5 dwarfs;][]{martins06}. The Gaia DR2 parallaxes have large errors.
By assuming  a conservative \BCK\ of $-4.38$ mag  
\citep[average of the values for O2-O8 stars,][]{martins06}, 
we obtained      \Mbol\ = $-8.51$ mag, $-8.48$ mag, and $-7.56$ mag.
Their massive nature is also confirmed by the finding of \citet{romanlopes09}.
They analyzed the brightest stars of the MSX B component, which
has \Ks\ magnitude similar to that of stars E5 and E7, 
and obtained the spectrum of an O3If$^+$ or O3-5V.

 Star E6 is a  dwarf  with \Aks\ = 0.20 mag;  
Gaia DR1 gives a proper motion of   
$\varpi = 0.74\pm0.29$ mas yr$^{-1}$  and a distance of $\approx 1.5$ kpc.
Gaia DR2   lists  $\varpi = 0.31\pm0.04$ mas yr$^{-1}$ 
and a distance of $\approx 2.9$ kpc.

Inferred parameters of detected early-type stars are given in {\bf Table \ref{table.mbolearly}}.

\subsection{Previous Spectroscopic Information from the Literature}

By searching  the catalog of \citet{skiff14},
we were able to collect spectral types of  63 OB stars 
(47 of which are above \Ks\ = 10 mag)
and 2 WR stars, 20 yellow stars, and  24 late-type stars  
located in the direction of G332.809$-$0.132.
Among the  24 late-type stars, 6 stars  are reported as class I stars
and are included in the catalog of class I stars detected 
in Gaia DR2 \citep{messineo19} and listed in Table 7.

The  63 OB stars are listed in Table \ref{obknown}.
All but two of the known OB stars have \Aks\ smaller than 0.4 mag, with a
mean of \Aks=0.18 mag and a $\sigma=0.07$ mag. Their parallactic distances 
ranges from 0.2 to 3.9 kpc. Fifteen stars out of the 63 
have Gaia DR2 distances from 2.0 to 4 kpc,
and among them seven have been previously reported as members of  
a population of massive stars at the distance of G332.809$-$0.132 
(see  Section \ref{associations and clusters?}).

The two WR stars are listed in Table \ref{wrrsgknown}.
SIMBAD reports WR 74 (P2) as a WN 7
detected by \citet{shara09} and \citet{vanderhucht01}. 
It is located on the Southern periphery of ``The Bridge'', and has a \Ks\ = 8.7 mag.
By assuming intrinsic $J-H$ = 0.02 mag and $H-$\Ks\ = 0.11 mag,  
and  \BCK\ = $-3.9$ mag from the work by \citet{crowther06},
we derived a moderate extinction, \Aks\ = 0.43 mag, and a dereddened \Ks\ of $+$8.40 mag,
and apparent bolometric magnitude $m_{bol}~=~4.5$ mag. 
For DM = 12.66, we obtained \Mk\ = $-4.3$ mag, and \Mbol\ = $-8.2$ mag.
In the work of \citet{hamann06}, for this WR, it is reported that DM = 13.0 mag and 
\Mbol\ = $-8.76$ mag, which, for DM = 12.66 mag, implies  
\Mbol\ = $-8.46$ mag and \Mk\ = $-4.56$ mag.
Its interstellar extinction agrees with the \Aks\ measurements
of star E4  (WN 8), { which has a spectrophotometric distance
of about 3 kpc;  therefore, it} supports the association with G332.809$-$0.132.
WN 7 stars in Westerlund 1 have brighter \Mk\ from  $-5.0$ to $-6.0$ mag. 
The Gaia  parallax from DR2 has high uncertainty ($0.14\pm0.05$ mas); 
a larger distance cannot be excluded.

$[$SMG2009$]$~1059$-$34 (P1) is another WR of WC 8 type \citep{shara09}.
By assuming  $J-H$ = 0.05 mag, $H-$\Ks\ = 0.38 mag, and  \BCK\ = $-3.6$ mag \citep{crowther06},
we estimate \Aks\ of about 1.7 mag, and with a DM = 12.66 mag and \Mbol\ = $-$6.57 mag.
\Mbol\ can span several orders of magnitudes in WC 8 types from $-$4.65 mag (WR135) to 
$-$8.5 mag (star in Westerlund 1).  Its parallax has not been measured.

\begin{sidewaystable*} 
\vspace*{+2cm} %;;;two columns
\caption{Known OB Stars \label{obknown}.}
{\tiny
\begin{tabular}{@{\extracolsep{-.1in}}rllrrrrrrrr rrrr llrrrlrlllll}
\hline
\hline
ID &            Ra      & Dec            & Iden     &$J$       &$H$       &\Ks       &  [8.0] &     Q1     & Q2     &  \Aks &        plx &                   DM         & VR    &DMv  &  Sp. Type$^a$   & Sp.(adopt) & DM(sp)$^b$ & DM(adopt) & \Mk$^b$   &2MASS-ID/Alias &   Ref$^c$ & Sp. year\\
   &    $($hh mm ss$)$  & $($deg mm ss$)$& (mag)    &(mag)     &(mag)     &(mag)     &  (mag) &  (mag)     & (mag)  &(mag)  &       (mas)&                   (mag)      &(\kms)  &(mag)&                &            & (mag)  &(mag)      & (mag) &               &        (RCW 103)&\\
\hline
    1   & 16 13 00.73  & $-$51 03 20.39  &   8.49   &   7.16   &   6.93   &   6.72   &   6.52   &  $-$0.14   &$-$0.11   & 0.35     & 0.30 $\pm$   0.05    & 12.43$[$12.14$$-$$12.76$]$    &  \nodata   &   \nodata &                     B4Ie  &           B3$-$5    I  &      12.64  & 12.43   & $-$6.06   &                     J16130073$-$5103203/LS~3508  &                 &                     2003   \\
    2   & 16 18 43.72  & $-$51 27 57.72  &   8.86   &   7.55   &   7.29   &   7.08   &   6.80   &  $-$0.12   &$-$0.29   & 0.36     & 0.23 $\pm$   0.04    & 12.96$[$12.62$$-$$13.37$]$    &  \nodata   &   \nodata &                A3II/A2Ib  &             A2  (I)  &    (13.91)  & 12.96   & $-$6.24   &               J16184372$-$5127577/LS~3544,Ly431  &             3  &                1971/1987   \\
    3   & 16 19 08.34  & $-$51 05 38.56  &   9.03   &   8.38   &   8.31   &   8.22   &   8.10   &  $-$0.08   &$-$0.17   & 0.18     & 4.66 $\pm$   0.04    &    6.64$[$6.62$$-$$6.66$]$    &  \nodata   &   \nodata &                      OB+  &                      &       \nodata &  6.64   &  1.40   &                   J16190833$-$5105385/HD~146522  &                 &                     1971   \\
    4   & 16 16 37.70  & $-$50 36 26.50  &   \nodata   &   8.53   &   8.44   &   8.35   &   8.32   &  $-$0.08   & 0.11   & 0.20     & 0.27 $\pm$   0.04    & 12.60$[$12.32$$-$$12.91$]$    &   \nodata    &   \nodata &                    A8$-$9V  &             A8    V  &       \nodata & 12.60   & $-$4.45   &           J16163769$-$5036264/LS~3529,HD~146058  &              1  &                     1978   \\
    5   & 16 13 49.36  & $-$50 52 12.56  &   9.29   &   8.58   &   8.42   &   8.39   &   8.25   &   0.09   &$-$0.17   & 0.16     & 0.82 $\pm$   0.05    & 10.36$[$10.24$$-$$10.48$]$    &  \nodata   &   \nodata &        O9II/B2V/B0$-$2I$-$II  &          O9$-$B2  (I)  &    (14.29)  & 10.36   & $-$2.13   &                 J16134935$-$5052125/CPD$-$50~9063  &                 &      1969/1977/1978/1987   \\
    6   & 16 17 23.73  & $-$50 51 53.49  &   9.41   &   8.92   &   8.71   &   8.66   &   8.50   &   0.11   &$-$0.19   & 0.19     & 2.79 $\pm$   0.04    &    7.75$[$7.72$$-$$7.79$]$    & $-$24.15    &  11.26  &                       A2  &                      &       \nodata &  7.75   &  0.72   &                   J16172373$-$5051534/HD~146184  &                 &                     1939   \\
    7   & 16 14 40.40  & $-$50 48 22.65  &   \nodata   &   9.17   &   8.97   &   8.74   &   7.35   &  $-$0.21   &$-$3.32   & 0.36     & 0.61 $\pm$   0.05    & 10.97$[$10.80$$-$$11.14$]$    &   \nodata    &   \nodata &                A9$-$F0IV$-$V  &             A9    V  &       \nodata & 10.97   & $-$2.59   &         J16144040$-$5048226/LS~3518,CD$-$50~10263  &              1  &                     1978   \\
    8   & 16 15 27.45  & $-$50 59 01.00  &   9.21   &   8.85   &   8.80   &   8.79   &   8.57   &   0.03   &$-$0.55   & 0.10     & 0.38 $\pm$   0.05    & 11.97$[$11.73$$-$$12.23$]$    &  27.00    &   \nodata &         B0.5III$-$IV/B2:Ve  &           B0$-$3    V  &      11.07  & 11.97   & $-$3.28   &   J16152744$-$5059010/HD~145828,HD~145828,Ly407  &            1,2  &                1969/1987   \\
    9   & 16 14 25.41  & $-$51 11 02.79  &   9.38   &   9.02   &   8.97   &   8.95   &   8.87   &   \nodata   &$-$0.16   & 0.11     & 0.77 $\pm$   0.05    & 10.50$[$10.36$$-$$10.64$]$    &  \nodata   &   \nodata &           B0III$-$B0.5III:  &             B0  III  &      13.15  & 10.50   & $-$1.66   &                   J16142541$-$5111027/HD~145637  &                 &                1961/1969   \\
   10   & 16 14 09.31  & $-$51 27 35.68  &   9.44   &   9.10   &   9.04   &   8.96   &   9.04   &  $-$0.07   & 0.36   & 0.17     & 0.63 $\pm$   0.05    & 10.91$[$10.76$$-$$11.07$]$    &  \nodata   &   \nodata &                 B3II$-$III  &           B0$-$3  III  &      12.26  & 10.91   & $-$2.12   &                   J16140930$-$5127356/HD~145579  &                 &                     1978   \\
   11   & 16 18 57.93  & $-$51 03 29.73  &   9.62   &   9.14   &   9.10   &   8.96   &   8.87   &  $-$0.20   &$-$0.07   & 0.23     & 0.50 $\pm$   0.04    & 11.38$[$11.23$$-$$11.55$]$    &  \nodata   &   \nodata &                B7$-$8Ib$-$II  &             B8 (II)  &    (15.69)  & 11.38   & $-$2.65   &                     J16185792$-$5103297/LS~3547  &                 &                     1978   \\
   12   & 16 14 53.58  & $-$51 19 01.05  &   9.37   &   9.10   &   9.06   &   9.05   &   8.95   &   0.03   &$-$0.24   & 0.09     & 3.33 $\pm$   0.06    &    7.37$[$7.33$$-$$7.41$]$    &  \nodata   &   \nodata &                  B2II/O8  &             O8  (I)  &    (14.48)  &  7.37   &  1.59   &                   J16145358$-$5119010/HD~145722  &                 &                1987/1987   \\
   13   & 16 17 03.84  & $-$51 00 43.56  &   9.56   &   9.27   &   9.17   &   9.17   &   9.03   &   0.08   &$-$0.28   & 0.11     & 0.80 $\pm$   0.07    & 10.41$[$10.23$$-$$10.61$]$    &  \nodata   &   \nodata &                 B3II$-$III  &             B3  III  &      11.60  & 10.41   & $-$1.35   &                   J16170384$-$5100435/HD~146125  &                 &                     1978   \\
   14   & 16 18 30.56  & $-$51 25 57.87  &   9.36   &   9.27   &   9.25   &   9.24   &   9.28   &  $-$0.01   & 0.15   & 0.10     & 2.00 $\pm$   0.22    &    8.46$[$8.24$$-$$8.71$]$    &  \nodata   &   \nodata &                  B7$-$8III  &             B8  III  &       \nodata &  8.46   &  0.68   &                   J16183055$-$5125578/HD~146373  &                 &                     1978   \\
   15   & 16 18 22.92  & $-$51 26 24.26  &   9.76   &   9.46   &   9.32   &   9.26   &   \nodata   &   0.02   & \nodata  & 0.18     & 0.34 $\pm$   0.05    & 12.15$[$11.88$$-$$12.46$]$    & $-$30.00    &   \nodata &                   B9IV$-$V  &             B9    V  &       9.14  & 12.15   & $-$3.07   & J16182291$-$5126242/LS~3540,CP$-$51~9232,CPD$-$51.9  &            1,2  &                     1978   \\
   16   & 16 16 22.28  & $-$50 26 22.74  &   \nodata   &   \nodata   &   \nodata   &   9.30   &   9.11   &   \nodata  & \nodata  & \nodata    & 0.38 $\pm$   0.04    & 11.93$[$11.72$$-$$12.16$]$    &  \nodata   &   \nodata &           B0.5III/B3III:  &           B0$-$3  III  &       \nodata &  \nodata  &  \nodata  &                     J16162228$-$5026227/LS~3526  &                 &           1961/1969/1987   \\
   17   & 16 15 15.53  & $-$50 30 35.55  &   9.89   &   9.52   &   9.38   &   9.30   &   9.18   &   0.01   &$-$0.13   & 0.19     & 1.51 $\pm$   0.04    &    9.06$[$9.00$$-$$9.12$]$    &  \nodata   &   \nodata &                B1III/B0V  &                      &       \nodata &  9.06   &  0.05   &                 J16151553$-$5030355/CPD$-$50~9104  &                 &                1969/1987   \\
   18   & 16 19 11.04  & $-$51 14 01.92  &   9.73   &   9.49   &   9.39   &   9.31   &   \nodata   &  $-$0.05   & \nodata  & 0.19     & 0.98 $\pm$   0.04    &   9.98$[$9.89$$-$$10.08$]$    &  \nodata   &   \nodata &                       A3  &                      &       \nodata &  9.98   & $-$0.86   &                   J16191104$-$5114019/HD~146523  &                 &                     1939   \\
   19   & 16 12 37.85  & $-$50 41 25.85  &  10.12   &   9.68   &   9.48   &   9.35   &   9.37   &  $-$0.02   & 0.37   & 0.26     & 1.34 $\pm$   0.04    &    9.32$[$9.25$$-$$9.39$]$    &  \nodata   &   \nodata &               B8$-$9II$-$III  &             B8  III  &       \nodata &  9.32   & $-$0.23   &                 J16123785$-$5041258/CPD$-$50~9031  &                 &                     1978   \\
   20   & 16 13 39.28  & $-$50 29 45.79  &  10.09   &   9.64   &   9.41   &   9.36   &   9.28   &   0.13   & 0.05   & 0.19     & 1.41 $\pm$   0.04    &    9.22$[$9.15$$-$$9.29$]$    &  \nodata   &   \nodata &                       A5  &                      &       \nodata &  9.22   & $-$0.05   &                 J16133928$-$5029457/CPD$-$50~9058  &                 &                     1939   \\
   21   & 16 13 58.11  & $-$50 54 58.69  &  10.11   &   9.69   &   9.46   &   9.40   &   9.38   &   0.12   & 0.24   & 0.20     & 1.44 $\pm$   0.06    &    9.17$[$9.08$$-$$9.27$]$    &  \nodata   &   \nodata &                       A7  &                      &       \nodata &  9.17   &  0.03   &                 J16135810$-$5054586/CPD$-$50~9066  &                 &                     1939   \\
   22   & 16 18 16.43  & $-$51 33 35.18  &   9.77   &   9.55   &   9.47   &   9.44   &   9.47   &   0.02   & 0.20   & 0.13     & 1.28 $\pm$   0.04    &    9.41$[$9.34$$-$$9.48$]$    &  \nodata   &   \nodata &                       A5  &                      &       \nodata &  9.41   & $-$0.10   &                   J16181643$-$5133351/HD~146340  &                 &                     1939   \\
   23   & 16 13 44.57  & $-$51 06 20.10  &  10.23   &   9.67   &   9.51   &   9.45   &   9.29   &   0.04   &$-$0.18   & 0.19     & 0.31 $\pm$   0.06    &                     $..$    & $-$40.00    &   \nodata &                 A0III$-$IV  &             A0  III  &       \nodata &  \nodata  &  \nodata  & J16134457$-$5106200/LS~3512,CP$-$50~9059,CPD$-$50.9  &            1,2  &                     1978   \\
   24   & 16 18 49.45  & $-$51 09 27.15  &   9.91   &   9.63   &   9.55   &   9.47   &   9.42   &  $-$0.07   & 0.02   & 0.18     &$-$7.43 $\pm$   1.22    &                     $..$    &  \nodata   &   \nodata &               B1III/A5II  &          B1$-$A5   II  &       \nodata &  \nodata  &  \nodata  &                 J16184944$-$5109271/CPD$-$50~9205  &                 &                1961/1987   \\
   25   & 16 16 40.84  & $-$50 39 20.62  &   \nodata   &   9.64   &   9.50   &   9.48   &   9.30   &   0.11   &$-$0.34   & 0.13     & 1.03 $\pm$   0.04    &    9.88$[$9.80$$-$$9.97$]$    &  \nodata   &   \nodata &                       A5  &                      &       \nodata &  9.88   & $-$0.53   &                   J16164083$-$5039206/HD~146074  &                 &                     1939   \\
   26   & 16 15 26.04  & $-$51 09 17.87  &  10.19   &   9.66   &   9.59   &   9.48   &   9.54   &  $-$0.12   & 0.31   & 0.21     & 0.34 $\pm$   0.04    & 12.17$[$11.93$$-$$12.45$]$    & $-$17.00    &   \nodata &                     B:I:  &           B0$-$3  (I)  &    (15.54)  & 12.17   & $-$2.90   & J16152603$-$5109178/CD$-$50~10282,CD$-$50~10282,Ly4  &            1,3  &                     1978   \\
   27   & 16 16 07.22  & $-$51 05 21.31  &   \nodata   &   9.73   &   9.55   &   9.49   &   9.38   &   0.06   &$-$0.04   & 0.19     & 1.88 $\pm$   0.04    &    8.60$[$8.56$$-$$8.64$]$    &  \nodata   &   \nodata &                  B2$-$3III  &             B2  III  &      12.44  &  8.60   &  0.70   &                 J16160722$-$5105213/CPD$-$50~9128  &                 &                     1969   \\
   28   & 16 14 11.81  & $-$51 14 51.32  &  10.03   &   9.61   &   9.55   &   9.50   &   9.38   &  $-$0.02   &$-$0.20   & 0.15     & 0.62 $\pm$   0.04    & 10.94$[$10.80$$-$$11.09$]$    & $-$26.00    &   \nodata &                       A5  &                      &       \nodata & 10.94   & $-$1.59   &   J16141180$-$5114513/LS~3514,CD$-$50~10251,Ly398  &            1,2  &                     1939   \\
   29   & 16 20 16.82  & $-$50 59 15.30  &  10.18   &   9.73   &   9.61   &   9.51   &   \nodata   &  $-$0.05   & \nodata  & 0.21     & 0.65 $\pm$   0.04    & 10.85$[$10.72$$-$$11.00$]$    &  \nodata   &   \nodata &            B1III/O9V/B2V  &          O9$-$B2    V  &      11.90  & 10.85   & $-$1.55   &                   J16201681$-$5059152/[L64]~444  &                 &           1969/1987/1993   \\
   30   & 16 20 00.43  & $-$51 07 32.32  &   9.96   &   9.73   &   9.49   &   9.57   &   9.75   &   0.38   & 0.65   & 0.06     & 1.90 $\pm$   0.05    &    8.57$[$8.52$$-$$8.63$]$    &   3.86    &  15.91  &                      OB$-$  &                      &       \nodata &  8.57   &  0.94   &                   J16200043$-$5107323/HD~146687  &                 &                     1964   \\
   31   & 16 14 53.41  & $-$51 13 37.40  &  10.34   &   9.79   &   9.69   &   9.59   &   9.36   &  $-$0.08   &$-$0.41   & 0.21     & 0.33 $\pm$   0.04    & 12.24$[$12.00$$-$$12.51$]$    & $-$61.00    &   \nodata &                    A8$-$9V  &             A8    V  &       \nodata & 12.24   & $-$2.86   &   J16145340$-$5113373/LS~3519,CD$-$50~10270,Ly401  &            1,3  &                     1978   \\
   32   & 16 18 36.36  & $-$51 17 16.72  &  10.27   &   9.78   &   9.67   &   9.59   &   9.60   &  $-$0.02   & 0.20   & 0.18     & 0.51 $\pm$   0.11    &                     $..$    &  \nodata   &   \nodata &              B2.5III/B2:  &             B2  III  &      12.55  &  \nodata  &  \nodata  &                     J16183636$-$5117167/LS~3542  &                 &                1969/1987   \\

 \hline
\end{tabular}
\begin{list}{}{}
\item {\bf Notes.} 
Identification numbers are followed by celestial coordinates,  $I,J,H$,\Ks, 
and [8.0] magnitudes; the $Q1$ and $Q2$ parameters; extinction \Aks, Gaia DR2 parallaxes (plx),  
derived parallactic distance moduli (DM);
Gaia DR2 spectroscopic radial velocities in the solar barycentric reference frame (VR);  
kinematically inferred
DMs (DMv) \citep{reid09}; spectral types, spectrophotometric DM(sp);
DMadopted; obtained absolute \Ks\ magnitudes (\Mk); 2MASS IDs/Alias;
references for the members of RCW 103; years of the collected spectral types. \\
\Aks\ are estimated by assuming $H-$\Ks\ = $-0.10$ mag and $J-$\Ks\ = $-0.01$ mag 
\citep[values for O stars by][]{martins06}. \\
Stars \#7, \#35, \#47, \#62, and \#63 have colors of free-free emitters \citep{messineo12}. \\
\item $^a$ Sp. type= spectral types collected from \citet{skiff14} or SIMBAD.
\item $^b$ Spectrophotometric DM=\Ks-\Aks+\Mk(sp) of OB and early A stars, DM(sp), 
are calculated with the \Mk(sp) values compiled by \citet[][and references therein]{messineo11}.
DM(sp) of stars reported as blue supergiants are listed between parentheses
because they are systematically larger  than the parallactic ones.\\
The column \Mk\ lists the absolute \Ks\ magnitudes obtained with the adopted DM, DM(adopt).
\item {\bf $^c$ References:} 1=\citet{melnik17}; 2=\citet{humphreys78}; 3=\citet{westerlund69}. 
\item $^d$ The velocity of LS$~3548$ is given in the work of \citet{melnik17}.
\end{list}
}
\end{sidewaystable*}

\addtocounter{table}{-1}
\begin{sidewaystable*} 
\caption{Known OB Stars \label{obknown}.}
{\tiny
\begin{center}
\begin{tabular}{@{\extracolsep{-.1in}}rllrrrrrrrr rrrr llrrrlrlllll}
\hline
\hline
ID &            Ra      & Dec            & Iden     &$J$       &$H$       &\Ks       &  [8.0] &     Q1     & Q2     &  \Aks & plx & DM & VR    &DMv  &  Sp. Type$^a$   & Sp.(adopt) & DM(sp)$^b$  & DM(adopt) & \Mk$^b$   &2MASS-ID/Alias &   Ref$^c$ &Sp. year\\
   &    $($hh mm ss$)$  & $($deg mm ss$)$& (mag)    &(mag)     &(mag)     &(mag)     &  (mag) &  (mag)     & (mag)  &(mag)  &       (mas)&                   (mag)      &(\kms)  &(mag)&                &            & (mag)  &(mag)      & (mag) &               &        (RCW 103)&\\
\hline

   33   & 16 13 17.45  & $-$51 23 40.59  &  10.75   &  10.22   &   9.91   &   9.60   &   8.28   &  $-$0.25   &$-$2.93   & 0.48     & 0.36 $\pm$   0.04    & 12.03$[$11.83$$-$$12.25$]$    &  \nodata   &   \nodata &                      OB$-$  &                      &       \nodata & 12.03   & $-$2.91   &                     J16131745$-$5123405/LS~3509  &                 &                     1971   \\
   34   & 16 18 40.65  & $-$51 26 09.36  &  10.02   &   9.86   &   9.74   &   9.65   &   \nodata   &  $-$0.03   & \nodata  & 0.20     & 2.62 $\pm$   0.05    &    7.89$[$7.84$$-$$7.93$]$    &  \nodata   &   \nodata &                     OB+e  &                      &       \nodata &  7.89   &  1.56   &                   J16184064$-$5126093/HD~146424  &                 &                1971/1976   \\
   35   & 16 19 24.13  & $-$51 11 37.66  &   9.84   &   9.75   &   9.71   &   9.68   &   9.47   &  $-$0.03   &$-$0.48   & 0.13     & 1.44 $\pm$   0.06    &    9.16$[$9.08$$-$$9.24$]$    &  \nodata   &   \nodata &                       B9  &                      &       \nodata &  9.16   &  0.39   &                   J16192412$-$5111376/HD~146575  &                 &                     1921   \\
   36   & 16 13 37.92  & $-$51 16 33.04  &  10.10   &   9.83   &   9.74   &   9.69   &   9.77   &  $-$0.02   & 0.37   & 0.16     & 0.85 $\pm$   0.04    & 10.28$[$10.17$$-$$10.39$]$    &  \nodata   &   \nodata &                   B8$-$9IV  &             B8   IV  &       9.92  & 10.28   & $-$0.75   &                   J16133791$-$5116330/HD~145489  &                 &                     1978   \\
   37   & 16 13 34.74  & $-$50 53 28.01  &  10.18   &   9.81   &   9.73   &   9.70   &   9.66   &   0.01   & 0.01   & 0.14     & 1.14 $\pm$   0.05    &    9.67$[$9.57$$-$$9.77$]$    &  \nodata   &   \nodata &                   B8$-$9II  &             B8 (II)  &    (16.52)  &  9.67   & $-$0.11   &                 J16133473$-$5053280/CPD$-$50~9055  &                 &                     1978   \\
   38   & 16 19 04.50  & $-$50 53 58.80  &  10.40   &   9.94   &   9.85   &   9.71   &   9.60   &  $-$0.17   &$-$0.04   & 0.25     & 0.45 $\pm$   0.04    & 11.59$[$11.42$$-$$11.78$]$    & $-$19.00$^d$    &   \nodata &                       A0  &                      &       \nodata & 11.59   & $-$2.13   &    J16190449$-$5053587/LS~3548,CP$-$50~9216,Ly435  &            1,3  &                     1939   \\
   39   & 16 17 29.67  & $-$50 36 17.19  &  10.46   &   9.93   &   9.81   &   9.72   &   9.30   &  $-$0.05   &$-$0.90   & 0.21     & 0.96 $\pm$   0.04    &  10.01$[$9.93$$-$$10.10$]$    &  \nodata   &   \nodata &              B0.5IV:/B0:  &             B0   IV  &      12.55  & 10.01   & $-$0.50   &                   J16172966$-$5036171/HD~146222  &                 &                1971/1987   \\
   40   & 16 15 01.59  & $-$51 23 50.72  &   9.98   &   9.77   &   9.75   &   9.75   &   9.74   &   0.02   & \nodata   & 0.08     & 0.94 $\pm$   0.05    &  10.08$[$9.97$$-$$10.19$]$    &  \nodata   &   \nodata &                       B8  &                      &       \nodata & 10.08   & $-$0.41   &                   J16150158$-$5123507/HD~145761  &                 &                     1921   \\
   41   & 16 13 38.25  & $-$50 35 59.01  &  10.06   &   9.86   &   9.79   &   9.77   &   9.64   &   0.03   &$-$0.27   & 0.12     & 0.78 $\pm$   0.05    & 10.45$[$10.32$$-$$10.58$]$    &  \nodata   &   \nodata &             B3$-$6III:(e?)  &           B3$-$5  III  &      13.12  & 10.45   & $-$0.80   &                   J16133824$-$5035590/HD~145488  &                 &                     1978   \\
   42   & 16 13 21.63  & $-$51 05 34.08  &  10.40   &  10.03   &   9.90   &   9.84   &   9.67   &   0.01   &$-$0.25   & 0.18     & 1.00 $\pm$   0.05    &   9.93$[$9.84$$-$$10.03$]$    &  \nodata   &   \nodata &               B7$-$8IIp:Si  &             B8 (II)  &    (16.62)  &  9.93   & $-$0.27   &                 J16132163$-$5105340/CPD$-$50~9049  &                 &                     1978   \\
   43   & 16 14 05.90  & $-$51 10 52.02  &  10.18   &   9.97   &   9.93   &   9.84   &   \nodata   &  $-$0.13   & \nodata  & 0.18     & 0.81 $\pm$   0.05    & 10.38$[$10.26$$-$$10.52$]$    &  \nodata   &   \nodata &                       A0  &                      &       \nodata & 10.38   & $-$0.72   &                 J16140589$-$5110520/CPD$-$50~9068  &                 &                     1939   \\
   44   & 16 18 45.58  & $-$50 44 16.98  &  10.18   &   9.93   &   9.87   &   9.84   &   9.81   &   0.01   & 0.01   & 0.12     & 2.48 $\pm$   0.04    &    8.00$[$7.97$$-$$8.04$]$    &  \nodata   &   \nodata &                       B9  &                      &       \nodata &  8.00   &  1.72   &                 J16184557$-$5044169/CPD$-$50~9203  &                 &                     1939   \\
   45   & 16 17 23.68  & $-$51 14 27.55  &  10.94   &  10.34   &  10.15   &   9.94   &   8.92   &  $-$0.21   &$-$2.33   & 0.35     & 0.30 $\pm$   0.04    & 12.44$[$12.17$$-$$12.76$]$    &  \nodata   &   \nodata &                       A5  &                      &       \nodata & 12.44   & $-$2.85   &                 J16172367$-$5114275/[M81]~I$-$505  &                 &                     1939   \\
   46   & 16 18 47.80  & $-$51 06 59.57  &   9.99   &  10.01   &   9.96   &   9.98   &  10.00   &   0.08   & 0.09   & 0.07     & 1.11 $\pm$   0.06    &    9.71$[$9.60$$-$$9.83$]$    &  \nodata   &   \nodata &                       em  &                      &       \nodata &  9.71   &  0.20   &                   J16184779$-$5106595/HD~146445  &                 &                     1981   \\
   47   & 16 15 37.39  & $-$51 12 51.75  &  10.40   &  10.13   &  10.07   &   9.98   &   9.89   &  $-$0.10   &$-$0.10   & 0.19     & 1.58 $\pm$   0.06    &    8.96$[$8.88$$-$$9.05$]$    &  \nodata   &   \nodata &                     B8$-$9  &                      &       \nodata &  8.96   &  0.83   &                 J16153739$-$5112517/CPD$-$50~9114  &                 &                     1978   \\
   48   & 16 16 55.77  & $-$50 35 45.56  &   \nodata   &  10.25   &  10.18   &  10.13   &  10.13   &  $-$0.01   & 0.13   & 0.15     & 0.93 $\pm$   0.04    & 10.09$[$10.00$$-$$10.17$]$    &  \nodata   &   \nodata &                       A5  &                      &       \nodata & 10.09   & $-$0.11   &                   J16165577$-$5035455/HD~146105  &                 &                     1939   \\
   49   & 16 18 48.17  & $-$50 42 29.82  &  10.60   &  10.44   &  10.35   &  10.22   &   \nodata   &  $-$0.14   & \nodata  & 0.23     & 1.01 $\pm$   0.17    &   9.92$[$9.59$$-$$10.30$]$    &  \nodata   &   \nodata &                   B7$-$9I:  &             B8  (I)  &    (16.95)  &  9.92   &  0.07   &                 J16184817$-$5042298/CPD$-$50~9207  &                 &                     1978   \\
   50   & 16 14 48.62  & $-$50 59 11.35  &  10.38   &  10.23   &  10.29   &  10.26   &  10.48   &  $-$0.10   & 0.55   & 0.09     & 1.09 $\pm$   0.07    &    9.75$[$9.63$$-$$9.88$]$    &  \nodata   &   \nodata &                       A0  &                      &       \nodata &  9.75   &  0.42   &                 J16144862$-$5059113/CPD$-$50~9088  &                 &                     1939   \\
   51   & 16 17 15.74  & $-$51 26 41.11  &  10.98   &  10.48   &  10.34   &  10.30   &  10.13   &   0.07   &$-$0.27   & 0.16     & 0.37 $\pm$   0.06    & 11.98$[$11.68$$-$$12.32$]$    &  \nodata   &   \nodata &                       A0  &                      &       \nodata & 11.98   & $-$1.84   &                  J16171574$-$5126411/HPD~Nor~34  &                 &                     1939   \\
   52   & 16 14 36.50  & $-$51 30 03.68  &   \nodata   &  10.50   &  10.37   &  10.34   &  10.22   &   0.07   &$-$0.16   & 0.14     & 0.28 $\pm$   0.04    & 12.56$[$12.28$$-$$12.88$]$    &  \nodata   &   \nodata &                        B  &                      &       \nodata & 12.56   & $-$2.36   &                     J16143650$-$5130036/LS~3516  &                 &                     1966   \\
   53   & 16 15 55.93  & $-$50 27 10.51  &  10.85   &  10.53   &  10.37   &  10.36   &  10.33   &   0.14   & 0.09   & 0.13     & 1.32 $\pm$   0.08    &    9.35$[$9.23$$-$$9.47$]$    &  \nodata   &   \nodata &                      OB$-$  &                      &       \nodata &  9.35   &  0.88   &                   J16155593$-$5027105/HD~145922  &                 &                     1971   \\
   54   & 16 14 00.67  & $-$51 12 59.07  &  10.76   &  10.45   &  10.43   &  10.37   &  10.34   &  $-$0.09   &$-$0.01   & 0.15     & 1.56 $\pm$   0.04    &    9.00$[$8.94$$-$$9.06$]$    &  \nodata   &   \nodata &                      B/A  &                      &       \nodata &  9.00   &  1.22   &                 J16140067$-$5112590/CPD$-$50~9067  &                 &                     1978   \\
   55   & 16 19 53.00  & $-$51 13 49.19  &  11.21   &   \nodata   &   \nodata   &  10.48   &  10.44   &   \nodata  & \nodata  & \nodata    & 0.38 $\pm$   0.04    & 11.94$[$11.74$$-$$12.15$]$    &  \nodata   &   \nodata &                       A0  &                      &       \nodata &  \nodata  &  \nodata  &                     J16195300$-$5113491/LS~3554  &                 &                     1939   \\
   56   & 16 15 18.42  & $-$50 55 37.52  &  11.00   &  10.62   &  10.53   &  10.49   &  10.76   &   0.03   & 0.84   & 0.14     & 0.29 $\pm$   0.06    &                     $..$    &  \nodata   &   \nodata &                   B4V/O9  &           B3$-$5    V  &      12.73  &  \nodata  &  \nodata  &             J16151842$-$5055375/[L64]~404,Ly404  &             3  &                1987/1987   \\
   57   & 16 18 17.00  & $-$51 15 26.94  &  10.65   &   \nodata   &  10.53   &  10.51   &  10.66   &   \nodata  & \nodata  & \nodata    & 1.35 $\pm$   0.05    &    9.31$[$9.23$$-$$9.39$]$    &  \nodata   &   \nodata &                       OB  &                      &       \nodata &  \nodata  &  \nodata  &                 J16181699$-$5115269/CPD$-$50~9192  &                 &                     1964   \\
   58   & 16 18 49.64  & $-$51 01 06.93  &  10.94   &  10.66   &  10.64   &  10.58   &  10.61   &  $-$0.08   & 0.16   & 0.14     & 0.96 $\pm$   0.08    &  10.03$[$9.85$$-$$10.22$]$    &  \nodata   &   \nodata &                       A0  &                      &       \nodata & 10.03   &  0.41   &             J16184964$-$5101069/[L64]~432,Ly432  &             3  &                     1939   \\
   59   & 16 15 51.09  & $-$51 37 29.81  &  11.50   &  10.95   &  10.80   &  10.66   &   \nodata   &  $-$0.11   & \nodata  & 0.27     & 0.68 $\pm$   0.04    & 10.76$[$10.65$$-$$10.88$]$    &  \nodata   &   \nodata &                       OB  &                      &       \nodata & 10.76   & $-$0.37   &                  J16155109$-$5137298/HPD~Nor~11  &                 &                     1964   \\
   60   & 16 15 56.02  & $-$50 27 24.03  &  10.88   &  10.71   &  10.69   &  10.68   &  10.45   &   \nodata   &$-$0.58   & 0.09     & 1.25 $\pm$   0.05    &    9.47$[$9.38$$-$$9.56$]$    &  \nodata   &   \nodata &                        B  &                      &       \nodata &  9.47   &  1.12   &                   J16155602$-$5027240/HD~145923  &                 &                     1966   \\
   61   & 16 17 44.56  & $-$51 01 53.86  &  11.61   &  11.06   &  10.86   &  10.82   &  10.39   &   0.13   &$-$0.93   & 0.17     & 0.36 $\pm$   0.04    & 12.03$[$11.81$$-$$12.26$]$    &  \nodata   &   \nodata &                      B$-$A  &                      &       \nodata & 12.03   & $-$1.38   &                     J16174456$-$5101538/LS~3537  &                 &                     1978   \\
   62   & 16 17 17.68  & $-$50 28 11.66  &   \nodata   &  14.34   &  12.30   &  10.96   &   8.27   &  $-$0.38   &$-$3.89   & 1.99     & \nodata    &                     $..$    &  \nodata   &   \nodata &                      OB:  &                      &       \nodata &  \nodata  &  \nodata  &               J16171767$-$5028116/[RC2017]~A~10  &                 &                     1971   \\
   63   & 16 16 16.61  & $-$51 12 40.73  &   \nodata   &  13.76   &  12.38   &  11.49   &   9.83   &  $-$0.23   &$-$2.20   & 1.36     &$-$0.36 $\pm$   0.73    &                     $..$    &  \nodata   &   \nodata &                       em  &                      &       \nodata & 12.66   & $-$2.53   &              J16161660$-$5112407/[RC2017]~B~139  &                 &                     2018   \\
\hline
\end{tabular}
\end{center}
}
\end{sidewaystable*} 

\subsection{Late-type Stars}
In Table \ref{tablelatelum}, we summarize some properties 
of the brightest observed late-type 
stars and of the six K-M I stars reported by \citet{skiff14}  and \citet{messineo19}
in the direction of G332.809$-$0.132, as listed in Table \ref{wrrsgknown}.

For four of the known K-M I stars Gaia  radial velocities\footnote{Gaia DR2  velocities
in  the solar barycentric reference frame were transformed into \Vlsr\
using     $(U_o, V_o, W_o)~=~(10.27, 15.32, 7.74)$ \kms. These are the same values
used in the fortran code that \citet{reid09} present to infer 
kinematic distances with their rotation curve.  } 
are available.
Stars P3 and P8 at \Aks\ = 0.16 mag $\sigma~=~0.1$ mag have \Vlsr\ = $-39$ \kms\ 
with $\sigma~=~2$ \kms, and stars P4 and P7 at  \Aks\ = $0.41$ mag with $\sigma~=~0.1$ mag 
have \Vlsr\ = $-53$ mag with $\sigma~=~7$ mag, which again indicate    kinematic distances 
of 2800 and 3490 kpc, respectively.
For stars P3 and P8, the distance inferred from their average Gaia DR2 parallaxes  is  
2628 kpc in good agreement with the kinematic ones, 
but for stars P4 and P7 the parallactic distances are anomalously small (918 pc).
From the radial velocities, we conclude that stars P3, P8, P4, and P7 
are associated with the Scutum-Crux arm
and   the G332.809$-$0.132 complex; by assuming a DM = 12.66 mag, 
we infer \Mbol\ from $-$4.5 mag (K3 I) to $-$6.90 mag (M2 I).
Star P5 has \Aks\ = 0.3 mag, Gaia MD = 9.29 mag, but its velocity is not available. 
Star P6 is an RSG (M5.5 I) at \Aks\ = 1.49 mag and \Mbol\ = $-8.30$ mag
when assuming the distance to  G332.809$-$0.132 
(its velocity is not available  and the Gaia DR2 parallax is of poor quality).

The 76 late-type stars  we observed are mostly  in the direction of the center of this complex; 
for a putative distance of 3.4 kpc, six of them were brighter than  \Mbol $ < -5.26$ mag 
(see Table \ref{tablelatelum}), with three highly probably RSGs owing to their large \ew(CO);
their \Aks\ ranges from 1.01 mag to 2.10 mag.
The high values of interstellar extinction are still compatible with
the \Aks\ of stars in  IRAS 16132$-$5039, which belongs to the 3.4 kpc complex.
However, we cannot exclude a larger distance; 
the near HESS 1640$-$465 source is reported at the likely distance of 6.5 kpc
by \citet{kargaltsev09}. Radial velocities are demanded for a clearer scenario.

\begin{table*}
\caption{The Most Luminous Late-type Targets \label{tablelatelum}.}
\begin{tabular}{@{\extracolsep{-.08in}}lllllllllrrrr}
\hline
\hline
ID &         Sp(rsg) & Sp(giant) &  \Aks             & BC[rsg] &   Ko                & \Mbol[rsg]             & \Mbol[giant] &  \Mbol     &   DM(adopt) & DM($\varpi$) &DM(\Vlsr) \\
   &                 &           &   [mag]           & [mag]   & [mag]               & [mag]                  &[mag]         &   [mag]    &  [mag]& [mag]   & [mag]\\
\hline
L02  &      M2   &    $..$   &   2.10$\pm$   0.02  &   2.80  &   2.31$\pm$   0.03  &  $-$7.55$\pm$   0.71  &    $..$   &  $-$7.60$\pm$   0.00  &  12.66$\pm$   0.50  &  $..$     &  $..$     &   \\
L03  &      M2   &    $..$   &   1.34$\pm$   0.02  &   2.80  &   3.61$\pm$   0.03  &  $-$6.25$\pm$   0.71  &    $..$   &  $-$6.20$\pm$   0.00  &  12.66$\pm$   0.50  &  $..$     &  $..$     &   \\
L08  &      M1   &    $..$   &   1.01$\pm$   0.02  &   2.73  &   4.57$\pm$   0.04  &  $-$5.36$\pm$   0.71  &    $..$   &  $-$5.25$\pm$   0.00  &  12.66$\pm$   0.50  &  $..$     &  $..$     &   \\
L09  &      K2   &     M1   &   1.15$\pm$   0.01  &   2.53  &   4.45$\pm$   0.02  &  $-$5.68$\pm$   0.71  &  $-$5.47$\pm$   0.71  &  $-$5.45$\pm$   0.00  &  12.66$\pm$   0.50  &  $..$     &  $..$     &   \\
L11  &      K5   &     M7   &   1.43$\pm$   0.01  &   2.66  &   4.47$\pm$   0.02  &  $-$5.53$\pm$   0.71  &  $-$4.97$\pm$   0.71  &  $-$5.37$\pm$   0.00  &  12.66$\pm$   0.50  &  $..$     &  $..$     &   \\
L18  &      K5   &     M6   &   1.67$\pm$   0.02  &   2.66  &   4.69$\pm$   0.02  &  $-$5.31$\pm$   0.71  &  $-$4.88$\pm$   0.71  &  $-$5.16$\pm$   0.00  &  12.66$\pm$   0.50  &  $..$     &  $..$     &   \\
L21  &      M2   &    $..$   &   1.33$\pm$   0.02  &   2.80  &   5.17$\pm$   0.04  &  $-$4.69$\pm$   0.71  &    $..$   &  $-$4.65$\pm$   0.00  &  12.66$\pm$   0.50  &  $..$     &  $..$     &   \\
L32  &      M1   &    $..$   &   1.37$\pm$   0.02  &   2.73  &   5.77$\pm$   0.02  &  $-$4.15$\pm$   0.71  &    $..$   &  $-$4.05$\pm$   0.00  &  12.66$\pm$   0.50  &  $..$     &  $..$     &   \\
L35  &      M1   &    $..$   &   2.13$\pm$   0.02  &   2.73  &   5.07$\pm$   0.02  &  $-$4.86$\pm$   0.71  &    $..$   &  $-$4.75$\pm$   0.00  &  12.66$\pm$   0.50  &  $..$     &  $..$     &   \\

\hline
       P3   &       K3   &      $..$   &   0.16$\pm$   0.01  &   2.53  &   5.63$\pm$   0.02  &  $-$4.50$\pm$   0.71  &    $..$    &    $..$    &  12.66$\pm$   0.50  &  12.23    &  12.13  & \\
       P4   &       M2   &      $..$   &   0.52$\pm$   0.10  &   2.80  &   2.96$\pm$   0.21  &  $-$6.90$\pm$   0.74  &    $..$    &    $..$    &  12.66$\pm$   0.50  &  10.17    &  12.51  & \\
       P5   &       M3   &      $..$   &   0.30$\pm$   0.02  &   2.84  &   3.67$\pm$   0.04  &  $-$2.78$\pm$   0.71  &    $..$    &    $..$    &   9.29$\pm$   0.50  &   9.29    &  $..$   & \\
       P6   &     M5.5   &      $..$   &   1.49$\pm$   0.16  &   3.02  &   1.33$\pm$   0.33  &  $-$8.30$\pm$   0.78  &    $..$    &    $..$    &  12.66$\pm$   0.50  &  $..$     &  $..$   & \\
       P7   &       M3   &      $..$   &   0.40$\pm$   0.15  &   2.84  &   3.44$\pm$   0.31  &  $-$6.38$\pm$   0.77  &    $..$    &    $..$    &  12.66$\pm$   0.50  &   9.92    &  12.80  & \\
       P8   &       M1   &      $..$   &   0.17$\pm$   0.02  &   2.73  &   4.53$\pm$   0.03  &  $-$5.39$\pm$   0.71  &    $..$    &    $..$    &  12.66$\pm$   0.50  &  11.68    &  12.26  & \\

\hline
\end{tabular}
\begin{list}{}{}
\item
{\it Top table:} Brightest late-type stars observed in this work. 
Identification numbers are as in Table \ref{table.latespectra}.
Stars brighter than \Mbol $< -5.26$ mag ( $L \ga 10^4$ \Lsun) 
or with measure \ew$> 45$ \AA\ are listed.
{\it Bottom table:} Known K-M I stars from \citet{skiff14}.
Identification numbers are as in Table \ref{wrrsgknown}.
Identification numbers (ID) are followed by spectral types estimated for RSGs and giants 
(Sp(rsg), Sp(giant)), interstellar extinction, \Aks, adopted bolometric correction, \BCK,
dereddened \Ks\ magnitudes, Ko, bolometric magnitudes obtained with for RSGs and for giants
(\Mbol[rsg], \Mbol[giant]),  bolometric magnitudes obtained by direct integration 
under the  SED  (\Mbol), adopted distance moduli,   DM(adopt),
distance moduli from Gaia, DM($\varpi$), and distance moduli from Gaia \Vlsr\, DM(\Vlsr). \\
\end{list}
\end{table*}

\begin{table*}
\caption{Candidate Free-free Emitters. \label{table.wr}}
{\tiny
\begin{tabular}{@{\extracolsep{-.08in}}rllrrrrrrrrrrrrrrrll}
\hline
\hline
ID &            Ra      & Dec            & Iden     &$J$       &$H$       &\Ks       &  [8.0] &     Q1     & Q2     &  \Aks & plx & DM & VR    &DMv  & DMadopt   & \Mk &VAR &2MASS-ID/Alias & Comments\\
   &    $($hh mm ss$)$  & $($deg mm ss$)$& (mag)    &(mag)     &(mag)     &(mag)     &  (mag) &  (mag)     & (mag)  &(mag)  &(mas)& (mag)&(\kms)  &(mag)&(mag)&(mag)&    &      &\\
\hline
    1   & 16 18 10.38  & $-$51 24 17.21  &  11.40   &   6.53   &   5.18   &   4.46   &   3.17   &   0.04   &$-$1.38   & 1.18     & 0.42 $\pm$   0.21    &   \nodata   &  \nodata   &   \nodata &  12.66  & $-$9.38  &$..$  &J16181037$-$5124172/IRAS~16143$-$5117  &    \\
    2   & 16 16 08.27  & $-$51 29 03.33  &   \nodata   &   7.30   &   5.59   &   4.61   &   2.71   &  $-$0.05   &$-$2.42   & 1.53     & 0.03 $\pm$   0.23    &   \nodata   &  \nodata   &   \nodata &  12.66  & $-$9.58  &$..$  &J16160826$-$5129033/IRAS~16123$-$5121  &    \\
    3   & 16 19 26.77  & $-$51 24 31.22  &  12.90   &   8.34   &   6.32   &   4.88   &   1.90   &  $-$0.58   &$-$4.55   & 2.09     & 0.16 $\pm$   0.17    &   \nodata   &  \nodata   &   \nodata &  12.66  & $-$9.87  &$..$  &J16192677$-$5124312/CGCS~6631  &C   \\
    4   & 16 14 00.75  & $-$51 30 39.81  &  11.40   &   6.97   &   5.88   &   5.13   &   3.85   &  $-$0.26   &$-$1.62   & 1.13     & 0.88 $\pm$   0.14    &   \nodata   &  \nodata   &   \nodata &  12.66  & $-$8.66  &VAR  &J16140074$-$5130398/IRAS~16102$-$5123  &    \\
    5   & 16 15 17.96  & $-$50 52 19.71  &  13.05   &   7.76   &   6.13   &   5.14   &   3.32   &  $-$0.14   &$-$2.30   & 1.52     & 0.69 $\pm$   0.19    &   \nodata   &  \nodata   &   \nodata &  12.66  & $-$9.04  &$..$  &J16151795$-$5052197/IRAS~16115$-$5044  &PN/cLBV   \\
    6   & 16 18 40.78  & $-$51 20 30.95  &  12.29   &   7.29   &   6.01   &   5.28   &   3.78   &  $-$0.05   &$-$2.04   & 1.17     & 0.13 $\pm$   0.21    &   \nodata   &  \nodata   &   \nodata &  12.66  & $-$8.55  &$..$  &J16184077$-$5120309  &    \\
    7   & 16 17 07.51  & $-$50 31 06.38  &  12.80   &   7.60   &   6.17   &   5.32   &   4.02   &  $-$0.09   &$-$1.22   & 1.32     & 0.35 $\pm$   0.19    &   \nodata   &  \nodata   &   \nodata &  12.66  & $-$8.66  &$..$  &J16170751$-$5031063  &    \\
    8   & 16 19 18.58  & $-$51 27 09.92  &  14.10   &   8.47   &   6.89   &   6.01   &   4.66   &   0.01   &$-$1.19   & 1.39     & 0.15 $\pm$   0.26    &   \nodata   &  \nodata   &   \nodata &  12.66  & $-$8.04  &$..$  &J16191858$-$5127099/IRAS~16154$-$5119  &    \\
    9   & 16 17 59.18  & $-$51 19 25.32  &   \nodata   &   8.81   &   7.04   &   6.01   &   4.12   &  $-$0.09   &$-$2.30   & 1.60     & \nodata    &   \nodata   &  \nodata   &   \nodata &  12.66  & $-$8.25  &VAR  &J16175917$-$5119253  &    \\
   10   & 16 17 05.61  & $-$51 28 15.67  &  15.70   &   8.97   &   7.11   &   6.07   &   4.56   &  $-$0.00   &$-$1.15   & 1.64     &$-$0.73 $\pm$   0.35    &   \nodata   &  \nodata   &   \nodata &  12.66  & $-$8.23  &$..$  &J16170560$-$5128156  &    \\
   11   & 16 17 53.80  & $-$50 43 48.62  &  11.10   &   7.74   &   6.79   &   6.19   &   5.02   &  $-$0.13   &$-$1.59   & 0.94     & 0.90 $\pm$   0.13    &  10.16    &  \nodata   &   \nodata &   \nodata & $-$4.91  &$..$  &J16175380$-$5043486  &    \\
   12   & 16 18 13.16  & $-$51 02 10.65  &  13.24   &   9.09   &   7.57   &   6.72   &   5.33   &  $-$0.00   &$-$1.37   & 1.35     &$-$0.38 $\pm$   0.26    &   \nodata   &  \nodata   &   \nodata &  12.66  & $-$7.29  &$..$  &J16181315$-$5102106  &    \\
   13   & 16 16 49.13  & $-$51 32 46.15  &   \nodata   &  10.35   &   8.17   &   6.80   &   4.52   &  $-$0.29   &$-$2.59   & 2.06     &$-$2.38 $\pm$   0.74    &   \nodata   &  \nodata   &   \nodata &  12.66  & $-$7.92  &$..$  &J16164913$-$5132461  &    \\
   14   & 16 15 24.42  & $-$51 15 53.99  &   \nodata   &  11.08   &   8.45   &   6.86   &   4.29   &  $-$0.23   &$-$2.70   & 2.40     &$-$0.21 $\pm$   0.51    &   \nodata   &  \nodata   &   \nodata &  12.66  & $-$8.20  &$..$  &J16152442$-$5115539/IRAS~16115$-$5108  &    \\
   15   & 16 17 23.41  & $-$50 45 26.15  &  16.40   &  13.67   &   9.54   &   7.00   &   3.24   &  $-$0.43   &$-$3.43   & 3.78     & \nodata    &   \nodata   &  \nodata   &   \nodata &  12.66  & $-$9.44  &   &J16172340$-$5045261/IRAS~16136$-$5038  &    \\
   16   & 16 15 06.76  & $-$50 35 49.94  &  15.81   &  10.43   &   8.53   &   7.11   &   4.19   &  $-$0.65   &$-$4.52   & 2.04     & 0.27 $\pm$   0.23    &   \nodata   &  \nodata   &   \nodata &  12.66  & $-$7.59  &$..$  &J16150676$-$5035499  &    \\
   17   & 16 15 03.54  & $-$51 35 26.35  &  15.76   &  10.68   &   8.70   &   7.57   &   5.67   &  $-$0.05   &$-$2.01   & 1.76     & 0.31 $\pm$   0.34    &   \nodata   &  \nodata   &   \nodata &  12.66  & $-$6.85  &$..$  &J16150354$-$5135263  &    \\
   18   & 16 12 11.56  & $-$50 53 18.38  &   \nodata   &  11.94   &   9.27   &   7.73   &   5.59   &  $-$0.10   &$-$1.56   & 2.36     & 1.94 $\pm$   3.40    &   \nodata   &  \nodata   &   \nodata &  12.66  & $-$7.29  &$..$  &J16121156$-$5053183  &    \\
   19   & 16 17 06.78  & $-$51 24 46.78  &   \nodata   &  10.92   &   8.94   &   7.84   &   6.29   &  $-$0.00   &$-$1.09   & 1.73     &$-$2.70 $\pm$   0.55    &   \nodata   &  \nodata   &   \nodata &  12.66  & $-$6.55  &$..$  &J16170678$-$5124467  &    \\
   20   & 16 17 52.51  & $-$50 44 41.75  &  12.17   &  10.51   &   9.50   &   8.49   &   5.53   &  $-$0.81   &$-$5.94   & 1.38     & 0.38 $\pm$   0.03    &  11.94    & $-$78.43    &  13.25  &   \nodata & $-$4.83  &$..$  &J16175251$-$5044417  &    \\
   21   & 16 14 13.82  & $-$51 30 39.40  &  13.60   &  10.33   &   9.28   &   8.62   &   6.97   &  $-$0.14   &$-$2.73   & 1.03     &$-$0.14 $\pm$   0.10    &   \nodata   &  \nodata   &   \nodata &  12.66  & $-$5.07  &$..$  &J16141381$-$5130393  &    \\
   22   & 16 18 14.74  & $-$51 32 04.72  &   \nodata   &  12.70   &  10.24   &   8.64   &   4.87   &  $-$0.42   &$-$6.11   & 2.37     & 0.54 $\pm$   1.05    &   \nodata   &  \nodata   &   \nodata &  12.66  & $-$6.39  &$..$  &J16181474$-$5132047/IRAS~16144$-$5124  &    \\
   23   & 16 12 46.71  & $-$50 48 00.56  &  13.84   &  12.61   &  10.35   &   8.69   &   6.25   &  $-$0.73   &$-$2.65   & 2.37     & 0.94 $\pm$   0.04    &  10.06    &  \nodata   &   \nodata &   \nodata & $-$3.74  &$..$  &J16124671$-$5048005  &    \\
   24   & 16 16 27.36  & $-$51 36 28.48  &   \nodata   &  12.88   &  10.46   &   8.78   &   5.15   &  $-$0.60   &$-$5.64   & 2.44     & \nodata    &   \nodata   &  \nodata   &   \nodata &  12.66  & $-$6.32  &$..$  &J16162735$-$5136284/MSX6C~G331.9057$-$00.6572  &AB?   \\
   25   & 16 16 25.44  & $-$50 58 31.04  &   \nodata   &  12.23   &  10.72   &   8.94   &   6.49   &  $-$1.68   &$-$3.31   & 2.29     & 0.59 $\pm$   0.04    &  11.03    &  \nodata   &   \nodata &   \nodata & $-$4.37  &$..$  &J16162543$-$5058310  &    \\
   26   & 16 15 58.26  & $-$50 52 45.37  &   \nodata   &  13.36   &  10.86   &   9.17   &   5.31   &  $-$0.54   &$-$6.21   & 2.47     & 0.11 $\pm$   0.86    &   \nodata   &  \nodata   &   \nodata &  12.66  & $-$5.96  &$..$  &J16155825$-$5052453/2MASS~J16155825$-$5052453  &AB?   \\
   27   & 16 17 44.25  & $-$50 51 09.87  &  12.88   &  11.40   &  10.19   &   9.22   &   7.69   &  $-$0.52   &$-$1.94   & 1.38     & 0.50 $\pm$   0.03    &  11.37    &  \nodata   &   \nodata &   \nodata & $-$3.53  &$..$  &J16174425$-$5051098  &    \\
   28   & 16 15 23.44  & $-$51 26 24.49  &  16.11   &  14.45   &  11.40   &   9.22   &   5.61   &  $-$0.86   &$-$4.50   & 3.11     & 0.30 $\pm$   0.09    &   \nodata   &  \nodata   &   \nodata &  12.66  & $-$6.55  &$..$  &J16152344$-$5126244  &    \\
   29   & 16 13 20.00  & $-$51 23 23.64  &  16.18   &  12.53   &  10.65   &   9.25   &   5.61   &  $-$0.64   &$-$6.51   & 2.01     & 0.13 $\pm$   0.18    &   \nodata   &  \nodata   &   \nodata &  12.66  & $-$5.42  &$..$  &J16132000$-$5123236/2MASS~J16132000$-$5123236  &AB?   \\
   30   & 16 16 35.42  & $-$50 40 45.99  &   \nodata   &  10.14   &   9.66   &   9.40   &   8.52   &  $-$0.01   &$-$1.62   & 0.48     & 0.18 $\pm$   0.05    &   \nodata   &  \nodata   &   \nodata &  12.66  & $-$3.74  &$..$  &J16163542$-$5040459  &    \\
   31   & 16 12 00.39  & $-$51 01 38.85  &  12.49   &  10.92   &  10.11   &   9.51   &   8.12   &  $-$0.28   &$-$2.32   & 0.91     & 0.46 $\pm$   0.03    &  11.55    &  $-$2.73    &   7.24  &   \nodata & $-$2.95  &$..$  &J16120038$-$5101388  &    \\
   32   & 16 15 00.75  & $-$51 38 33.97  &   \nodata   &  14.12   &  11.33   &   9.58   &   6.41   &  $-$0.36   &$-$3.99   & 2.61     &$-$0.92 $\pm$   0.66    &   \nodata   &  \nodata   &   \nodata &  12.66  & $-$5.69  &$..$  &J16150075$-$5138339/2MASS~J16150075$-$5138339  &AB?   \\
   33   & 16 14 10.41  & $-$51 29 32.30  &  16.45   &  14.35   &  11.50   &   9.77   &   7.62   &  $-$0.26   &$-$1.18   & 2.61     & 0.03 $\pm$   0.16    &   \nodata   &  \nodata   &   \nodata &  12.66  & $-$5.50  &$..$  &J16141041$-$5129323  &    \\
   34   & 16 18 39.78  & $-$50 46 46.77  &  15.68   &  14.22   &  11.62   &   9.81   &   7.38   &  $-$0.66   &$-$2.15   & 2.62     & 0.42 $\pm$   0.07    &  11.74    &  \nodata   &   \nodata &   \nodata & $-$4.55  &$..$  &J16183978$-$5046467  &    \\
   35   & 16 16 11.73  & $-$50 38 08.75  &   \nodata   &  13.67   &  11.74   &   9.99   &   8.23   &  $-$1.21   &$-$1.04   & 2.38     & 0.18 $\pm$   0.15    &   \nodata   &  \nodata   &   \nodata &  12.66  & $-$5.05  &$..$  &J16161172$-$5038087  &    \\
\hline
\end{tabular}
}
\begin{list}{}
\item Candidate massive stars (\Ks $<9$ mag) with free-free emission. For a few targets SIMBAD provides information is found as annotated.
AB? = Asymptotic Giant Branch Star candidate;  C = carbon star; PN = planetary nebula; cLBV= candidate LBV in the present work.
\item Columns are as in Table \ref{obknown}. 
\end{list}
\end{table*}

\begin{table*}
\caption{Candidate Normal OB Stars. \label{table.ob}}
{\tiny
\begin{tabular}{@{\extracolsep{-.08in}}rllrrrrrrrrrrrrrrrll}
\hline
\hline
ID &            Ra(J2000)      & Dec(J2000)         & Iden     &$J$       &$H$       &\Ks     &  [8.0] &     Q1 & Q2    &  \Aks & plx & DM & VR    &DMv  & DMadopt   & \Mk &VAR &2MASS-ID/Alias & Comments\\
   &    $($hh mm ss$)$  & $($deg mm ss$)$& (mag)    &(mag)     &(mag)     &(mag)     &  (mag) &  (mag) & (mag)  &(mag)  &(mas)& (mag)&(\kms)  &(mag)&(mag)&(mag)&    &      &\\
\hline
    1   & 16 16 55.83  & $-$51 32 12.26  &   \nodata   &   6.54   &   5.28   &   4.63   &   3.96   &   0.10   & 0.11   & 1.07     & 0.17 $\pm$   0.29    &   \nodata   &  \nodata   &   \nodata &  12.66  & $-$9.10  &$..$  &J16165583$-$5132122  &    \\
    2   & 16 16 42.73  & $-$50 21 29.69  &   \nodata   &   6.29   &   5.20   &   4.66   &   4.25   &   0.12   & 0.55   & 0.93     & 1.17 $\pm$   0.19    &   \nodata   &  \nodata   &   \nodata &  12.66  & $-$8.93  &$..$  &J16164272$-$5021296  &    \\
    3   & 16 17 51.61  & $-$50 34 39.46  &  10.86   &   6.83   &   5.61   &   4.99   &   4.32   &   0.12   & 0.01   & 1.03     & 1.67 $\pm$   0.16    &   \nodata   &  \nodata   &   \nodata &  12.66  & $-$8.70  &VAR  &J16175160$-$5034394  &    \\
    4   & 16 19 47.26  & $-$50 57 45.72  &   8.86   &   5.85   &   5.31   &   5.04   &   4.61   &   0.05   &$-$0.37   & 0.50     & 0.19 $\pm$   0.04    &   \nodata   &  \nodata   &   \nodata &  12.66  & $-$8.11  &$..$  &J16194726$-$5057457/TYC~8324$-$1718$-$1  &    \\
    5   & 16 19 25.80  & $-$51 19 47.51  &  11.44   &   7.06   &   5.80   &   5.13   &   4.32   &   0.06   &$-$0.27   & 1.09     & 1.25 $\pm$   0.20    &   \nodata   &  \nodata   &   \nodata &  12.66  & $-$8.61  &$..$  &J16192580$-$5119475  &    \\
    6   & 16 18 00.98  & $-$51 09 19.27  &  14.78   &   8.07   &   6.26   &   5.28   &   4.27   &   0.03   & 0.09   & 1.57     &$-$0.24 $\pm$   0.23    &   \nodata   &  \nodata   &   \nodata &  12.66  & $-$8.95  &$..$  &J16180098$-$5109192  &    \\
    7   & 16 15 06.28  & $-$51 12 12.76  &  12.09   &   7.44   &   6.04   &   5.30   &   4.47   &   0.08   &$-$0.11   & 1.20     & 0.09 $\pm$   0.20    &   \nodata   &  \nodata   &   \nodata &  12.66  & $-$8.56  &$..$  &J16150628$-$5112127  &    \\
    8   & 16 13 25.20  & $-$50 55 19.93  &  12.96   &   7.77   &   6.21   &   5.35   &   4.40   &   0.01   &$-$0.14   & 1.37     & 0.77 $\pm$   0.20    &   \nodata   &  \nodata   &   \nodata &  12.66  & $-$8.68  &$..$  &J16132519$-$5055199  &    \\
    9   & 16 17 58.45  & $-$50 26 56.50  &  12.82   &   7.97   &   6.25   &   5.38   &   4.46   &   0.14   & 0.13   & 1.43     & 0.70 $\pm$   0.19    &   \nodata   &  \nodata   &   \nodata &  12.66  & $-$8.71  &$..$  &J16175844$-$5026565  &    \\
   10   & 16 19 36.95  & $-$51 15 29.28  &  14.68   &   8.41   &   6.57   &   5.60   &   4.43   &   0.10   &$-$0.35   & 1.56     & 0.66 $\pm$   0.22    &   \nodata   &  \nodata   &   \nodata &  12.66  & $-$8.62  &$..$  &J16193694$-$5115292  &    \\
   11   & 16 17 18.77  & $-$51 01 36.43  &   \nodata   &   8.75   &   6.74   &   5.61   &   4.15   &  $-$0.03   &$-$0.77   & 1.77     & 0.10 $\pm$   0.32    &   \nodata   &  \nodata   &   \nodata &  12.66  & $-$8.82  &$..$  &J16171876$-$5101364  &    \\
   12   & 16 15 20.52  & $-$51 06 20.91  &  14.47   &   8.66   &   6.85   &   5.82   &   4.43   &  $-$0.03   &$-$0.90   & 1.61     &$-$0.22 $\pm$   0.33    &   \nodata   &  \nodata   &   \nodata &  12.66  & $-$8.45  &$..$  &J16152052$-$5106209  &    \\
   13   & 16 13 51.52  & $-$50 28 53.75  &   \nodata   &  10.50   &   7.60   &   5.95   &   4.00   &  $-$0.08   &$-$0.69   & 2.54     & 1.87 $\pm$   1.66    &   \nodata   &  \nodata   &   \nodata &  12.66  & $-$9.25  &$..$  &J16135151$-$5028537  &    \\
   14   & 16 13 46.57  & $-$50 52 57.95  &  17.54   &   9.26   &   7.16   &   6.00   &   4.55   &   0.02   &$-$0.64   & 1.82     &$-$0.43 $\pm$   0.34    &   \nodata   &  \nodata   &   \nodata &  12.66  & $-$8.48  &$..$  &J16134657$-$5052579  &    \\
   15   & 16 19 10.20  & $-$51 20 30.50  &  11.18   &   7.86   &   6.62   &   6.01   &   5.70   &   0.14   & 1.01   & 1.03     & 1.02 $\pm$   0.16    &   \nodata   &  \nodata   &   \nodata &  12.66  & $-$7.68  &$..$  &J16191020$-$5120304  &    \\
   16   & 16 19 39.13  & $-$51 02 59.16  &  12.28   &   7.97   &   6.75   &   6.10   &   \nodata   &   0.06   & \nodata  & \nodata    &$-$0.02 $\pm$   0.15    &   \nodata   &  \nodata   &   \nodata &  12.66  &  \nodata &$..$  &J16193912$-$5102591  &    \\
   17   & 16 13 28.78  & $-$50 42 55.18  &  17.48   &   9.60   &   7.45   &   6.31   &   5.09   &   0.09   & 0.01   & 1.82     & \nodata $\pm$   0.45    &   \nodata   &  \nodata   &   \nodata &  12.66  & $-$8.17  &$..$  &J16132878$-$5042551  &    \\
   18   & 16 18 20.03  & $-$50 50 20.73  &  10.12   &   7.61   &   6.83   &   6.43   &   5.80   &   0.05   &$-$0.51   & 0.70     & 0.27 $\pm$   0.13    &   \nodata   &  \nodata   &   \nodata &  12.66  & $-$6.93  &$..$  &J16182003$-$5050207  &    \\
   19   & 16 12 47.18  & $-$50 59 45.21  &  10.99   &   7.86   &   6.98   &   6.46   &   5.92   &  $-$0.07   &$-$0.03   & 0.85     & 0.61 $\pm$   0.10    &  10.96    &  \nodata   &   \nodata &   \nodata & $-$5.36  &$..$  &J16124717$-$5059452  &    \\
   20   & 16 17 23.27  & $-$50 30 19.65  &  11.39   &   7.99   &   7.03   &   6.49   &   5.93   &  $-$0.02   & 0.02   & 0.89     & 1.02 $\pm$   0.15    &   \nodata   &  \nodata   &   \nodata &  12.66  & $-$7.06  &$..$  &J16172326$-$5030196  &    \\
   21   & 16 18 19.24  & $-$51 23 39.79  &   9.11   &   7.26   &   6.82   &   6.53   &   6.20   &  $-$0.10   &$-$0.13   & 0.50     & 0.20 $\pm$   0.05    &   \nodata   &  \nodata   &   \nodata &  12.66  & $-$6.63  &$..$  &J16181923$-$5123397/TYC~8323$-$2185$-$1  &    \\
   22   & 16 14 52.17  & $-$51 20 29.92  &  12.56   &   8.59   &   7.25   &   6.58   &   6.10   &   0.13   & 0.73   & 1.12     & 0.42 $\pm$   0.16    &   \nodata   &  \nodata   &   \nodata &  12.66  & $-$7.20  &VAR  &J16145216$-$5120299  &    \\
   23   & 16 12 34.67  & $-$51 16 42.24  &  10.94   &   7.98   &   7.18   &   6.71   &   6.17   &  $-$0.05   &$-$0.16   & 0.78     & 0.42 $\pm$   0.10    &   \nodata   &  \nodata   &   \nodata &  12.66  & $-$6.73  &$..$  &J16123467$-$5116422  &    \\
   24   & 16 18 45.09  & $-$51 15 50.01  &   9.99   &   7.76   &   7.12   &   6.75   &   6.30   &  $-$0.03   &$-$0.19   & 0.63     & 0.28 $\pm$   0.11    &   \nodata   &  \nodata   &   \nodata &  12.66  & $-$6.54  &$..$  &J16184508$-$5115500  &    \\
   25   & 16 14 03.22  & $-$50 58 20.95  &  14.35   &   9.25   &   7.75   &   6.96   &   6.07   &   0.09   &$-$0.12   & 1.28     & 0.20 $\pm$   0.21    &   \nodata   &  \nodata   &   \nodata &  12.66  & $-$6.98  &$..$  &J16140322$-$5058209  &    \\
   26   & 16 19 20.50  & $-$51 02 54.18  &  14.07   &   9.26   &   7.87   &   7.08   &   \nodata   &  $-$0.03   & \nodata  & \nodata    & 0.28 $\pm$   0.18    &   \nodata   &  \nodata   &   \nodata &  12.66  &  \nodata &$..$  &J16192049$-$5102541  &    \\
   27   & 16 17 04.50  & $-$51 15 19.68  &  10.59   &   8.21   &   7.54   &   7.17   &   6.67   &  $-$0.00   &$-$0.30   & 0.64     & 0.31 $\pm$   0.09    &   \nodata   &  \nodata   &   \nodata &  12.66  & $-$6.13  &$..$  &J16170449$-$5115196  &    \\
   28   & 16 17 22.84  & $-$51 32 42.86  &  14.71   &   9.59   &   8.03   &   7.22   &   6.18   &   0.11   &$-$0.45   & 1.32     &$-$0.96 $\pm$   0.31    &   \nodata   &  \nodata   &   \nodata &  12.66  & $-$6.76  &$..$  &J16172283$-$5132428  &    \\
   29   & 16 13 54.73  & $-$50 47 41.68  &   \nodata   &  10.93   &   8.57   &   7.28   &   5.85   &   0.04   &$-$0.22   & 2.02     &$-$0.22 $\pm$   0.91    &   \nodata   &  \nodata   &   \nodata &  12.66  & $-$7.40  &$..$  &J16135473$-$5047416  &    \\
   30   & 16 19 13.72  & $-$50 51 13.89  &  14.00   &   9.70   &   8.17   &   7.38   &   6.68   &   0.12   & 0.45   & 1.29     & 0.18 $\pm$   0.17    &   \nodata   &  \nodata   &   \nodata &  12.66  & $-$6.57  &$..$  &J16191372$-$5051138  &    \\
   31   & 16 17 21.71  & $-$51 34 39.26  &  16.65   &  10.26   &   8.42   &   7.41   &   6.16   &   0.02   &$-$0.53   & 1.60     & 0.15 $\pm$   0.37    &   \nodata   &  \nodata   &   \nodata &  12.66  & $-$6.85  &$..$  &J16172170$-$5134392  &    \\
   32   & 16 16 04.91  & $-$51 36 33.96  &   \nodata   &   8.18   &   7.69   &   7.44   &   6.96   &   0.05   &$-$0.56   & 0.46     & 0.23 $\pm$   0.07    &   \nodata   &  \nodata   &   \nodata &  12.66  & $-$5.68  &$..$  &J16160490$-$5136339  &    \\
   33   & 16 14 25.74  & $-$51 21 41.00  &  10.28   &   8.35   &   7.83   &   7.51   &   7.23   &  $-$0.07   & 0.09   & 0.55     & 0.23 $\pm$   0.07    &   \nodata   &  \nodata   &   \nodata &  12.66  & $-$5.70  &$..$  &J16142574$-$5121410  &    \\
   34   & 16 14 35.41  & $-$51 21 44.74  &   \nodata   &   8.74   &   7.93   &   7.53   &   7.17   &   0.09   & 0.25   & 0.70     & 0.41 $\pm$   0.08    &   \nodata   &  \nodata   &   \nodata &  12.66  & $-$5.83  &$..$  &J16143540$-$5121447  &    \\
   35   & 16 19 30.74  & $-$51 20 07.67  &  14.19   &   9.66   &   8.25   &   7.54   &   6.87   &   0.15   & 0.30   & 1.18     &$-$0.07 $\pm$   0.19    &   \nodata   &  \nodata   &   \nodata &  12.66  & $-$6.30  &$..$  &J16193074$-$5120076  &    \\
   36   & 16 19 30.44  & $-$51 23 39.89  &  15.97   &  10.07   &   8.46   &   7.55   &   6.62   &  $-$0.04   & 0.03   & 1.44     & 0.06 $\pm$   0.27    &   \nodata   &  \nodata   &   \nodata &  12.66  & $-$6.55  &$..$  &J16193043$-$5123398  &    \\
   37   & 16 17 08.28  & $-$51 27 43.04  &   9.79   &   8.29   &   7.84   &   7.63   &   7.32   &   0.07   &$-$0.15   & 0.42     & 0.11 $\pm$   0.07    &   \nodata   &  \nodata   &   \nodata &  12.66  & $-$5.45  &$..$  &J16170828$-$5127430  &    \\
   38   & 16 15 10.94  & $-$51 36 18.55  &  11.93   &   9.01   &   8.13   &   7.70   &   7.10   &   0.08   &$-$0.30   & 0.76     &$-$0.04 $\pm$   0.10    &   \nodata   &  \nodata   &   \nodata &  12.66  & $-$5.72  &$..$  &J16151093$-$5136185  &    \\
   39   & 16 14 21.90  & $-$51 36 03.24  &  13.19   &   9.62   &   8.55   &   7.98   &   7.39   &   0.04   & 0.07   & 0.95     & 0.25 $\pm$   0.12    &   \nodata   &  \nodata   &   \nodata &  12.66  & $-$5.63  &$..$  &J16142189$-$5136032  &    \\
   40   & 16 14 34.10  & $-$50 33 01.83  &   \nodata   &  11.41   &   9.17   &   8.00   &   6.96   &   0.13   & 0.63   & 1.87     & \nodata    &   \nodata   &  \nodata   &   \nodata &  12.66  & $-$6.53  &$..$  &J16143410$-$5033018  &    \\
   41   & 16 13 08.16  & $-$50 51 28.93  &   \nodata   &  11.98   &   9.46   &   8.05   &   6.45   &  $-$0.01   &$-$0.36   & 2.19     & \nodata    &   \nodata   &  \nodata   &   \nodata &  12.66  & $-$6.80  &$..$  &J16130815$-$5051289  &    \\
   42   & 16 13 13.15  & $-$51 08 37.11  &   \nodata   &  12.58   &   9.65   &   8.05   &   7.12   &   0.05   & 2.02   & 2.49     & \nodata    &   \nodata   &  \nodata   &   \nodata &  12.66  & $-$7.10  &   &J16131315$-$5108371  &    \\
   43   & 16 15 33.81  & $-$51 12 09.05  &  16.15   &  10.59   &   8.94   &   8.06   &   7.13   &   0.08   &$-$0.00   & 1.41     & 0.50 $\pm$   0.28    &   \nodata   &  \nodata   &   \nodata &  12.66  & $-$6.01  &$..$  &J16153381$-$5112090  &    \\
   44   & 16 15 40.77  & $-$51 20 59.33  &  14.21   &  10.05   &   8.78   &   8.10   &   7.08   &   0.06   &$-$0.80   & 1.11     &$-$0.02 $\pm$   0.18    &   \nodata   &  \nodata   &   \nodata &  12.66  & $-$5.67  &$..$  &J16154076$-$5120593  &    \\
   45   & 16 20 08.17  & $-$50 48 30.83  &  10.87   &   8.96   &   8.44   &   8.11   &   7.73   &  $-$0.08   &$-$0.17   & 0.56     & 0.28 $\pm$   0.06    &  12.52    &  \nodata   &   \nodata &   \nodata & $-$4.97  &$..$  &J16200817$-$5048308  &    \\
   46   & 16 14 09.48  & $-$50 58 25.93  &  13.84   &   9.88   &   8.72   &   8.14   &   7.48   &   0.11   &$-$0.04   & 0.98     & 0.33 $\pm$   0.14    &   \nodata   &  \nodata   &   \nodata &  12.66  & $-$5.49  &$..$  &J16140948$-$5058259  &    \\
   47   & 16 17 41.39  & $-$50 28 25.59  &  14.85   &  10.24   &   8.91   &   8.22   &   7.39   &   0.10   &$-$0.21   & 1.14     & 0.84 $\pm$   0.19    &   \nodata   &  \nodata   &   \nodata &  12.66  & $-$5.58  &$..$  &J16174139$-$5028255  &    \\
   48   & 16 19 24.80  & $-$50 57 51.65  &   9.70   &   8.64   &   8.39   &   8.22   &   7.92   &  $-$0.06   &$-$0.38   & 0.32     & 0.31 $\pm$   0.04    &  12.36    &  \nodata   &   \nodata &   \nodata & $-$4.46  &$..$  &J16192480$-$5057516/TYC~8324$-$913$-$1  &    \\
   49   & 16 16 19.50  & $-$51 07 27.09  &   \nodata   &  10.37   &   8.98   &   8.25   &   7.52   &   0.09   & 0.14   & 1.19     &$-$0.52 $\pm$   0.23    &   \nodata   &  \nodata   &   \nodata &  12.66  & $-$5.60  &$..$  &J16161949$-$5107270  &    \\
   50   & 16 15 52.56  & $-$51 37 56.41  &  11.30   &   9.29   &   8.73   &   8.37   &   7.84   &  $-$0.11   &$-$0.49   & 0.60     & \nodata    &   \nodata   &  \nodata   &   \nodata &  12.66  & $-$4.89  &   &J16155255$-$5137564  &    \\
   51   & 16 17 30.77  & $-$50 31 16.89  &  13.12   &   9.80   &   8.90   &   8.41   &   7.86   &   0.01   &$-$0.08   & 0.82     & 0.51 $\pm$   0.12    &   \nodata   &  \nodata   &   \nodata &  12.66  & $-$5.07  &$..$  &J16173077$-$5031168  &    \\
   52   & 16 18 23.87  & $-$50 40 12.31  &  13.47   &   9.93   &   8.93   &   8.42   &   7.69   &   0.07   &$-$0.45   & 0.87     & 0.20 $\pm$   0.12    &   \nodata   &  \nodata   &   \nodata &  12.66  & $-$5.11  &$..$  &J16182386$-$5040123  &    \\
   53   & 16 18 51.47  & $-$51 17 32.70  &  11.83   &   9.47   &   8.80   &   8.43   &   7.90   &   0.02   &$-$0.39   & 0.63     & 0.08 $\pm$   0.08    &   \nodata   &  \nodata   &   \nodata &  12.66  & $-$4.86  &$..$  &J16185147$-$5117326  &    \\
   54   & 16 17 02.59  & $-$50 27 09.75  &  12.18   &   9.70   &   8.96   &   8.54   &   7.88   &  $-$0.02   &$-$0.63   & 0.70     & 0.41 $\pm$   0.08    &  11.80    &  \nodata   &   \nodata &   \nodata & $-$3.96  &$..$  &J16170258$-$5027097  &    \\
   55   & 16 13 34.20  & $-$50 59 14.17  &  13.28   &  10.03   &   9.03   &   8.55   &   8.07   &   0.15   & 0.19   & 0.83     & 0.27 $\pm$   0.10    &   \nodata   &  \nodata   &   \nodata &  12.66  & $-$4.94  &$..$  &J16133419$-$5059141  &    \\
   56   & 16 14 33.05  & $-$51 28 30.90  &   \nodata   &  13.69   &  10.47   &   8.59   &   6.56   &  $-$0.16   &$-$0.35   & 2.85     & 0.45 $\pm$   0.06    &  11.61    &  \nodata   &   \nodata &   \nodata & $-$5.87  &$..$  &J16143304$-$5128309  &    \\
   57   & 16 18 13.56  & $-$51 18 26.02  &   \nodata   &  11.47   &   9.67   &   8.59   &   7.45   &  $-$0.15   &$-$0.20   & 1.66     & 0.43 $\pm$   0.03    &  11.68    &  \nodata   &   \nodata &   \nodata & $-$4.75  &$..$  &J16181355$-$5118260  &    \\
   58   & 16 17 29.97  & $-$50 30 56.67  &  13.19   &   9.99   &   9.09   &   8.59   &   7.96   &   \nodata   &$-$0.31   & 0.83     & 0.39 $\pm$   0.13    &   \nodata   &  \nodata   &   \nodata &  12.66  & $-$4.90  &$..$  &J16172997$-$5030566  &    \\
   59   & 16 18 01.77  & $-$51 27 01.22  &   \nodata   &  12.20   &   9.93   &   8.67   &   7.02   &  $-$0.00   &$-$0.93   & 1.97     &$-$2.09 $\pm$   0.51    &   \nodata   &  \nodata   &   \nodata &  12.66  & $-$5.96  &$..$  &J16180176$-$5127012  &    \\
   60   & 16 19 52.70  & $-$50 54 15.03  &  10.78   &   9.32   &   8.91   &   8.73   &   8.45   &   0.06   &$-$0.15   & 0.38     &$-$0.31 $\pm$   0.11    &   \nodata   &  \nodata   &   \nodata &  12.66  & $-$4.31  &$..$  &J16195270$-$5054150  &    \\
   61   & 16 20 18.78  & $-$51 04 58.69  &  10.31   &   9.18   &   8.91   &   8.76   &   8.63   &  $-$0.00   & 0.06   & 0.30     & 0.34 $\pm$   0.04    &  12.13    &  \nodata   &   \nodata &   \nodata & $-$3.68  &$..$  &J16201878$-$5104586  &    \\

\hline
\end{tabular}
}
\begin{list}{}
\item Columns are as in Table \ref{obknown}.
\end{list}
\end{table*}

\addtocounter{table}{-1}
\begin{table*}
\caption{Continuation of Table \ref{table.ob}}
{\tiny
\begin{tabular}{@{\extracolsep{-.08in}}rllrrrrrrrrrrrrrrrll}
\hline
\hline
ID &            Ra(J2000)      & Dec(J2000)         & Iden     &$J$       &$H$       &\Ks     &  [8.0] &     Q1 & Q2    &  \Aks & plx & DM & VR    &DMv  & DMadopt   & \Mk &VAR &2MASS-ID/Alias & Comments\\
   &    $($hh mm ss$)$  & $($deg mm ss$)$& (mag)    &(mag)     &(mag)     &(mag)     &  (mag) &  (mag) & (mag)  &(mag)  &(mas)& (mag)&(\kms)  &(mag)&(mag)&(mag)&    &      &\\
\hline

   62   & 16 19 42.80  & $-$51 14 26.33  &  11.84   &   9.66   &   9.10   &   8.79   &   8.45   &   0.02   &$-$0.05   & 0.54     & 0.19 $\pm$   0.07    &   \nodata   &  \nodata   &   \nodata &  12.66  & $-$4.41  &$..$  &J16194279$-$5114263  &    \\
   63   & 16 14 44.01  & $-$50 44 03.89  &  10.94   &   9.49   &   9.06   &   8.85   &   8.52   &   0.04   &$-$0.22   & 0.41     & 0.29 $\pm$   0.06    &  12.47    &  \nodata   &   \nodata &   \nodata & $-$4.03  &$..$  &J16144401$-$5044038  &    \\
   64   & 16 18 08.57  & $-$50 28 55.06  &  10.28   &   9.40   &   9.09   &   8.93   &   8.78   &   0.04   & 0.06   & 0.32     & 0.48 $\pm$   0.04    &  11.45    &  \nodata   &   \nodata &   \nodata & $-$2.84  &$..$  &J16180856$-$5028550/TYC~8319$-$705$-$1  &    \\
   65   & 16 17 50.93  & $-$51 36 34.22  &  10.67   &   9.52   &   9.18   &   8.95   &   8.79   &  $-$0.06   & 0.16   & 0.40     & 0.24 $\pm$   0.05    &  12.85    &  \nodata   &   \nodata &   \nodata & $-$4.30  &$..$  &J16175093$-$5136342  &    \\
   66   & 16 12 09.88  & $-$50 58 08.36  &   \nodata   &  10.37   &   9.44   &   9.00   &   \nodata   &   0.13   & \nodata  & \nodata    & 0.39 $\pm$   0.03    &  11.90    &  \nodata   &   \nodata &   \nodata &  \nodata &$..$  &J16120988$-$5058083  &    \\
   67   & 16 19 46.48  & $-$51 02 57.04  &  12.80   &  10.30   &   9.55   &   9.03   &   \nodata   &  $-$0.18   & \nodata  & \nodata    &$-$0.06 $\pm$   0.09    &   \nodata   &  \nodata   &   \nodata &  12.66  &  \nodata &$..$  &J16194647$-$5102570  &    \\
   68   & 16 19 45.37  & $-$51 02 59.65  &  12.82   &  10.31   &   9.62   &   9.12   &   \nodata   &  $-$0.20   & \nodata  & \nodata    & \nodata    &   \nodata   &  \nodata   &   \nodata &  12.66  &  \nodata &$..$  &J16194536$-$5102596  &    \\
   69   & 16 17 35.20  & $-$50 31 26.34  &  11.49   &   9.78   &   9.34   &   9.13   &   8.80   &   0.08   &$-$0.24   & 0.40     &$-$0.64 $\pm$   0.28    &   \nodata   &  \nodata   &   \nodata &  12.66  & $-$3.92  &$..$  &J16173520$-$5031263  &    \\
   70   & 16 15 39.41  & $-$51 40 22.31  &  11.78   &   9.93   &   9.42   &   9.14   &   8.86   &   0.01   & 0.03   & 0.50     & 0.21 $\pm$   0.06    &   \nodata   &  \nodata   &   \nodata &  12.66  & $-$4.02  &$..$  &J16153940$-$5140223  &    \\
   71   & 16 13 07.79  & $-$50 56 13.43  &  16.13   &  11.44   &   9.95   &   9.14   &   8.46   &   0.02   & 0.44   & 1.30     &$-$0.48 $\pm$   0.23    &   \nodata   &  \nodata   &   \nodata &  12.66  & $-$4.82  &$..$  &J16130778$-$5056134  &    \\
   72   & 16 19 37.87  & $-$51 18 25.60  &  11.28   &   9.82   &   9.40   &   9.22   &   8.96   &   0.09   &$-$0.09   & 0.37     & 0.18 $\pm$   0.05    &   \nodata   &  \nodata   &   \nodata &  12.66  & $-$3.81  &$..$  &J16193787$-$5118255  &    \\
   73   & 16 16 10.12  & $-$51 38 01.68  &   \nodata   &   9.87   &   9.48   &   9.27   &   9.14   &   0.02   & 0.24   & 0.39     & 0.16 $\pm$   0.05    &   \nodata   &  \nodata   &   \nodata &  12.66  & $-$3.78  &$..$  &J16161012$-$5138016  &    \\
   74   & 16 18 41.46  & $-$51 01 22.62  &  14.06   &  10.83   &   9.80   &   9.29   &   8.88   &   0.11   & 0.42   & 0.87     & 0.24 $\pm$   0.10    &   \nodata   &  \nodata   &   \nodata &  12.66  & $-$4.24  &$..$  &J16184146$-$5101226  &    \\
   75   & 16 17 33.89  & $-$50 34 39.21  &  13.45   &  10.57   &   9.71   &   9.30   &   8.97   &   0.12   & 0.39   & 0.73     & 0.54 $\pm$   0.10    &  11.22    &  \nodata   &   \nodata &   \nodata & $-$2.66  &$..$  &J16173389$-$5034392  &    \\
   76   & 16 18 23.90  & $-$50 39 01.70  &  13.93   &  10.72   &   9.79   &   9.30   &   8.80   &   0.04   & 0.06   & 0.83     & 0.08 $\pm$   0.12    &   \nodata   &  \nodata   &   \nodata &  12.66  & $-$4.19  &$..$  &J16182390$-$5039016  &    \\
   77   & 16 19 01.12  & $-$51 10 13.29  &  10.28   &   9.65   &   9.41   &   9.31   &   9.20   &   0.05   & 0.07   & 0.25     & 0.67 $\pm$   0.04    &  10.78    &  \nodata   &   \nodata &   \nodata & $-$1.73  &$..$  &J16190111$-$5110132/TYC~8324$-$972$-$1  &    \\
   78   & 16 15 16.33  & $-$51 06 55.28  &  12.98   &  10.44   &   9.70   &   9.33   &   8.94   &   0.10   & 0.05   & 0.65     & 0.14 $\pm$   0.11    &   \nodata   &  \nodata   &   \nodata &  12.66  & $-$3.98  &$..$  &J16151633$-$5106552  &    \\
   79   & 16 15 57.33  & $-$51 10 06.57  &  12.33   &  10.29   &   9.68   &   9.37   &   9.07   &   0.05   & 0.10   & 0.56     & 0.13 $\pm$   0.07    &   \nodata   &  \nodata   &   \nodata &  12.66  & $-$3.85  &$..$  &J16155732$-$5110065  &    \\
   80   & 16 19 18.73  & $-$51 01 11.04  &  11.67   &  10.14   &   9.63   &   9.39   &   9.29   &   0.06   & 0.47   & 0.46     & 0.35 $\pm$   0.04    &  12.12    &  \nodata   &   \nodata &   \nodata & $-$3.18  &$..$  &J16191873$-$5101110  &    \\
   81   & 16 17 29.36  & $-$50 30 31.98  &  14.56   &  10.95   &   9.95   &   9.45   &   9.09   &   0.09   & 0.53   & 0.86     & 0.22 $\pm$   0.14    &   \nodata   &  \nodata   &   \nodata &  12.66  & $-$4.07  &$..$  &J16172935$-$5030319  &    \\
   82   & 16 18 40.94  & $-$50 31 58.69  &  14.78   &  11.07   &  10.04   &   9.47   &   8.89   &   0.02   & 0.03   & 0.93     & 0.24 $\pm$   0.17    &   \nodata   &  \nodata   &   \nodata &  12.66  & $-$4.12  &$..$  &J16184094$-$5031586  &    \\
   83   & 16 17 21.14  & $-$50 47 29.44  &  13.77   &  10.82   &   9.96   &   9.48   &   9.07   &   0.01   & 0.26   & 0.80     & 0.36 $\pm$   0.10    &   \nodata   &  \nodata   &   \nodata &  12.66  & $-$3.98  &$..$  &J16172114$-$5047294  &    \\
   84   & 16 14 44.60  & $-$50 43 30.86  &  11.16   &  10.00   &   9.66   &   9.50   &   9.31   &   0.04   &$-$0.02   & 0.33     & 0.32 $\pm$   0.05    &  12.30    &  \nodata   &   \nodata &   \nodata & $-$3.13  &$..$  &J16144459$-$5043308  &    \\
   85   & 16 16 24.75  & $-$51 09 55.20  &   \nodata   &  11.21   &  10.18   &   9.53   &   8.76   &  $-$0.12   &$-$0.41   & 1.01     & 0.26 $\pm$   0.04    &  12.68    &  \nodata   &   \nodata &   \nodata & $-$4.15  &$..$  &J16162474$-$5109551  &    \\
   86   & 16 13 55.24  & $-$51 26 05.77  &   \nodata   &  12.54   &  10.60   &   9.54   &   8.54   &   0.06   & 0.31   & 1.67     &$-$0.15 $\pm$   0.19    &   \nodata   &  \nodata   &   \nodata &  12.66  & $-$4.79  &$..$  &J16135523$-$5126057  &    \\
   87   & 16 14 59.86  & $-$51 22 45.23  &  12.81   &  10.50   &   9.85   &   9.55   &   9.34   &   0.10   & 0.39   & 0.56     & 0.32 $\pm$   0.10    &   \nodata   &  \nodata   &   \nodata &  12.66  & $-$3.67  &$..$  &J16145985$-$5122452  &    \\
   88   & 16 17 20.82  & $-$50 50 27.56  &  10.48   &   9.84   &   9.64   &   9.59   &   9.51   &   0.09   & 0.05   & 0.19     & 0.29 $\pm$   0.04    &  12.46    &  \nodata   &   \nodata &   \nodata & $-$3.06  &$..$  &J16172081$-$5050275/TYC~8323$-$805$-$1  &    \\
   89   & 16 15 21.10  & $-$51 13 20.56  &  13.88   &  10.94   &  10.10   &   9.60   &   9.00   &  $-$0.08   &$-$0.28   & 0.82     &$-$0.78 $\pm$   0.72    &   \nodata   &  \nodata   &   \nodata &  12.66  & $-$3.88  &$..$  &J16152109$-$5113205  &    \\
   90   & 16 16 02.48  & $-$51 12 07.58  &   \nodata   &  11.89   &  10.38   &   9.60   &   8.84   &   0.12   & 0.22   & 1.27     &$-$1.28 $\pm$   0.40    &   \nodata   &  \nodata   &   \nodata &  12.66  & $-$4.33  &$..$  &J16160248$-$5112075  &    \\
   91   & 16 16 51.98  & $-$51 34 04.62  &   \nodata   &  10.24   &   9.81   &   9.63   &   9.47   &   0.10   & 0.21   & 0.38     & 3.25 $\pm$   0.76    &   \nodata   &  \nodata   &   \nodata &  12.66  & $-$3.41  &$..$  &J16165198$-$5134046  &    \\
   92   & 16 13 48.19  & $-$51 26 16.85  &  12.88   &  10.57   &   9.96   &   9.65   &   9.56   &   0.05   & 0.70   & 0.56     &$-$1.73 $\pm$   0.46    &   \nodata   &  \nodata   &   \nodata &  12.66  & $-$3.57  &$..$  &J16134818$-$5126168  &    \\
   93   & 16 17 19.75  & $-$51 33 59.28  &  12.43   &  10.69   &  10.01   &   9.65   &   8.40   &   0.04   &$-$2.33   & 0.62     & 0.10 $\pm$   0.06    &   \nodata   &  \nodata   &   \nodata &  12.66  & $-$3.63  &$..$  &J16171975$-$5133592  &    \\
   94   & 16 20 14.18  & $-$50 51 43.60  &  11.91   &  10.32   &   9.88   &   9.66   &   9.42   &   0.06   &$-$0.00   & 0.41     & 0.20 $\pm$   0.05    &   \nodata   &  \nodata   &   \nodata &  12.66  & $-$3.41  &$..$  &J16201417$-$5051435  &    \\
   95   & 16 18 35.60  & $-$51 15 46.06  &  10.57   &   9.92   &   9.81   &   9.67   &   9.37   &  $-$0.12   &$-$0.55   & 0.25     & 0.69 $\pm$   0.04    &  10.72    &  \nodata   &   \nodata &   \nodata & $-$1.30  &$..$  &J16183560$-$5115460/TYC~8323$-$2582$-$1  &    \\
   96   & 16 17 33.34  & $-$51 20 47.19  &  11.84   &  10.32   &   9.91   &   9.69   &   9.60   &   0.02   & 0.39   & 0.41     & 0.15 $\pm$   0.05    &   \nodata   &  \nodata   &   \nodata &  12.66  & $-$3.38  &$..$  &J16173333$-$5120471  &    \\
   97   & 16 14 59.48  & $-$50 37 30.89  &   \nodata   &  12.38   &  10.62   &   9.71   &   9.00   &   0.13   & 0.79   & 1.48     &$-$0.71 $\pm$   0.61    &   \nodata   &  \nodata   &   \nodata &  12.66  & $-$4.43  &$..$  &J16145947$-$5037308  &    \\
   98   & 16 20 16.11  & $-$50 55 28.32  &  12.54   &  10.57   &  10.02   &   9.71   &   9.55   &  $-$0.02   & 0.43   & 0.54     & 0.05 $\pm$   0.06    &   \nodata   &  \nodata   &   \nodata &  12.66  & $-$3.49  &$..$  &J16201610$-$5055283  &    \\
   99   & 16 12 32.85  & $-$50 55 32.67  &  11.17   &  10.18   &   9.89   &   9.72   &   9.53   &  $-$0.02   &$-$0.06   & 0.33     & 0.58 $\pm$   0.04    &  11.06    &  \nodata   &   \nodata &   \nodata & $-$1.67  &$..$  &J16123285$-$5055326/TYC~8323$-$451$-$1  &    \\
  100   & 16 13 24.35  & $-$51 27 41.89  &   \nodata   &  12.60   &  10.73   &   9.76   &   8.91   &   0.13   & 0.57   & 1.57     &$-$0.11 $\pm$   0.58    &   \nodata   &  \nodata   &   \nodata &  12.66  & $-$4.47  &$..$  &J16132435$-$5127418  &    \\
  101   & 16 16 46.59  & $-$51 06 49.36  &   \nodata   &  11.68   &  10.40   &   9.77   &   9.04   &   0.14   &$-$0.04   & 1.07     & 0.28 $\pm$   0.23    &   \nodata   &  \nodata   &   \nodata &  12.66  & $-$3.96  &$..$  &J16164658$-$5106493  &    \\
  102   & 16 17 35.03  & $-$50 28 46.67  &  10.46   &   9.94   &   9.85   &   9.77   &   9.74   &  $-$0.04   & 0.09   & 0.18     & 0.52 $\pm$   0.04    &  11.29    &  \nodata   &   \nodata &   \nodata & $-$1.70  &$..$  &J16173502$-$5028466/TYC~8319$-$954$-$1  &    \\
  103   & 16 20 06.54  & $-$50 52 12.85  &  10.61   &   9.98   &   9.87   &   9.78   &  10.13   &  $-$0.05   & 1.15   & 0.20     & 0.04 $\pm$   0.15    &   \nodata   &  \nodata   &   \nodata &  12.66  & $-$3.08  &$..$  &J16200654$-$5052128/TYC~8324$-$1509$-$1  &    \\
  104   & 16 17 32.55  & $-$51 30 33.25  &  10.87   &  10.17   &   9.93   &   9.79   &   9.78   &  $-$0.01   & 0.34   & 0.28     & 0.47 $\pm$   0.23    &   \nodata   &  \nodata   &   \nodata &  12.66  & $-$3.15  &$..$  &J16173255$-$5130332/TYC~8323$-$1405$-$1  &    \\
  105   & 16 16 08.14  & $-$51 37 07.93  &   \nodata   &  10.43   &  10.03   &   9.81   &   9.71   &  $-$0.01   & 0.35   & 0.41     &$-$0.41 $\pm$   0.06    &   \nodata   &  \nodata   &   \nodata &  12.66  & $-$3.26  &$..$  &J16160814$-$5137079  &    \\
  106   & 16 16 51.77  & $-$51 14 28.46  &   \nodata   &  10.77   &  10.18   &   9.82   &   9.49   &  $-$0.05   & 0.05   & 0.60     & 0.05 $\pm$   0.07    &   \nodata   &  \nodata   &   \nodata &  12.66  & $-$3.44  &$..$  &J16165176$-$5114284  &    \\
  107   & 16 15 23.49  & $-$51 38 42.21  &  13.43   &  11.01   &  10.24   &   9.84   &   9.48   &   0.06   & 0.20   & 0.69     & 0.04 $\pm$   0.07    &   \nodata   &  \nodata   &   \nodata &  12.66  & $-$3.51  &$..$  &J16152348$-$5138422  &    \\
  108   & 16 19 59.03  & $-$51 10 21.00  &  12.52   &  10.57   &  10.11   &   9.85   &   9.34   &   \nodata   &$-$0.65   & 0.46     & 0.18 $\pm$   0.05    &   \nodata   &  \nodata   &   \nodata &  12.66  & $-$3.27  &$..$  &J16195903$-$5110210  &    \\
  109   & 16 18 32.21  & $-$51 13 16.21  &  16.81   &  12.02   &  10.60   &   9.85   &   8.89   &   0.07   &$-$0.44   & 1.22     &$-$0.57 $\pm$   0.26    &   \nodata   &  \nodata   &   \nodata &  12.66  & $-$4.03  &$..$  &J16183221$-$5113162  &    \\
  110   & 16 16 28.28  & $-$51 05 49.79  &   \nodata   &  10.76   &  10.18   &   9.88   &   9.62   &   0.02   & 0.20   & 0.55     &$-$0.10 $\pm$   0.08    &   \nodata   &  \nodata   &   \nodata &  12.66  & $-$3.33  &$..$  &J16162828$-$5105497  &    \\
  111   & 16 15 03.31  & $-$51 32 09.88  &  10.42   &  10.04   &   9.91   &   9.91   &   9.79   &   0.13   &$-$0.17   & 0.12     & 0.59 $\pm$   0.04    &  11.04    &  \nodata   &   \nodata &   \nodata & $-$1.25  &$..$  &J16150330$-$5132098/TYC~8323$-$2297$-$1  &    \\
  112   & 16 13 35.95  & $-$51 05 11.59  &   \nodata   &  12.13   &  10.68   &   9.91   &   9.49   &   0.06   & 1.08   & 1.25     &$-$0.01 $\pm$   0.15    &   \nodata   &  \nodata   &   \nodata &  12.66  & $-$3.99  &$..$  &J16133595$-$5105115  &    \\
  113   & 16 16 48.56  & $-$50 23 13.24  &   \nodata   &  10.53   &  10.12   &   9.93   &   9.55   &   0.05   &$-$0.41   & 0.39     & 0.23 $\pm$   0.05    &   \nodata   &  \nodata   &   \nodata &  12.66  & $-$3.12  &$..$  &J16164856$-$5023132  &    \\
  114   & 16 17 05.06  & $-$50 47 25.74  &  14.01   &  11.20   &  10.37   &   9.94   &   \nodata   &   0.06   & \nodata  & \nodata    & 0.20 $\pm$   0.09    &   \nodata   &  \nodata   &   \nodata &  12.66  &  \nodata &$..$  &J16170505$-$5047257  &    \\
  115   & 16 17 03.36  & $-$51 10 25.14  &  16.92   &  12.09   &  10.66   &   9.94   &   9.23   &   0.14   & 0.22   & 1.19     & 0.73 $\pm$   0.29    &   \nodata   &  \nodata   &   \nodata &  12.66  & $-$3.91  &$..$  &J16170335$-$5110251  &    \\
  116   & 16 17 03.33  & $-$50 40 12.71  &  12.90   &  10.85   &  10.24   &   9.95   &   9.55   &   0.09   &$-$0.18   & 0.54     & 0.18 $\pm$   0.06    &   \nodata   &  \nodata   &   \nodata &  12.66  & $-$3.25  &$..$  &J16170332$-$5040127  &    \\
  117   & 16 18 48.06  & $-$51 06 41.59  &  12.77   &  10.80   &  10.24   &   9.96   &   9.60   &   0.06   &$-$0.11   & 0.51     & 0.12 $\pm$   0.06    &   \nodata   &  \nodata   &   \nodata &  12.66  & $-$3.21  &$..$  &J16184805$-$5106415  &    \\
  118   & 16 19 22.85  & $-$51 22 49.52  &  10.67   &  10.19   &  10.05   &   9.96   &   9.87   &  $-$0.03   &$-$0.02   & 0.21     & 0.84 $\pm$   0.04    &  10.30    &  \nodata   &   \nodata &   \nodata & $-$0.55  &$..$  &J16192284$-$5122495/TYC~8324$-$2022$-$1  &    \\
  119   & 16 15 29.17  & $-$51 40 13.23  &  12.72   &  10.94   &  10.29   &   9.96   &   9.60   &   0.06   & \nodata   & 0.58     & 0.10 $\pm$   0.24    &   \nodata   &  \nodata   &   \nodata &  12.66  & $-$3.27  &$..$  &J16152916$-$5140132  &    \\
  120   & 16 13 18.66  & $-$51 04 43.97  &  10.87   &  10.27   &  10.07   &   9.97   &   9.89   &   0.02   & 0.08   & 0.23     & 0.85 $\pm$   0.04    &  10.29    &  \nodata   &   \nodata &   \nodata & $-$0.54  &$..$  &J16131865$-$5104439/TYC~8323$-$686$-$1  &    \\
  121   & 16 16 06.16  & $-$51 36 33.70  &   \nodata   &  10.57   &  10.18   &   9.98   &   9.89   &   0.02   & 0.35   & 0.39     & 0.11 $\pm$   0.04    &   \nodata   &  \nodata   &   \nodata &  12.66  & $-$3.07  &$..$  &J16160615$-$5136336  &    \\
  122   & 16 13 37.15  & $-$51 22 01.96  &  11.94   &  10.64   &  10.25   &   9.99   &   9.87   &  $-$0.08   & 0.31   & 0.44     & 0.36 $\pm$   0.07    &   \nodata   &  \nodata   &   \nodata &  12.66  & $-$3.10  &$..$  &J16133715$-$5122019  &    \\

\hline
\end{tabular}
}
\end{table*}

\subsection{Candidate  Evolved Early-type Stars }
\label{earlyselection}

As seen in the previous sections, G332.809$-$0.132 hosts a number of massive transitional 
objects and WRs, as expected for being a region rich in remnants.
Stars E2 (Ofpe/WN 9), E4 (WN 8), and WR74 (WN 7) have \Ks\ of 6.05, 7.43, 
and 8.80 mag and \Aks\ = 0.94, 0.47, and 0.43 mag, respectively.
We searched for other candidate massive stars with free-free excess and  
that are brighter than 
\Ks\ = 10.0 mag, which for a DM = 12.66 mag and \Aks\ = 0.5 mag roughly corresponds 
to the observed magnitudes for an O9 dwarf \citep{martins06}, 
as shown in Table \ref{cutmag}, as described in Sect. 5.6.

\begin{table}
\caption{ \label{cutmag} Apparent Magnitudes of O9 and O5 Stars}
\begin{tabular}{rrrrrr}
\hline
\hline
Sp. & Class& \Ks$_{\rm abs}$& \Aks&DM & \Ks\\
    &      &(mag)           &(mag)&(mag)&(mag)\\
\hline
 O5 &  I &$-$5.52 & 0.40 &12.66 & 7.54\\
 O9 &  I &$-$5.52 & 0.40 &12.66 & 7.54\\
 O5 &  I &$-$5.52 & 1.60 &12.66 & 8.74\\
 O9 &  I &$-$5.52 & 1.60 &12.66 & 8.74\\
 O5 &  I &$-$5.52 & 2.20 &12.66 & 9.34\\
 O9 &  I &$-$5.52 & 2.20 &12.66 & 9.34\\
 O5 &  I &$-$5.52 & 0.40 &13.70 & 8.58\\
 O9 &  I &$-$5.52 & 0.40 &13.70 & 8.58\\
 O5 &  I &$-$5.52 & 1.60 &13.70 & 9.78\\
 O9 &  I &$-$5.52 & 1.60 &13.70 & 9.78\\
 O5 &  I &$-$5.52 & 2.20 &13.70 &10.38\\
 O9 &  I &$-$5.52 & 2.20 &13.70 &10.38\\
 O5 &  V &$-$4.39 & 0.40 &12.66 & 8.67\\
 O9 &  V &$-$3.28 & 0.40 &12.66 & 9.78\\
 O5 &  V &$-$4.39 & 1.60 &12.66 & 9.87\\
 O9 &  V &$-$3.28 & 1.60 &12.66 &10.98\\
 O5 &  V &$-$4.39 & 2.20 &12.66 &10.47\\
 O9 &  V &$-$3.28 & 2.20 &12.66 &11.58\\
 O5 &  V &$-$4.39 & 0.40 &13.70 & 9.71\\
 O9 &  V &$-$3.28 & 0.40 &13.70 &10.82\\
 O5 &  V &$-$4.39 & 1.60 &13.70 &10.91\\
 O9 &  V &$-$3.28 & 1.60 &13.70 &12.02\\
 O5 &  V &$-$4.39 & 2.20 &13.70 &11.51\\
 O9 &  V &$-$3.28 & 2.20 &13.70 &12.62\\
 \hline
\end{tabular}
\end{table}

We adopted the criteria described in Section \ref{sample} and \citet[][]{messineo12}, 
which should yield a fraction larger than 80\% of the Wolf-Rayet star population,
and retrieved  35 targets (other than the two known WRs), 
which are listed in Table \ref{table.wr},
and illustrated in Fig.\  \ref{free.fig}.
Six of them are reported  in the literature as possible late-type stars 
(one carbon, four candidate asymptotic giant stars (AGBs) and one planetary nebula, PN).
Thirty-four (out of 35) objects are   in Gaia DR2 with 
\Vlsr\ measurements available only for two stars
(radial velocities are fitted with templates spectra with \Teff $< 10000$ k).
This suggests that   83\% of the sample consists 
of genuine windy massive stars with free-free emission.
Among the spectroscopically known OB stars listed in Table \ref{obknown}, 
there are five stars
photometrically classified as  free-free emitters. They all have emission lines.

Having excluded those stars  classified as late-type stars and  free-free emitters
based on their 2MASS and GLIMPSE magnitudes \citep{messineo12},
by using the classical $J-H$ versus $H-$\Ks\ diagram with 2MASS datapoints 
and $I-J$ versus $J-$\Ks\ diagram of DENIS datapoints,  
184 other likely OB stars are found. 
SIMBAD provides a spectral type for  46 of them, with 37 being early types, 
8 (17\%) yellow stars, and one (2\%) late-type.
For  179 (out of  184) selected OB stars, 
Gaia DR2 parallaxes  exist, but only 22 Gaia DR2 radial velocities have been measured with  
6 of them associated with SIMBAD F-types and one C-type.
This confirms the early nature of the bulk of the sample.
By excluding those already known or with Gaia DR2 velocities (e.g., cooler than 10,000 K), 
we remain with
122   likely new OB stars, which are listed in  Table \ref{table.ob}.
In Fig.\  \ref{free.fig}, these targets are shown along with a 
comparison sample of known early-type stars in the inner Galaxy.

\subsubsection{Notes on IRAS  16115$-$5044}
IRAS 16115$-$5044 is a candidate free-free emitter that is marked in SIMBAD as a PN.
IRAS  16115$-$5044 was spectroscopically 
observed in the $JHK$ bands during the IRAS survey for late-type stars by \citet{weldrake03}. 
In this conference  proceeding contribution, the authors 
estimated a B4I spectral type  from the $H$ band. 
\citet{suarez06} also concluded that this source is not a PN, but a young star.
The $K$ band spectrum in Fig. 1 of by \citet{weldrake03} -- with \brgamma\ in emission 
with a P-Cygni profile and a FWHM velocity of about 300 \kms,  Fe II   at 2.091 \um,
Mg II at 2.138 and 2.144 \um, and Na I at 2.206 and 2.209 \um\ in emission -- 
recalls the spectra of  the LBV Pistol (11800 K),
of  G24.73+0.69 (12000 K), and  G26.47+0.02 (17000 K) \citep{clark03,najarro09}.
By assuming a B4 type (14000 K), the 2MASS magnitudes
$J=7.76$ mag, $H=6.13$ mag, and \Ks$=5.14$ mag, and 
the extinction curve n.3 of \citet{messineo05},
we estimate \Aks($J-H$)=1.4 mag (\Aks($H-K$)=1.5 mag).
The Gaia DR2 parallax ($\varpi=0.69 \pm 0.19$ mas yr$^{-1}$) 
has a fractional error larger than 27\%, and the Gaia distance ranges 
from 1.1 to 2 kpc. For a distance of 1.5 kpc, 
by integrating under the stellar energy distribution (SED) 
and extrapolating with a  blackbody curve,
we infer  a \BCK=$-1.16\pm0.49$ mag and  
\Mbol=$-8.3$ mag (log$_{10}\frac{\Lstar}{\Lsun}=5.2)$, 
which confirms a supergiant class. 
The \Aks\ value is suggestive of a larger distance. 
For a distance of 3.4 kpc,
\Mbol=$-10.08$ mag (log$_{10}\frac{\Lstar}{\Lsun}=5.9)$.
A more detailed study of this source will be presented in a forthcoming paper.

\subsection{Discussion of ``The Bridge'' Massive Stars}
\label{Bridgediscussion}

A 2MASS $J-$\Ks\ versus \Ks\ magnitude diagram for objects  in the 
direction of ``the Bridge'' is shown in Fig.\ \ref{cmdmassSKIFF.fig}. 
The diagram  shows a foreground main sequence at $J-$\Ks$\approx$0.3 mag  
($0< J-$\Ks$ < 0.4$ mag, \Aks $\approx 0.15$ mag)  
and a more reddened main sequence at  about $J-$\Ks$\approx$ 1.0 mag 
($0.7 < J-$\Ks$< 1.2$ mag, \Aks $\approx 0.55$ mag). 
The first main sequence  is highly likely to be  dominated by stars in the
Sagittarius-Carina arm, and the redder one  by stars in  
the Scutum-Crux arm ($-46-52$ \kms\ clouds), as outlined in Fig. \ref{whereareyou}.

We have located 229 early-type stars projected toward the Bridge area
(122 OB candidates + 35 candidate free-free emitters + 7 OB observed + 63
previously known OB stars and  2 WRs).
All but two  previously known OBA stars appear to be located 
on the bluer main sequence at $0<J-$\Ks$< 0.4$ mag. 
Using the best  39 quality Gaia DR2 parallaxes of  
OBA stars with $10 <$ \Ks $ < 7$ mag, we measured
an average distance  of  980 pc (DM $=~9.96\pm0.04$ mag).
When plotting the  candidate  early-type stars in Table \ref{table.ob}, 
it is also possible to recognize a group of stars on the second  
main sequence at $J-$\Ks$\approx$ 1.0 mag; 
seven stars  at \Aks\ = 0.13 mag yield  
an average parallactic distance of 1570 pc (DM = $10.98\pm0.11$ mag), while the 
three OB candidates 
at \Aks\ = 0.49 mag an average parallactic distance of about 2680 pc (DM = $12.14 \pm 0.12$ mag).
The latter Gaia distance can be attributed to the Scutum-Crux arm, which hosts 
the G332.809$-$0.132 complex.
This Gaia distance estimate is slightly shorter than that reported in 
\citet[3.4 kpc,][]{rahman10},  and it agrees better with the \ion{H}{I} spectrophotometric 
and kinematic distance of IRAS $16132-5039$ of \citet[][]{corti16},
as stated in Section \ref{region1}.

{In the color-magnitude diagram (CMD), E6  (\Aks\ = 0.2 mag) is located 
on the red edge of the first main sequence 
which is  populated by stars in the Sagittarius-Carina arm.
However, its   Gaia DR2 parallax ($0.31\pm0.04$ mas) suggests a distance of 2.9 kpc (Scutum-Crux arm),
still consistent with a dwarf.}
WR 74 (P2) is located just above the second main-sequence peak with  \Aks\ = 0.43 mag 
and most probably belongs to G332.809$-$0.132 in the Scutum-Crux arm, 
as well as the new WN 8 (E4, \Aks\ = 0.5 mag).
The other observed early-type stars have \Aks\ from 0.74 mag to 1.55 mag.
The spectrophotometric properties of star E2 (Ofpe/WN 9, \Aks\ = 0.9 mag) 
are  compatible with  the 3 kpc kinematic distance  
of the G332.809$-$0.132 molecular complex. 
The  stellar cluster [DBS2003]160-161 (IRAS $16132-5039$) 
is also  at a similar distance, but its members are still embedded 
with  \Aks\ = 1.2-1.7 mag \citep{romanlopes04}.
Therefore, it appears that stars with \Aks\ from 0.4 mag to 1.6 mag may belong to 
the  G332.809$-$0.132 molecular complex of the Scutum-Crux arm. 
If we consider the early-type stars with $J-$\Ks\ color 
from 0.7 mag to 3.0 mag, i.e., with \Aks\ from 0.4 mag to 1.6 mag, 
in the ``Bridge'' we count  108 stars ( =  83 OB candidates +16 free-free +6 OB observed + 3
previously known OB stars and WRs)  stars  brighter than 
\Ks $<10$ mag (O9 V at 3.4 kpc) that are the potential ionizing stars of the 
G332.809$-$0.132 molecular complex. 
These numbers along with the sampled \Aks\
and the expected magnitudes in Table \ref{cutmag}, suggest that we have identified 
most of the supergiants in the region, and that those stars may already account
for about 50\% of the ionizing flux measured by WMAP (see Section \ref{region1}). 
In Table \ref{cutmag} the apparent magnitudes of an O5 and O9 stars
are calculated for \Aks=0.4, 1.6, 2.2 mag, and 3.4 and 5.5 kpc.
These early-type stars are sparsely  distributed on ``The Bridge'', 
but rare in the northwestern sector (i.e., toward the direction of SNRs $G332.0+00.2$
and Kes 32, which are reported at a larger distance of 6-7 kpc), as seen 
in the Fig. \ref{g332map}.
Note that in our exercise we consider stars brighter than \Ks\ = 10 mag.
As demonstrated by \cite{romanlopes09}, the IRAS $16132-5039$ region contains 
obscured massive stars, with only the two brightest around this magnitude.
The actual number of Lyman continuum emitters is greater than estimated.
The $G332.0+00.2$ complex is unquestionably rich in evolved massive stars.

The complex also hosts  a number of K-M I stars, with six of them listed by \citet{skiff14}.
We are able to confirm five of those as associated with the  $G332.0+00.2$ complex.
By looking at the spatial distribution, we can see that they  are  located at the periphery 
of ``The Bridge''. A few other candidate RSGs were identified in the central region.

\section{How Does Our Analysis Compare to Previous Knowledge on Stellar
Associations and Clusters? } 
\label{associations and clusters?}

\begin{figure}
\begin{center}
\resizebox{1\hsize}{!}{\includegraphics[angle=0]{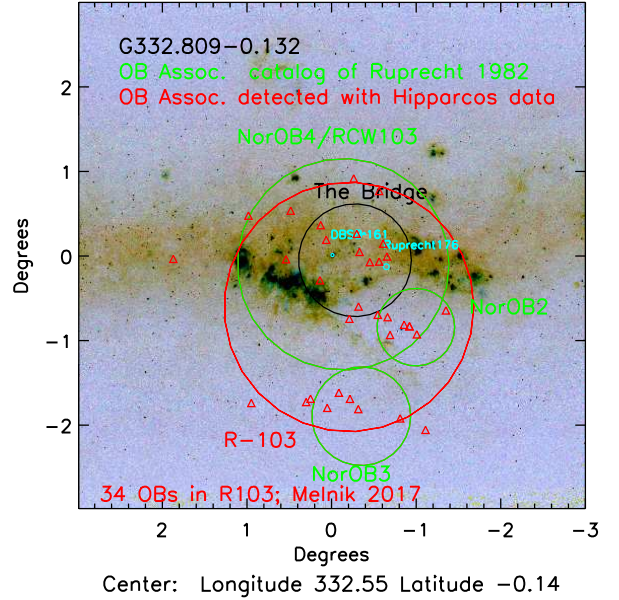}}
\end{center}
\caption{
\label{figbicabig} 
MSX composite image of ``The Bridge'' (area encircled by the black circle).
The green circles show the Nor OB2, Nor OB3, and Nor OB4 associations 
listed by \citet{ruprecht82}. The red circle show the
RCW 103 association following the partitions of Blaha and Humphreys 
listed in \citet{melnik95}; the 34 OB
members are marked with red triangles  \citep{melnik17}.}
\end{figure}

The kind referee has suggested considering  known stellar clusters projected toward
this line of sight, and so here we report on our search in the literature.
It has long been known that  this area of the sky was rich in massive stars and explosive 
supernovae. Historical information, though, was not correlated, and
several authors reported independently about the beauty of this region and
its content of massive stars.
As shown in  Section \ref{Bridgediscussion}, 
this line of sight is quite complex because it crosses three spiral arms;
furthermore, the morphology of the complex at mid-infrared light suggests
strong  patchiness of interstellar extinction.

In the case of  G332.809$-$0.132, the extended \HH\ region 
identified in the WMAP radio data (see §5) lies  within 
the area of the Nor OB4/RCW 103 association \citep{alter70,ruprecht82}.
A sketch of the No rOB associations is shown in Fig. \ref{figbicabig}. 
This OB association  is 150\arcmin\ in size and encloses ``The Bridge'' area here considered,
which comprises its central portion (diameter of 80\arcmin).
The Nor OB4 association is also mentioned by \citet{ambartsumian89} 
as related to a region of recent star formation containing stellar associations.
There are no associated articles before 1969. 
Nor stands for Norma \citep{ruprecht82}, as this line of sight is a tangent point 
to the Norma spiral  arm 
at longitudes 332$^\circ$ (see our Fig. \ref{whereareyou} and 
also the list of tangent points provided by \citet{englmaier99}).

\citet{westerlund69} reports about 36 OB stars in the neighborhood
of the emission nebula RCW 103 ($331^\circ<l<334^\circ$, $-2^\circ<b<+1^\circ$).
\citet{humphreys78} lists 26 OB stars in the same area.
Their radial velocities  are quite uncertain
(e.g., 40 \kms\ variations are measured toward the same star) and 
vary from +27 to $-80$ \kms. 
Thirty-four OB stars and two stellar kinematic subgroups  (six and seven stars) 
were  identified  in RCW 103 by \citet[][]{melnik95}, who assigned 
a kinematic distance of 3.2 kpc    
($\Vlsr= -48$ \kms\ with  $\sigma= 30.5$ \kms). 
Identified members are listed in \citet{melnik17} and 
their locations are shown in Fig. \ref{figbicabig}.
Only 9 of these 34 OB members fall in the central area we 
analyzed (``The Bridge''). These members are marked in Table \ref{obknown}, 
have \Aks$ < 0.4$ mag and are among the 15 most distant OB stars previously
known (parallactic distances from 2 to 4 kpc) --
with  average proper motions 
$\mu_{ra}=  -2.026$ mas yr$^{-1}$ with $\sigma$=1.5   mas yr$^{-1}$,
$\mu_{dec}= -2.934$ mas yr$^{-1}$ with $\sigma=1.5$   mas yr$^{-1}$.  
A recent search for star formations and infrared OB stars 
in the direction of this WMAP extended complex 
was  carried out by \citet[][]{rahman13}, with two detected  
overdensities (RMM34a and RMM34b) of  6\arcmin\ and 17\arcmin\ 
in size, located on the left-side bright 8 \um\ emission  east of the sled. 

Ruprecht~176 and DBSB~161 are the only two known open clusters 
enclosed in ``The Bridge'' area  \citep[][]{kharchenko16}. 
DBSB~161 has already been described already above (see Section \ref{Bridgediscussion}); 
it is located in the center of ``The Bridge'' at about 3 kpc, 
is still embedded (\Aks\ = 1.2-1.7 mag),
and rich in massive O stars \citep{romanlopes04,corti16}.\\
Ruprecht~176 is an older known open cluster.
\citet{kharchenko16} estimate a distance of 2.7 kpc, E(B-V)=0.48 mag 
(\Aks $\approx$ 0.13 mag), and an age of 580 Myr.
\citet[][]{sampedro17} provides 438 individual members 
based on the motions of stars in the 
UCAC4 catalog,  E(B-V)=0.69 mag (\Aks=0.2 mag), a distance of 1.5 kpc, and age of 288 Myr.
Newer Gaia data from DR2, using 130 members, have yielded
a most likely  distance (the mode of the likelihood) of  2.7 kpc (from 2.1 to 3.7 kpc), 
proper motions $\mu_{ra}=-2.644$ mas yr$^{-1}$ with $\sigma$=0.13 mas yr$^{-1}$,
and $\mu_{dec}= -2.984$ mas yr$^{-1}$ with $\sigma= 0.078$ mas yr$^{-1}$;  
$\varpi=0.335$ mas with $\sigma=0.070$ mas,
and a radius of 2\farcm34 \citep{cantat18}. With Gaia DR2,
\Av=1.68 mag (\Aks $\approx 0.14$ mag) and the age is lowered to 150 Myr \citep{bossini19}.
Its parallactic distance, proper motions, and \Aks\
are consistent with those  of RCW 103 given by \citet{melnik17}, 
and  identical to values we measured for OBs on the second main sequences of the CMD  
in Fig. \ref{cmdmassSKIFF.fig}, however, using redder stars at \Aks=0.4 mag. 
This indicates differential extinction and the need for parallaxes and velocities
to assign distances. 
This brings the open cluster Ruprecht~176  into the Scutum-Crux arm, perhaps
in its front side.

\begin{figure*}[t]
\begin{center}
\resizebox{0.31\hsize}{!}{\includegraphics[angle=0]{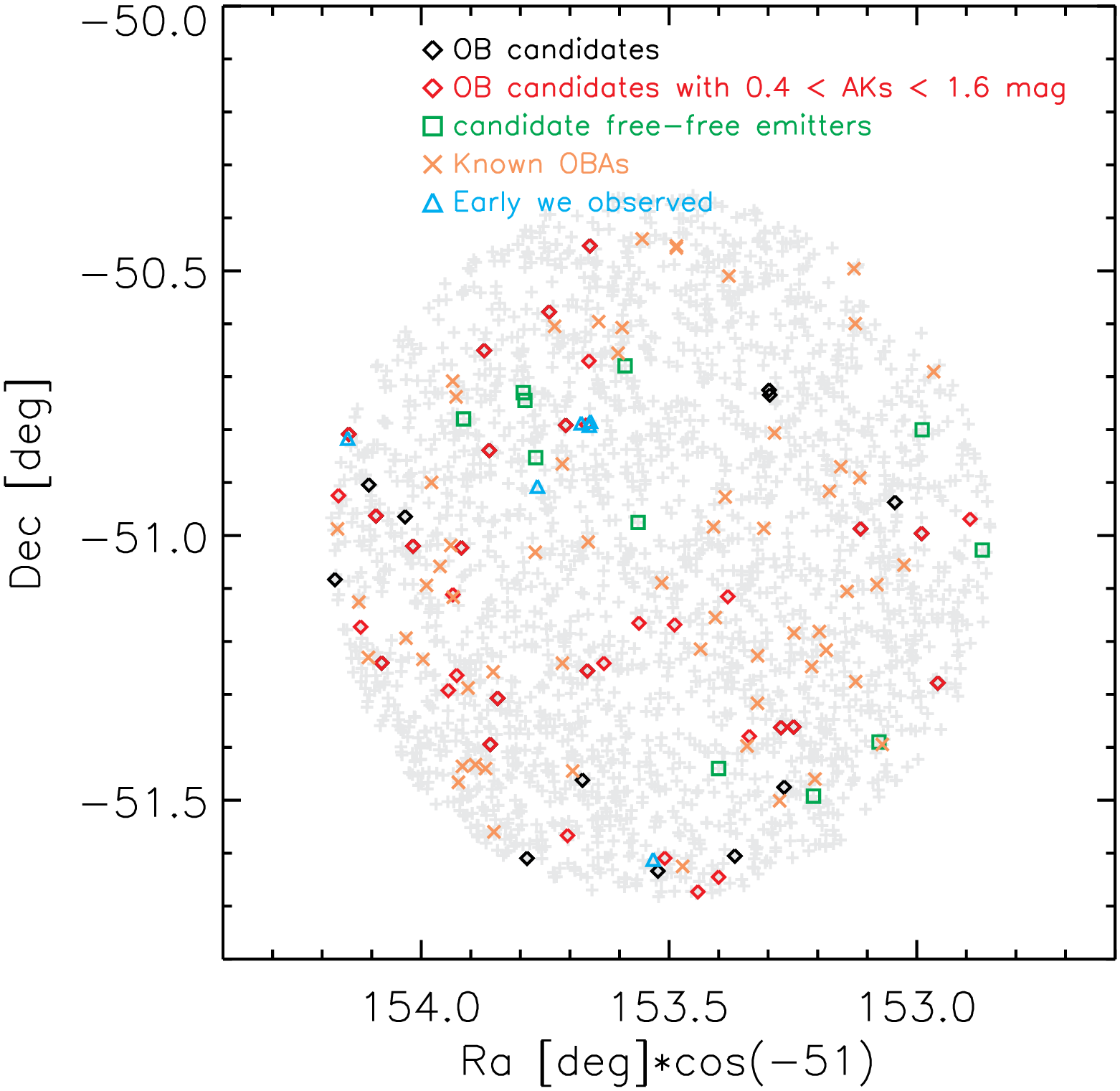}}
\resizebox{0.31\hsize}{!}{\includegraphics[angle=0]{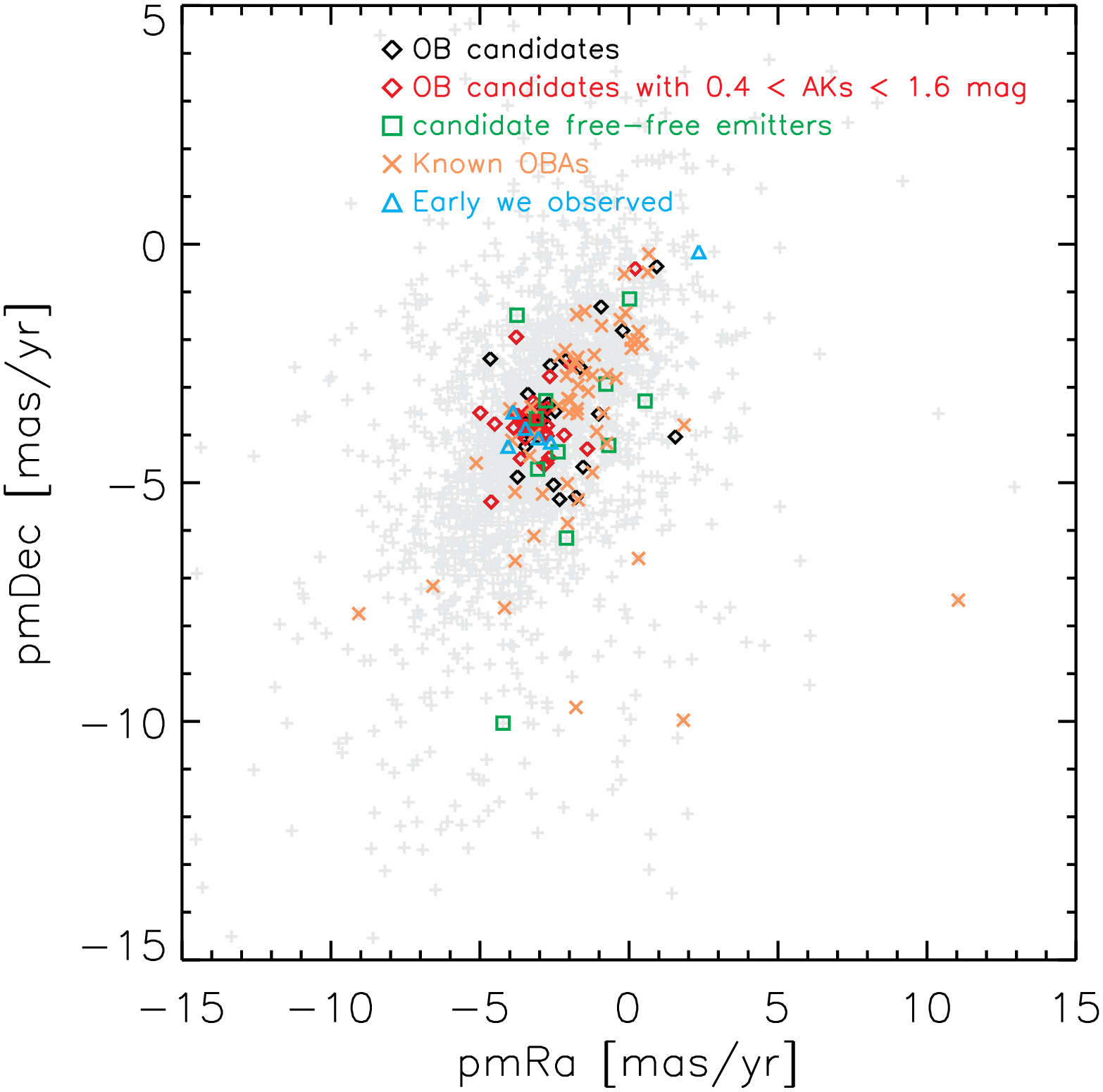}}
\resizebox{0.31\hsize}{!}{\includegraphics[angle=0]{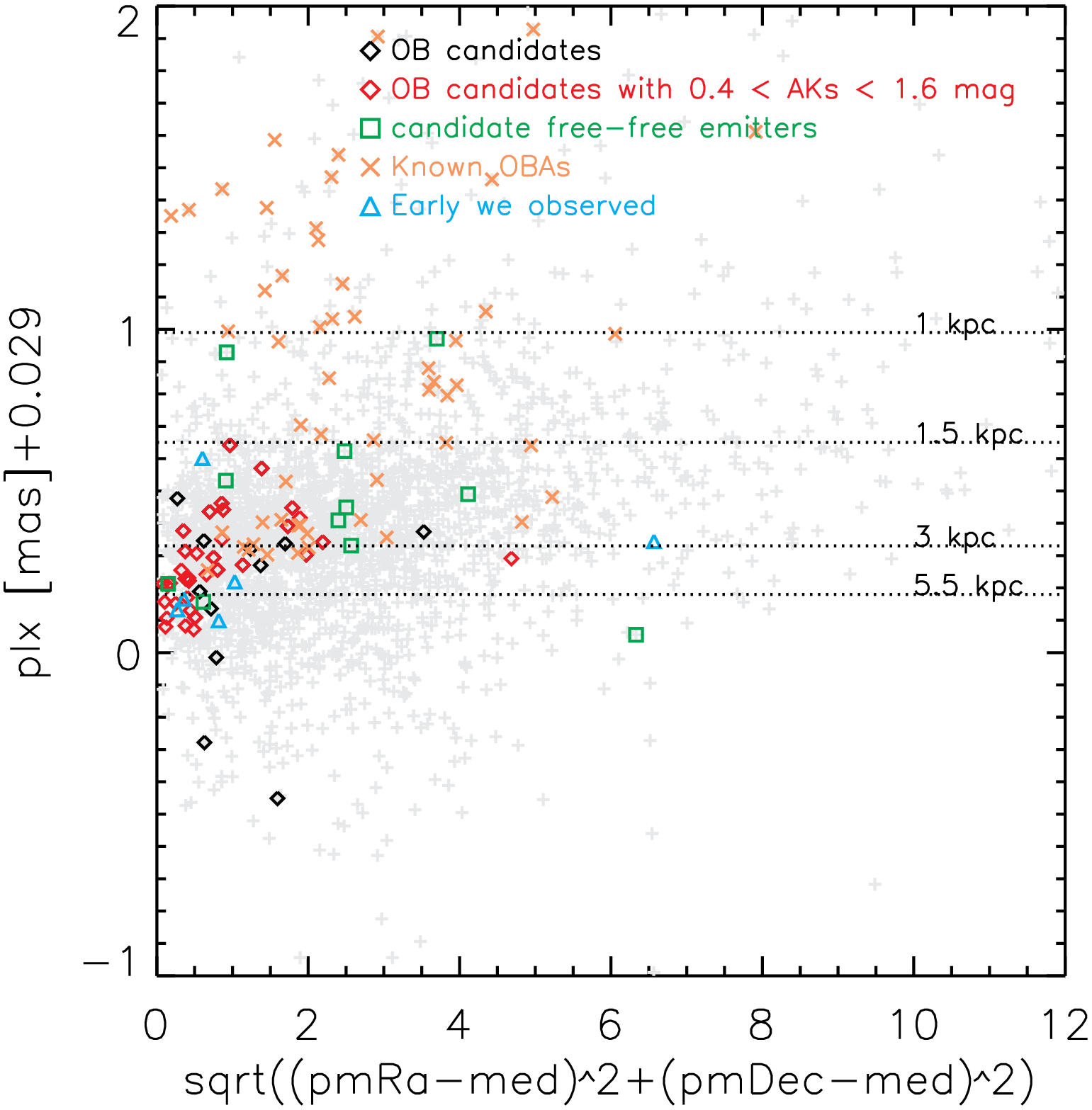}}
\end{center}
\caption{ \label{propermotion} {\it Left panel:} Spatial distribution of 
stars in ``The Bridge''.  {\it Middle panel:} Proper motions.
{\it Right panel:} $\varpi$ values versus the distance in proper motion space
from the median point of OB candidates. Distance values of 1, 1.5, 3, and 5.5 kpc are indicated with horizontal lines.
In the three panels, only  stars  with UWE value within the limits of 
\citet{gaiaplx2} are plotted.
Gray plus signes indicate  2MASS datapoints (\Ks $<10$ mag) in the Bridge.
Black diamonds are the OB candidates from Table \ref{table.ob},
and in red those with $0.4 <$\Aks$< 1.6$ mag.
Green squares mark the candidate free-free emitters from Table \ref{table.wr}.
Orange crosses indicate the previously known early-type stars from 
Table \ref{obknown}, cyan triangles the early-type stars observed in this work.
}

\end{figure*}

\citet{bica19} present a list of $\sim 35$  candidate clusters toward the ``Bridge''. 
The bulk of them consists of tiny 
regions (radius $< 1$\arcmin) of strong diffuse mid-infrared emission, 
whose positions do not spatially correlate
with those of selected OB stars. 
Those could represent very young stellar objects, the results of recently triggered
star formation.
Several of them may also reside in the background $-90$ \kms\ molecular cloud 
(5 kpc, Norma arm); for example,
GLIMPSE 77 is located in the northern edge 
of ``The Bridge'', has an \Aks=0.9 mag, but a spectrophotometric distance  of  
5.2 kpc \citep{messineo18,morales13} and belongs to the Norma arm;
 VVV CL065 is another embedded cluster in the $-95$ \kms\ cloud \citep{morales13}.
$[$MCM2005b$]$ 76, VVV CL066, $[$MCM2005b$]$ 74, VVV CL064 
are also embedded, but associated
with molecular clumps at $-50$ \kms\ \citep{morales13}. 
In conclusion, the G332.809$-$0.132 complex is made of molecular clouds  at $\approx-$50 
\kms\  in the Scutum-crux arm; but, behind it, a conspicuous amount of
clouds  accumulates with $V_{lsr}$ at $\approx -$90 \kms\ because of the tangent point to the  
Norma arm.
Plus, there is a stellar population (not associated with clouds) belonging to the nearest
spiral arm (Sagittarius).

In the present work, we have analyzed about  100 
candidate early-type stars selected over ``The Bridge'' area,
that are sparsely distributed, i.e.,  not in clusters, with \Ks$<10$ mag 
(O9 at 3.4 kpc)  by aiming to detect evolved massive stars.
The large size and patchiness of the interstellar extinction hamper
a precise knowledge of the stratification of the various stellar populations. 
In Fig. \ref{propermotion}, we display the Gaia DR2
proper motions and parallaxes of the candidate OB stars and show that
the range of extinction $0.4 <$ \Aks $< 1.6$ mag  samples stars at 3 kpc and beyond.
An average $\mu_{ra}=-3.11$ mas yr$^{-1}$ with $\sigma$=0.91 mas yr$^{-1}$,
and $\mu_{dec}= -3.68$ mas yr$^{-1}$ with $\sigma= 0.82$ mas yr$^{-1}$ are measured. 
These proper motions are  consistent with that of the massive stars with free-free emission;
indeed, for IRAS 16115$-$5044,   $\mu_{ra}=-3.51\pm0.33$ mas yr$^{-1}$
and  $\mu_{dec}=-3.17\pm0.19$ mas yr$^{-1}$;
for P2 (WR 74), $\mu_{ra}=-3.47\pm0.07$ mas yr$^{-1}$ and $\mu_{dec}=-3.86\pm0.05$ mas yr$^{-1}$;
for E2 (Ofpe/WN 9), $\mu_{ra}=-2.63\pm0.19$ mas yr$^{-1}$ and $\mu_{dec}=-4.15\pm0.14$ mas yr$^{-1}$;
for E4 (WN 8), interestingly,  
$\mu_{ra}=-5.49\pm0.74$ mas yr$^{-1}$ and $\mu_{dec}=-6.41\pm0.68$ mas yr$^{-1}$,
but errors are still large.

Gaia DR4 will  release spectroscopic spectral types and improved parallaxes.
With this information, it will  be possible to assign type, motion, and distance
to each individual bright massive star.

\section{CS 78/NGC 6334}
\label{region2}

\begin{figure}[!]
\begin{center}
\resizebox{0.9\hsize}{!}{\includegraphics[angle=0]{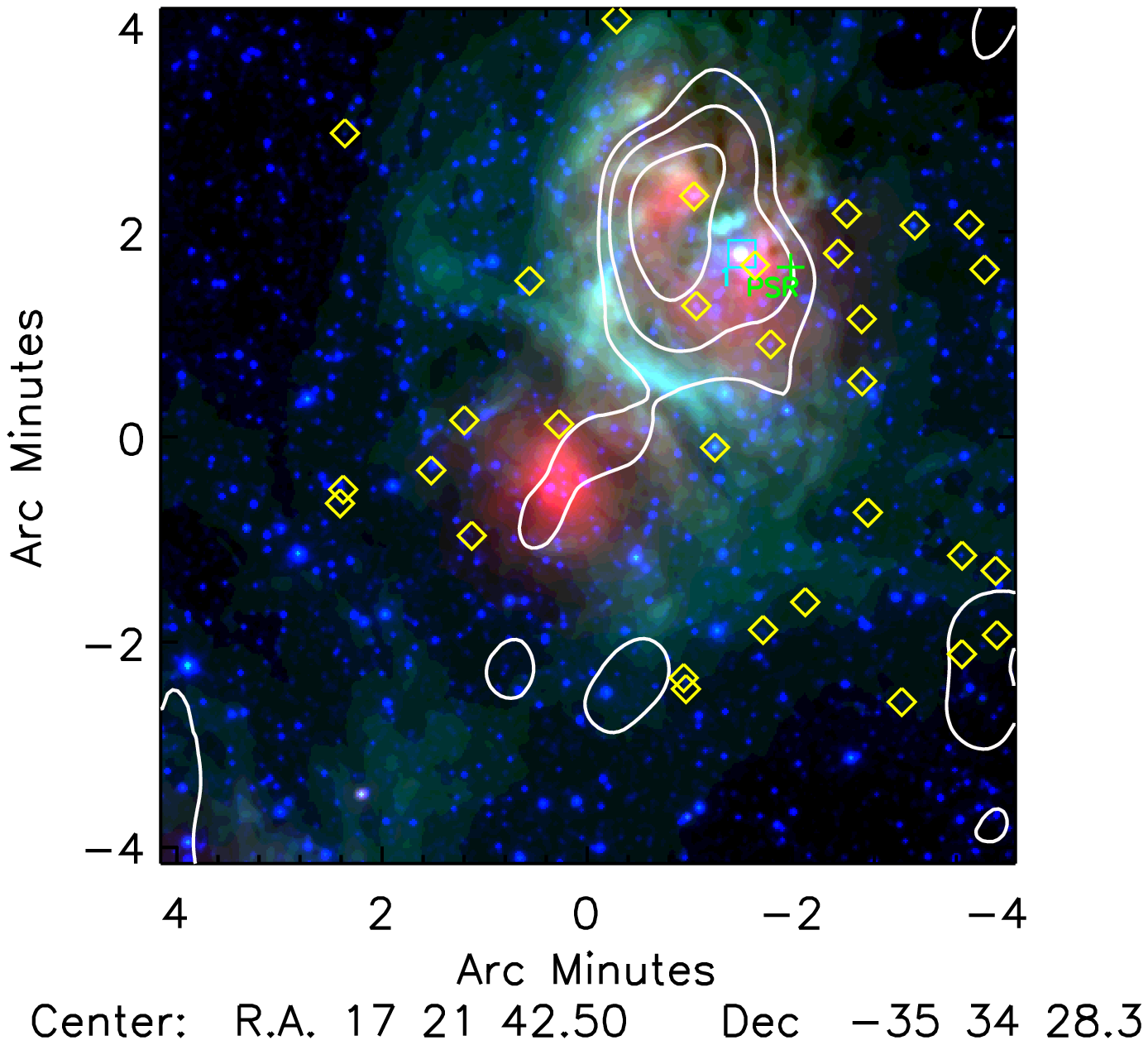}}
\resizebox{0.9\hsize}{!}{\includegraphics[angle=0]{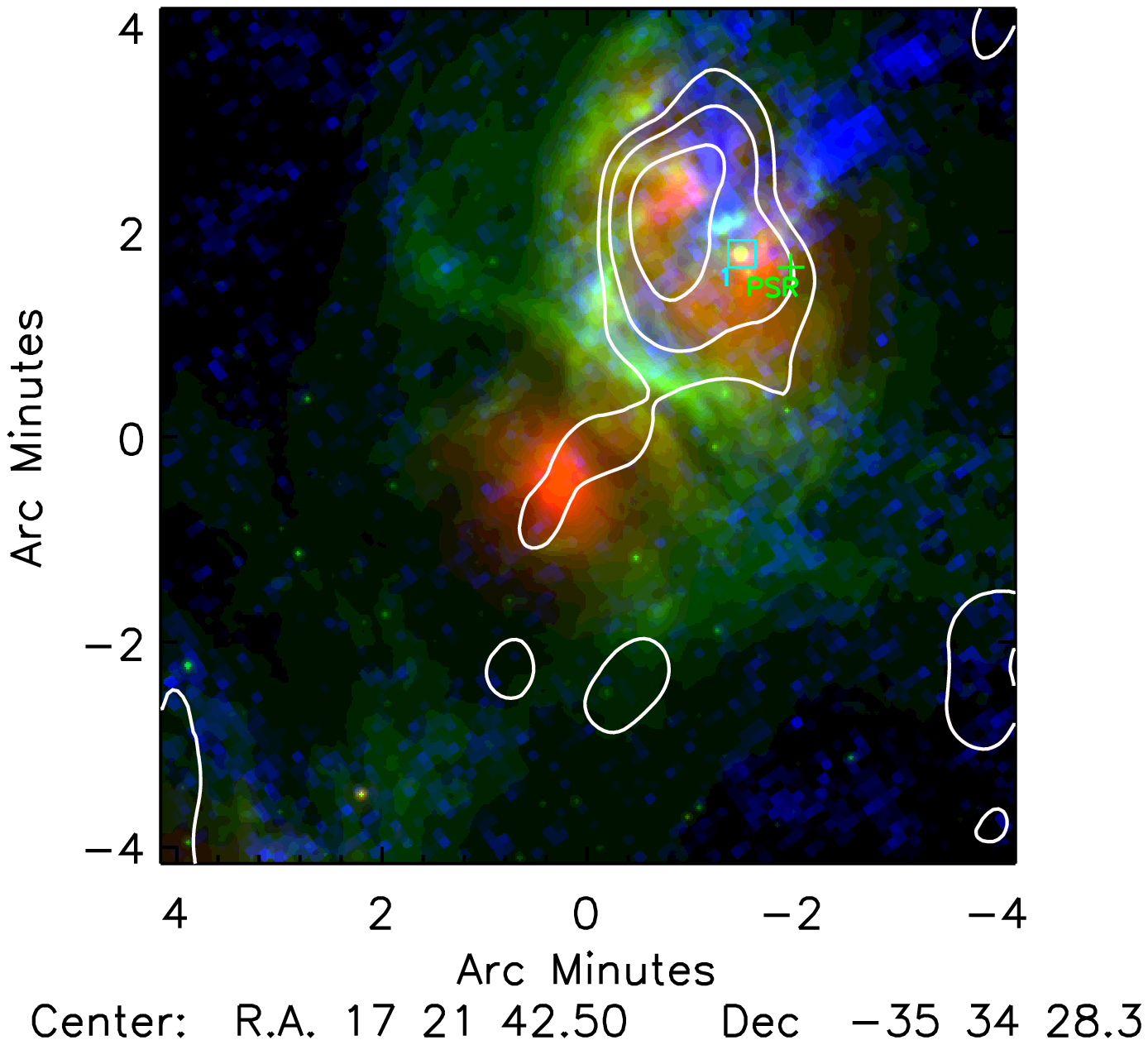}}
\end{center}
\caption{\label{mapbethoven} {\it Upper panel:} Composite image of CS 78; in blue the GLIMPSE 3.6 \um\ image, 
in green the GLIMPSE 8.0  \um, in red the MIPSGAL 24.0 \um. NVSS contours are overplotted  in white color.
The green cross marks the position of the pulsar B1718$-$35, the cyan square the star 2MASS J17213513$-$3532415.
Yellow diamonds marks the GLIMPSE young stellar objects identified by \citet{willis13}.
{\it Upper panel:} Composite image of CS 78; in blue the ATLASGAL image, 
in green the GLIMPSE 8.0  \um, in red the MIPSGAL 24.0 \um. NVSS contours are overplotted  in white color.
The green cross marks the position of the pulsar B1718$-$35, the cyan square the star 2MASS J17213513$-$3532415.
}
\end{figure}

We observed the bright infrared star 2MASS J17213513$-$3532415 
located at the center of the dust shell CS 78 and with a nearby pulsar.
This nebula is listed among the bubbles detected by GLIMPSE \citep[][]{churchwell07},
and it coincides with the 4\arcmin\ radio nebula, G351.70+0.66, detected by NVSS  
at 1.4 GHz \citep{condon98},  as shown in Fig.\ \ref{mapbethoven}.
This bubble is at the periphery  of a large giant molecular complex
NGC 6334 \citep[e.g.,][]{russeil16}. 

\subsection{The Nearby Pulsar B1718$-$35 and Distance to  CS 78}
The bright star  is located a few arcsecs away from
the B1718$-$35 pulsar \citep{minter08}. This is an energetic
young pulsar (spin down $4.5 \times  10^{34}$ erg s$^{-1}$,  age = 176 kyr) 
at a distance of about 4.6 kpc \citep{manchester05}; the
previously credited distance was 6.4 kpc \citep{gaensler00}. 
\citet{minter08} detected OH absorption against the PSR B1718$-$35. 
There are only two other pulsars with such absorption.  
The OH radial velocity (from $-1.5$ to $-1.9$ \kms) is consistent with
the radial velocity of the NGC 6334 complex \citep{brooks01}. 
OH absorption may arise in the interaction region of a molecular cloud 
with an SNR or an \HH\ region. Minter concluded that the continuum object 
along the line of sight toward PSR B1718$-$35 was an \HH\ region associated 
with the NGC 6334 complex. 
Indeed, for CS 78 (their G351.69+0.67), using radio continuum 
data \citet{russeil16} measured a thermal source ($\alpha=1.2$).
with a systemic \Vlsr=+3 \kms.\\
With $H_\alpha$, radio data and OB spectrophotometric distances, 
\citet{russeil16} and \citet{russeil17} concluded
that velocities from zero to $-11$ \kms\ belongs to the NGC 6334 complex.
In $H_\alpha$, they detected a foreground cloud at +6 \kms\ and a background at $-25$ \kms 
($\approx 3.7$ kpc); with CO data, larger velocities at $-$40 \kms\  $-$125 \kms\  
were also seen along the same line of sight, which correspond to distances of 
4.67 and 6.97 kpc \citep{reid09}. Therefore, the pulsar B1718$-$35
is likely located in the $-$40 \kms\  cloud.
\citet{fukui17} also published a new CO map of the NGC 6334 complex.
These authors detected the main component centered at $-4$ \kms\ (NGC 6334), 
a secondary (shifted) component at $-16$ \kms, a foreground component at 
+6 \kms\ in the local gas close to the sun, and an additional component at 
$-25$ \kms\ more distant than NGC 6334.

\subsection{The Star 2MASS J17213513$-$3532415}
The Gaia parallax ($0.81\pm0.22$ mas s$^{-1}$)
has a fractional error still larger than 20\%, and suggests a distance
from 0.9 kpc to 1.5 kpc, that  matches well that of the NGC 6334.

The 2MASS catalog reports high-quality measurements;
furthermore, the 2MASS photometry is in excellent agreement with the  
DENIS photometry (\Ks\ = 6.73 and 6.78  mag, $J$ = 10.13 and 10.11 mag).
Within the measurement errors, we can exclude variability \citep[e.g.,][]{messineo04}.
By assuming an intrinsic  $J-K = -0.02$ mag, and adopting the extinction ratios of 
\citet{messineo05}, we estimate   \Aks($JH$) = 1.54 mag  and \Aks($J$\Ks) = 1.81 mag.
By assuming a distance of 1.35 kpc (see next Section), and 
an absolute \Ks\ magnitude of  $-5.41$ mag, from  the dereddened flux densities, 
we estimate  a \BCK\ of $-0.75$ mag and \Mbol= $-6.16$ mag, i.e., a luminosity 
 log$_{10}\frac{\Lstar}{\Lsun}=4.36$.
Dwarfs of B3-B5 types have  \Mbol\  from $-4.0$ to $-2.2$ mag,
Ia supergiants  (B3-B5) have \Mbol\ $\approx -6.5$  mag,
Ib supergiants  (B3-B5) have \Mbol\ $\approx -5.8$ mag, and
a B3 giant  has \Mbol $\approx -4.5$ mag 
\citep{humphreys84,lejeune01,koornneef83,messineo11}.

\begin{figure}[t]
\begin{center}
\resizebox{0.99\hsize}{!}{\includegraphics[angle=0]{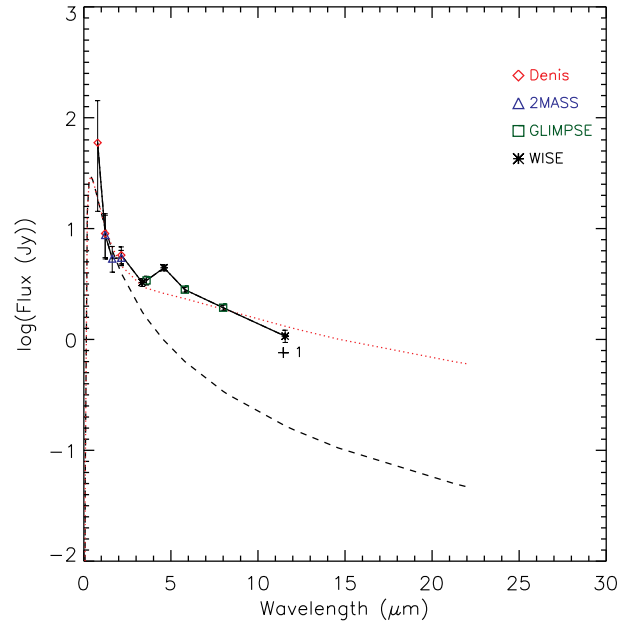}}
\end{center}
\caption{\label{bethovensed} Stellar energy distribution (SED) of star E1. 
The long-dashed curve marks a blackbody curve, $F_\nu$, of 21,500 K.
The red dotted line is a modified blackbody curve, where a second blackbody of
650 K and a free-free emission has been added.
Magnitudes from DENIS are marked with diamonds, from 2MASS with triangles, 
from GLIMPSE with squares, from MSX with crosses, and from WISE with asterisks. 
}
\end{figure}

We conclude, therefore, that star E1 is a B supergiant.
Although, higher-resolution spectra are needed to characterize the stellar 
atmosphere of star E1; its SED shows 
an excess at mid-infrared wavelengths (Fig.\ \ref{bethovensed}), 
most likely due to a combination of free-free emission 
plus a dust component ($\approx 650$ K).

There is  ample ongoing discussion about the classification of B supergiants 
with emission lines  \citep[e.g.,][]{humphreys17,smith18, clark05}. 
Spectroscopically, candidate LBVs and B[e] supergiants can appear similar.
LBVs can only be  classified with the help of  long-term monitoring in order to 
verify their photometric variability 
and the presence of sporadic luminosity outbursts. B[e] stars are  characterized by 
warm dust emitting longward of 1 \um\ \citep[e.g.,][]{clark99,humphreys17}.
Known LBVs have initial masses from $\approx30$ to $\approx$100 \Msun,
luminosities from 200,000 \Lsun\ to 5,000,000 \Lsun, 
and are located on the instability strip 
near the Eddington limit. B[e] stars  are  fainter  
(typically from 30,000 to 200,000 \Lsun, 15-40 \Msun) 
with only narrow overlap in luminosities with the LBVs \citep[e.g.,][]{humphreys17,smith15}.
It is unclear if those low-luminosity stars, like 2MASS J17213513$-$3532415, 
are  analogous to LBVs, or if they come from a different formation  scenario
(e.g., mass transfer in binary systems and stellar mergers)
\citep[e.g.,][]{smith15}.

\subsection{ CS 78 and Its Ionizing Source}
CS 78 is likely associated  with the NGC 6334 complex at about 1.35 kpc 
\citep{wu14,minter08,russeil16}.
The bright 2MASS J17213513$-$3532415 is located in the core of this bubble.  

The morphology and high extinction of the bubble 
suggest that 2MASS J17213513$-$3532415  is the main 
ionizing source of the CS 78 bubble.  
The  radio continuum flux at 1.4 GHz detected toward the nebula by NVSS is
89.8 mJy. By assuming  a typical temperature of 7500 K  for an optically thin 
thermal nebula \citep{martinhernandez03}, we calculated  a number  of Lyman 
continuum photons (N$_L$) of  $10^{45.88}$. When we consider an NVSS flux of 124 mJy 
(which includes two fainter NVSS sources at the edge of the nebula), we calculated   
N$_L$ = $10^{46.02}$. This number of photons is expected from a single B2-B3 
supergiant \citep{panagia73}, and this further confirms the luminosity class; 
for  example, a B3 III star would only emit $10^{44.3}$. 
2MASS J17213513$-$3532415, therefore, is the ionizing star of bubble CS 78.

The bubble is flagged as an incomplete ring \citep{churchwell07},
and is shown in Fig. \ref{mapbethoven}. The 8 \um\ emission  is due to PAH emission located 
at the edge of the photodissociation region (NVSS radio continuum source), 
which is filled with 24 \um\ emission (emission by heated dust).  
The entire NGC 6334 complex has been analyzed with GLIMPSE data to locate young 
stellar objects by \citet{willis13}. A few are located in the direction of CS 78, 
indicating that star formation is still ongoing in the shell. 
  
\section{Conclusive Considerations}

We searched for massive OB stars in 1 of 
the 14 most prominent \HH\ region complexes 
of the Milky Way detected  with data from the Planck satellite 
\citep{rahman10}. 
Of these, the G332.809$-$0.132 complex is
relatively nearby (about 3 kpc away), and particularly rich in high-energy sources.
Three SNRs and one TeV emitter, HESS J1616$-$508,
are found along its  line of sight.  This associated molecular
cloud complex  comprises the Nor OB4/RCW 103 association.
We focused our analysis on the central 80 pc of Nor OB4, which encloses the three SNRs. 
This is a rare coincidence; indeed,  no SNRs among those listed in
\citet{green14} are associated with the giant \HH\ regions of 
\citet{conti04}.
Forty-nine SNRs (17\%)  have their centroids projected on one of the 14 WMAP sources  \citep[Table1 of][]{rahman10}
(from 0 to 10 SNRs). Only 11 SNRs (4\%) are projected on one of the corresponding 40 SF complexes 
identified at mid-infrared in Table 2 of \citet[][]{rahman10} (from zero to three SNRs). 
G332.809$-$0.132 is the only  complex in Table 2 of Rahman with 3 SNRs.

Those three SNRs are found in the area covered by Nor OB4, providing
 evidence for a  rich population
of  evolved massive stars and transitional objects toward this direction. 
In the molecular cloud GMC G23.3-0.3 (80 pc in size), which hosts four SNRs,
\citet{messineo14a} detected one rare candidate LBV  
in apparent isolation along with other evolved massive stars.
Stellar clusters may have dissolved already, or the most massive 
stars may have been ejected, or have even been born in isolation.
As noticed by \citet{smith15}, with a few exceptions 
(e.g Eta Car is in the Tr16 cluster)
LBVs are mostly found in isolation: 
AG Car, HR Car, Hen 3-519, HD160529, MWC 930  are located 
more than 20 pc away from an O-type star. 
It is clear that to find those spectacular stars we must look for 
bright stars in extended \HH\ regions.

Its richness in SNRs and the shaped mid-infrared emission 
of G332.809$-$0.132
are suggestive of feedback from massive stars.  In its core,
we   found more than 200 early-type stars, about 110 early-type stars 
brighter than \Ks\ = 10 mag (O9) and with $J-$\Ks\ color 
from 0.7 mag to 3.0 mag. 
By assuming supergiants, this number alone could account for  50\% 
of the number of Lyman continuum photons measured towards G332.809$-$0.132.
Therefore, we confirm that this region is rich in massive stars and 
transitional objects.
Indeed,  two  WR stars were already known in the direction 
of this complex, WR74 (WN 7) and $[$SMG2009$]$~1059$-$34 (WC 8).
Among the free-free candidates,
IRAS 16115$-$5044 is reclassify as a candidate LBV.
By targeting bright stars with colors typical of free-free emitters,
we have added spectroscopic detections of another WR, a WN 8 star, 
and  a rare transitional
star, an Ofpe/WN 9. Those three WR stars have spectrophotometric
properties consistent with those of supergiants at a distance of about 3.4 kpc --
\Mbol\ = $-$6.57 (WC 8), $-$8.16 (WN 7), $-$9.16 (WN 8). 
For the  Ofpe/WN 9 star, \Mbol\ = $-$10.45 mag.
Ofpe/WN 9 stars are transitional objects,  perhaps 
dormant LBVs \citep{martins07}.
For 3.4 kpc IRAS 16115$-$5044 has \Mbol\ = $-$8.8 mag.

The spatial distribution of classical LBVs is expected to be statistically 
similar to that of  WR stars \citep{aadland18}.
Within the Nor OB4 association, we measured  a typical separation of 3--5 pc 
between candidate free-free emitters and the closest OB candidates.

Another rare  transitional object is reported here as  the central 
ionizing star of CS 78. For an assumed distance of 1.35 kpc,
2MASS J17213513$-$3532415 has   \Mbol$\approx -6.16$ mag; 
with this faint magnitude it is likely to be a B[e] star, 
and not an LBV. Since, their formation, final fate, 
and relation to the high-mass LBVs are highly uncertain \citep{smith18}, 
it is of primary importance to increase the
number of these peculiar  B supergiants.

\begin{appendix}
\def\thefigure{\thesection}
\def\thetable{\thesection}

\section{Photometric Data}
\label{appendix.photometry}

We provide  collected photometric information on the spectroscopically detected stars in the direction of 
G332.809$-$0.132 and  CS 78 in Table \ref{table.phot}1.

\begin{sidewaystable*} 
\vspace*{+0.5cm} %;;;two columns
\caption{\label{table.phot} 1. Infrared Counterparts of the Spectroscopically Observed Targets.}  
{\tiny
\begin{tabular}{@{\extracolsep{-.08in}}r|rrrr|rrr|rrrr|rrrr|rrrr|rrr|rrrrr|rrrr|r}
\hline 
\hline 
    &  \multicolumn{4}{c}{\rm 2MASS}   &\multicolumn{3}{c}{\rm DENIS} &   \multicolumn{4}{c}{\rm GLIMPSE}   &  \multicolumn{4}{c}{\rm MSX}& \multicolumn{4}{c}{\rm WISE} & \multicolumn{3}{c}{\rm NOMAD} & \multicolumn{5}{c}{\rm GAIA-DR2} \\ 
\hline 
 {\rm ID}& {\rm 2MASS-ID} &  {\it J} & {\it H} & {\it $K_{\rm S}$}  &
 {\it I} & {\it J} & {\it $K_{\rm S}$} & { [3.6]} & { [4.5]} & { [5.8]} & { [8.0]} &
 {\it A} & {\it D} & {\it C}& {\it D} &{\it W1} &{\it W2} & {\it W3} &{\it W4} & {\it B} &{\it V} & {\it R} & {\it G} & VR &{ plx} & { pmRa} & { pmDec} &  \\ 
\hline 
  & & {\rm (mag)}   &    {\rm (mag)}    & {\rm (mag)}     & {\rm (mag)} &{\rm (mag)}  & {\rm (mag)}  & 
 {\rm (mag)}  & {\rm (mag)} &{\rm (mag)} &{\rm (mag)}  & {\rm (mag)}&{\rm (mag)}&{\rm (mag)}&{\rm (mag)}&{\rm (mag)}&{\rm (mag)}& 
 {\rm (mag)}& {\rm (mag)}& {\rm (mag)}&{\rm (mag)}&{\rm (mag)}&{\rm (mag)}& (\kms)& (mas) & (mas/yr) & (mas/yr) \\ 
\hline                       

                            E1       &   J17213513-3532415  & 10.13 &  8.30 &  6.78 & 14.68 & 10.11 &  6.73 &  5.48 & \nodata &  4.48 &  4.29 & \nodata & \nodata & \nodata & \nodata &  5.62 &  4.43 &  4.05 & \nodata & \nodata& \nodata&  17.61 &     16.11   &      $..$    &     0.81$\pm$     0.22 &   $-$0.38$\pm$   0.36 &   $-$2.77$\pm$   0.24 &  \\
                            E2       &   J16174291-5054251  &  7.64 &  6.63 &  6.05 & 10.84 &  7.17 &  5.96 &  5.84 &  6.74 &  5.23 &  5.07 &  5.03 & \nodata & \nodata & \nodata &  5.72 &  5.25 &  5.12 &  3.92 & \nodata& \nodata& \nodata&     12.37   &      $..$    &     0.57$\pm$     0.14 &   $-$2.63$\pm$   0.19 &   $-$4.15$\pm$   0.14 &  \\
                            E3       &   J16174204-5056372  &  7.78 &  6.68 &  6.08 & 11.40 &  7.43 &  5.62 &  6.70 &  6.04 &  5.43 &  5.37 &  5.22 & \nodata & \nodata & \nodata &  5.58 &  5.17 &  5.43 &  4.99 & \nodata&  17.02 &  13.69 &     12.85   &      $..$    &     0.88$\pm$     0.19 &   $-$2.23$\pm$   0.25 &   $-$5.31$\pm$   0.20 &  \\
                            E4       &   J16201606-5058076  &  8.43 &  7.78 &  7.43 & 10.69 &  8.58 &  7.47 &  7.11 &  6.87 &  6.68 &  6.56 & \nodata & \nodata & \nodata & \nodata &  6.94 &  6.73 &  5.34 &  0.39 &  13.70 &  13.24 &  12.60 &     12.08   &      $..$    &     1.32$\pm$     0.56 &   $-$5.49$\pm$   0.74 &   $-$6.41$\pm$   0.68 &  \\
                            E5       &   J16170220-5047031  & 11.97 & 10.65 &  9.73 & 15.64 & \nodata & \nodata & \nodata & \nodata & \nodata & \nodata &  1.40 &  0.30 &  0.05 & $-$2.51 & \nodata & \nodata & \nodata & \nodata & \nodata& \nodata&  17.21 &     17.77   &      $..$    &     0.19$\pm$     0.18 &   $-$4.06$\pm$   0.33 &   $-$4.24$\pm$   0.20 &  \\
                            E6       &   J16200902-5048588  & 10.43 & 10.23 & 10.10 & 11.23 & 10.53 & 10.06 &  9.99 &  9.94 &  9.84 &  9.44 & \nodata & \nodata & \nodata & \nodata & \nodata & \nodata & \nodata & \nodata &  12.90 &  11.98 &  11.40 &     12.06   &      $..$    &     0.31$\pm$     0.04 &    2.34$\pm$   0.06 &   $-$0.16$\pm$   0.04 &  \\
                            E7       &   J16170923-5047147  & 13.26 & 11.47 & 10.15 & 15.82 & 13.24 & 10.15 &  8.79 &  8.22 &  7.69 &  7.08 & \nodata & \nodata & \nodata & \nodata &  8.73 &  7.91 &  5.77 & \nodata &  17.43 &  16.91 &  16.42 &     18.40   &      $..$    &     0.10$\pm$     0.23 &   $-$3.03$\pm$   0.39 &   $-$4.06$\pm$   0.27 &  \\
                            E8       &   J16170300-5047308  & 11.54 & 10.68 & 10.29 & 14.46 & 11.59 & 10.34 &  9.97 &  9.92 &  9.80 & \nodata & \nodata & \nodata & \nodata & \nodata & \nodata & \nodata & \nodata & \nodata & \nodata&  16.88 &  16.03 &     16.00   &      $..$    &     0.07$\pm$     0.10 &   $-$3.89$\pm$   0.17 &   $-$3.52$\pm$   0.11 &  \\
                            L1       &   J16172975-5055396  &  5.58 &  4.47 &  3.87 &  9.00 &  5.74 &  3.66 & \nodata & \nodata &  3.65 &  3.45 &  3.52 &  3.51 &  3.36 & \nodata &  3.69 &  3.29 &  3.47 &  2.61 &  13.91 &  13.50 &  11.86 &     10.18   &      $..$    &     1.73$\pm$     0.20 &   $-$0.41$\pm$   0.34 &   $-$3.58$\pm$   0.21 &  \\
                            L2       &   J16202085-5053372  &  9.31 &  6.04 &  4.41 & \nodata &  9.20 &  3.86 & \nodata &  3.31 &  2.61 & \nodata &  1.53 & $-$0.10 & $-$0.32 & $-$0.86 &  3.12 &  2.01 & \nodata & $-$1.26 & \nodata& \nodata& \nodata&     19.04   &      $..$    &     0.74$\pm$     0.52 &   $-$0.54$\pm$   0.77 &   $-$3.26$\pm$   0.53 &  \\
                            L3       &   J16165866-5104016  &  8.44 &  6.10 &  4.95 & \nodata & \nodata & \nodata &  4.27 &  4.51 &  3.96 &  4.01 &  3.60 &  2.41 &  2.11 & \nodata &  4.39 &  3.74 &  2.92 &  1.74 & \nodata& \nodata& \nodata&     16.78   &      $..$    &    $-$0.42$\pm$     0.44 &   $-$3.71$\pm$   1.13 &   $-$1.84$\pm$   0.85 &  \\
                            L4       &   J16181449-5056134  &  8.04 &  6.06 &  5.09 & 14.86 &  8.05 &  4.18 & \nodata &  4.42 &  3.72 & \nodata &  3.48 &  2.61 &  2.32 & \nodata &  4.82 &  3.33 &  2.89 &  1.98 & \nodata& \nodata& \nodata&     15.94   &      $..$    &     0.11$\pm$     0.27 &   $-$3.53$\pm$   0.48 &   $-$3.88$\pm$   0.29 &  \\
                            L5       &   J16182588-5102055  &  8.38 &  6.27 &  5.16 & 15.48 &  8.45 &  4.41 &  4.40 &  4.70 &  4.00 &  4.05 &  3.54 &  2.29 &  2.12 &  1.60 &  4.63 &  3.88 &  2.99 &  1.77 & \nodata& \nodata& \nodata&     16.19   &      $..$    &     0.37$\pm$     0.27 &   $-$3.53$\pm$   0.44 &   $-$2.79$\pm$   0.29 &  \\
                            L6       &   J16173092-5058351  &  6.46 &  5.61 &  5.32 &  8.89 &  6.06 &  4.41 &  5.22 &  6.67 &  5.17 &  5.14 &  5.07 & \nodata & \nodata & \nodata &  5.27 &  5.05 &  5.11 &  4.02 &  12.28 &  10.08 &   9.40 &      9.21   &    $-$24.24$\pm$     0.37  &     0.61$\pm$     0.04 &   $-$0.77$\pm$   0.07 &   $-$1.70$\pm$   0.05 &  \\
                            L7       &   J16170875-5102172  &  6.55 &  5.77 &  5.43 &  8.73 &  6.14 &  4.88 &  5.22 & \nodata &  5.23 &  5.23 &  5.13 & \nodata & \nodata & \nodata &  5.57 &  5.38 &  5.36 & \nodata &  12.14 &  10.33 &   9.45 &      9.31   &    $-$42.46$\pm$     0.27  &     0.56$\pm$     0.04 &   $-$1.21$\pm$   0.07 &   $-$2.51$\pm$   0.04 &  \\
                            L8       &   J16165202-5105335  &  8.46 &  6.51 &  5.58 & \nodata & \nodata & \nodata &  5.63 &  5.85 &  4.70 &  4.45 &  4.20 &  3.51 &  3.47 & \nodata &  5.42 &  4.49 &  3.95 &  3.08 & \nodata& \nodata& \nodata&     15.33   &      $..$    &     0.19$\pm$     0.27 &   $-$2.80$\pm$   0.49 &   $-$1.75$\pm$   0.28 &  \\
                            L9       &   J16171876-5101364  &  8.75 &  6.74 &  5.61 & \nodata &  8.74 &  5.04 & \nodata & \nodata &  4.43 &  4.15 &  3.67 &  2.65 &  2.46 & \nodata &  5.03 &  3.91 &  3.00 &  1.89 & \nodata& \nodata& \nodata&     16.48   &      $..$    &     0.10$\pm$     0.32 &   $-$3.68$\pm$   0.61 &   $-$3.88$\pm$   0.31 &  \\
                           L10       &   J16165827-5050474  &  6.98 &  6.00 &  5.61 & \nodata & \nodata & \nodata & \nodata &  6.12 &  5.40 &  5.40 &  5.31 & \nodata & \nodata & \nodata &  5.59 &  5.47 &  5.50 &  5.31 &  13.11 &  12.27 &  10.05 &     10.32   &    $-$19.60$\pm$     0.56  &     0.54$\pm$     0.08 &    1.24$\pm$   0.13 &   $-$2.43$\pm$   0.08 &  \\
                           L11       &   J16171805-5057292  &  9.55 &  7.05 &  5.90 & 16.83 &  9.43 &  5.41 &  6.97 & \nodata &  4.79 &  4.59 &  4.62 & \nodata & \nodata & \nodata &  5.28 &  4.81 &  4.66 &  3.98 & \nodata& \nodata& \nodata&     17.47   &      $..$    &    \nodata$\pm$     0.25 &   $-$4.32$\pm$   0.49 &   $-$4.57$\pm$   0.28 &  \\
                           L12       &   J16164958-5051172  &  8.51 &  6.73 &  5.93 & \nodata & \nodata & \nodata &  5.34 &  6.08 &  5.36 &  5.34 &  4.95 & \nodata & \nodata & \nodata &  5.32 &  5.41 &  5.35 &  5.03 & \nodata& \nodata&  15.43 &     14.57   &      $..$    &     0.34$\pm$     0.14 &   $-$2.89$\pm$   0.29 &   $-$3.62$\pm$   0.16 &  \\
                           L13       &   J16170473-5059417  &  7.49 &  6.38 &  6.00 &  9.82 &  7.64 &  5.54 &  6.07 &  6.13 &  5.80 &  5.74 &  5.63 & \nodata & \nodata & \nodata &  5.87 &  5.91 &  5.68 &  5.14 & \nodata& \nodata& \nodata&     11.28   &     $-$7.18$\pm$     0.25  &     0.67$\pm$     0.14 &   $-$2.67$\pm$   0.20 &   $-$4.80$\pm$   0.15 &  \\
                           L14       &   J16161669-5114598  &  8.31 &  6.81 &  6.11 & \nodata & \nodata & \nodata &  5.88 &  6.35 &  5.59 &  5.58 &  5.49 & \nodata & \nodata & \nodata &  5.86 &  5.72 &  5.07 &  3.38 & \nodata& \nodata& \nodata&     13.83   &      $..$    &    $-$0.46$\pm$     0.18 &   $-$3.77$\pm$   0.26 &   $-$3.97$\pm$   0.17 &  \\
                           L15       &   J16190855-5045090  &  9.12 &  7.11 &  6.11 & 16.38 &  9.09 &  5.95 &  6.10 &  5.91 &  5.12 &  4.97 &  4.79 & \nodata & \nodata & \nodata &  5.65 &  5.21 &  4.46 &  3.30 & \nodata& \nodata& \nodata&     16.90   &      $..$    &     0.85$\pm$     0.29 &    0.58$\pm$   0.44 &   $-$1.45$\pm$   0.30 &  \\
                           L16       &   J16222857-5049280  &  9.42 &  7.36 &  6.33 & 16.45 &  9.36 &  6.24 &  5.57 &  6.09 &  5.18 &  5.06 &  4.92 & \nodata & \nodata & \nodata &  5.72 &  5.28 &  4.67 &  2.64 & \nodata& \nodata& \nodata&     17.02   &      $..$    &    $-$1.46$\pm$     0.39 &   $-$4.48$\pm$   0.52 &   $-$3.93$\pm$   0.33 &  \\
                           L17       &   J16172446-5103320  &  7.74 &  6.67 &  6.33 &  9.78 &  7.58 &  6.26 &  6.42 & \nodata &  6.15 &  6.12 &  6.23 & \nodata & \nodata & \nodata &  6.12 &  6.26 &  6.21 & \nodata &  13.78 &  12.78 &  11.22 &     11.16   &    $-$43.41$\pm$     0.23  &     0.54$\pm$     0.09 &   $-$2.27$\pm$   0.15 &   $-$1.57$\pm$   0.10 &  \\
                           L18       &   J16193796-5057298  & 10.46 &  7.75 &  6.36 & 16.54 & 10.35 &  6.16 &  5.48 &  5.92 &  4.95 &  4.88 & \nodata & \nodata & \nodata & \nodata &  5.64 &  4.96 &  4.29 &  1.80 &  17.76 & \nodata&  15.63 &     19.60   &      $..$    &    $-$0.56$\pm$     0.63 &   $-$2.64$\pm$   1.33 &   $-$3.75$\pm$   0.93 &  \\
                           L19       &   J16191621-5102176  & 10.26 &  7.64 &  6.38 & \nodata & 10.28 &  6.34 &  5.30 &  6.63 &  4.89 &  4.77 &  5.35 & \nodata & \nodata & \nodata &  5.65 &  5.24 &  4.17 &  2.42 & \nodata& \nodata& \nodata&     19.05   &      $..$    &    $-$0.53$\pm$     0.46 &   $-$6.00$\pm$   0.82 &   $-$4.79$\pm$   0.51 &  \\
                           L20       &   J16215861-5105148  & 10.04 &  7.66 &  6.39 & \nodata & 10.15 &  6.46 &  6.66 &  5.54 &  5.16 &  4.94 &  5.03 & \nodata & \nodata & \nodata &  6.06 &  5.35 &  4.51 &  3.43 & \nodata& \nodata& \nodata&     19.31   &      $..$    &     0.22$\pm$     0.68 &    0.04$\pm$   1.14 &   $-$4.75$\pm$   0.79 &  \\
                           L21       &   J16173873-5059383  &  9.97 &  7.63 &  6.50 & 17.60 &  9.95 &  6.43 &  6.03 &  5.86 &  5.46 &  5.33 &  5.12 & \nodata & \nodata & \nodata &  5.78 &  5.58 &  4.75 &  3.36 & \nodata& \nodata& \nodata&     18.10   &      $..$    &     0.03$\pm$     0.31 &   $-$3.66$\pm$   0.56 &   $-$5.63$\pm$   0.34 &  \\
                           L22       &   J16184330-5056401  &  9.40 &  7.44 &  6.50 & 15.45 &  9.35 &  6.42 &  6.69 &  6.00 &  5.64 &  5.48 &  4.98 & \nodata & \nodata & \nodata &  5.92 &  5.75 &  4.71 &  3.64 & \nodata& \nodata& \nodata&     16.30   &      $..$    &     0.19$\pm$     0.25 &   $-$4.63$\pm$   0.38 &   $-$4.82$\pm$   0.24 &  \\
                           L23       &   J16181315-5102106  &  9.09 &  7.57 &  6.72 & 13.24 &  8.90 &  6.59 &  6.71 &  5.85 &  5.59 &  5.33 &  4.97 & \nodata & \nodata & \nodata &  6.15 &  5.77 &  4.73 &  3.61 & \nodata& \nodata& \nodata&     16.44   &      $..$    &    $-$0.38$\pm$     0.26 &   $-$2.99$\pm$   0.41 &   $-$4.11$\pm$   0.27 &  \\
                           L24       &   J16215678-5101145  &  9.19 &  7.51 &  6.78 & 14.02 &  9.13 &  6.73 &  6.83 &  6.60 &  6.29 &  6.22 & \nodata & \nodata & \nodata & \nodata &  6.34 &  6.46 &  6.33 &  3.64 & \nodata& \nodata& \nodata&     15.33   &      $..$    &    $-$0.12$\pm$     0.23 &   $-$2.49$\pm$   0.30 &   $-$2.99$\pm$   0.22 &  \\
                           L25       &   J16170351-5055026  &  9.25 &  7.55 &  6.88 & 13.41 &  9.22 &  6.96 &  6.83 &  6.72 &  6.41 &  6.37 & \nodata & \nodata & \nodata & \nodata &  6.52 &  6.60 &  6.49 &  6.39 &  17.80 & \nodata&  15.63 &     14.96   &      $..$    &     0.29$\pm$     0.17 &   $-$4.86$\pm$   0.28 &   $-$4.34$\pm$   0.18 &  \\
                           L26       &   J16222509-5048358  & 10.27 &  8.04 &  6.95 & \nodata & 10.19 &  6.94 &  6.63 &  6.17 &  5.78 &  5.59 &  5.46 & \nodata & \nodata & \nodata &  6.28 &  5.99 &  5.07 &  3.66 & \nodata& \nodata& \nodata&     18.36   &      $..$    &    $-$0.89$\pm$     0.36 &   $-$4.97$\pm$   0.52 &   $-$5.02$\pm$   0.34 &  \\
                           L27       &   J16180145-5055588  &  9.42 &  7.69 &  6.97 & 13.87 &  9.33 &  6.98 &  6.68 &  6.70 &  6.38 &  6.36 & \nodata & \nodata & \nodata & \nodata &  6.64 &  6.64 &  6.04 &  4.06 & \nodata& \nodata&  18.28 &     15.26   &      $..$    &     0.10$\pm$     0.19 &   $-$1.29$\pm$   0.27 &   $-$2.46$\pm$   0.20 &  \\
                           L28       &   J16173617-5055236  &  9.40 &  7.71 &  7.03 & 13.55 &  9.27 &  7.01 &  6.87 &  6.79 &  6.52 &  6.48 & \nodata & \nodata & \nodata & \nodata &  6.68 &  6.74 &  6.78 & \nodata & \nodata& \nodata&  17.26 &     15.06   &      $..$    &     0.13$\pm$     0.16 &   $-$3.34$\pm$   0.24 &   $-$1.99$\pm$   0.17 &  \\
                           L29       &   J16161970-5114017  &  9.56 &  7.84 &  7.07 & \nodata & \nodata & \nodata &  6.73 &  6.64 &  6.39 &  6.43 &  5.53 & \nodata & \nodata & \nodata &  6.58 &  6.57 &  5.92 &  3.86 & \nodata& \nodata& \nodata&     15.65   &      $..$    &    $-$0.00$\pm$     0.20 &   $-$3.67$\pm$   0.28 &   $-$2.59$\pm$   0.19 &  \\
                           L30       &   J16174539-5058020  &  9.43 &  7.82 &  7.11 & 13.69 &  9.37 &  7.14 &  6.82 &  6.95 &  6.61 &  6.68 & \nodata & \nodata & \nodata & \nodata &  6.73 &  6.82 &  4.97 &  3.03 & \nodata& \nodata& \nodata&     15.06   &      $..$    &     0.17$\pm$     0.17 &   $-$4.58$\pm$   0.24 &   $-$4.06$\pm$   0.17 &  \\
                           L31       &   J16173534-5057091  &  8.15 &  7.39 &  7.14 &  9.75 &  8.14 &  7.08 &  7.00 &  7.09 &  6.99 &  6.91 &  6.33 & \nodata & \nodata & \nodata &  6.91 &  7.03 &  6.53 &  4.77 &  12.69 &  11.72 &  10.73 &     10.77   &    $-$57.69$\pm$     0.74  &     0.57$\pm$     0.09 &   $-$0.75$\pm$   0.14 &   $-$3.88$\pm$   0.11 &  \\
                           L32       &   J16170651-5100058  & 10.67 &  8.25 &  7.14 & 17.83 & 10.57 &  7.12 &  6.88 &  6.47 &  6.17 &  6.13 & \nodata & \nodata & \nodata & \nodata &  6.43 &  6.41 &  6.28 &  5.57 & \nodata& \nodata& \nodata&     18.46   &      $..$    &    $-$0.34$\pm$     0.35 &   $-$4.89$\pm$   0.62 &   $-$4.58$\pm$   0.41 &  \\
                           L33       &   J16192478-5108209  &  9.75 &  8.00 &  7.18 & 14.74 &  9.72 &  7.17 &  6.82 &  6.86 &  6.59 &  6.57 & \nodata & \nodata & \nodata & \nodata &  6.77 &  6.81 &  6.44 &  4.10 & \nodata& \nodata&  20.90 &     16.06   &      $..$    &     0.15$\pm$     0.18 &   $-$3.56$\pm$   0.28 &   $-$1.87$\pm$   0.20 &  \\
                           L34       &   J16164196-5106055  &  9.94 &  8.05 &  7.20 & \nodata & \nodata & \nodata &  6.78 &  6.79 &  6.56 &  6.57 & \nodata & \nodata & \nodata & \nodata &  6.56 &  6.63 &  7.03 & \nodata & \nodata& \nodata&  19.67 &     16.45   &      $..$    &    $-$0.02$\pm$     0.22 &   $-$3.42$\pm$   0.46 &   $-$2.87$\pm$   0.23 &  \\
                           L35       &   J16202293-5053290  & 12.17 &  8.83 &  7.20 & \nodata & 11.97 &  7.12 &  6.70 & \nodata &  5.63 &  5.74 & \nodata & \nodata & \nodata & \nodata &  6.15 &  6.01 &  5.71 & \nodata & \nodata& \nodata& \nodata&    \nodata  &      $..$    &    \nodata&    $..$   &    $..$   &  \\
                           L36       &   J16165014-5112236  & 10.08 &  8.15 &  7.28 & \nodata & \nodata & \nodata &  7.14 &  6.92 &  6.62 &  6.52 & \nodata & \nodata & \nodata & \nodata &  6.91 &  6.89 &  6.73 &  5.13 &  20.47 & \nodata&  19.81 &     16.96   &      $..$    &    $-$0.32$\pm$     0.28 &   $-$4.63$\pm$   0.49 &   $-$4.39$\pm$   0.28 &  \\
                           L37       &   J16185780-5052293  & 10.23 &  8.25 &  7.29 & 16.42 & 10.17 &  7.30 &  7.00 &  6.90 &  6.53 &  6.53 & \nodata & \nodata & \nodata & \nodata &  6.83 &  6.86 &  5.96 &  4.01 & \nodata& \nodata& \nodata&     17.32   &      $..$    &     0.72$\pm$     0.26 &   $-$3.26$\pm$   0.41 &   $-$4.40$\pm$   0.27 &  \\
                           L38       &   J16164375-5111132  & 10.31 &  8.26 &  7.33 & \nodata & \nodata & \nodata &  7.00 &  6.79 &  6.48 &  6.45 &  6.39 & \nodata & \nodata & \nodata &  6.72 &  6.69 &  6.36 &  4.95 & \nodata& \nodata& \nodata&     17.59   &      $..$    &    \nodata&    $..$   &    $..$   &  \\
                           L39       &   J16164698-5108434  & 10.01 &  8.22 &  7.39 & \nodata & \nodata & \nodata &  7.00 &  7.07 &  6.80 &  6.76 &  6.17 & \nodata & \nodata & \nodata &  6.87 &  6.91 &  6.60 &  4.55 & \nodata& \nodata& \nodata&     16.22   &      $..$    &    $-$0.27$\pm$     0.24 &   $-$4.31$\pm$   0.34 &   $-$1.52$\pm$   0.23 &  \\

\hline
\end{tabular}
}
\begin{list}{}{}
\item[{\bf Notes.}] Identification numbers are as in Tables \ref{obs.early},
\ref{table.latespectra}.~ 
($^a$) 2MASS upper limits were replaced with VVV photometry.
Stars with $11.5 <$ \Ks $<10.5$ are, however, in the nonlinear regime of the VIRCAM camera.
\end{list}
\end{sidewaystable*}

\addtocounter{table}{-1}
\begin{sidewaystable*} 
\vspace*{+0.5cm} %;;;two columns
\caption{ 1. Continuation of Table \ref{table.phot}1.}  
{\tiny
\begin{tabular}{@{\extracolsep{-.08in}}r|rrrr|rrr|rrrr|rrrr|rrrr|rrr|rrrrr|rrrr|r}
\hline 
\hline 
    &  \multicolumn{4}{c}{\rm 2MASS}   &\multicolumn{3}{c}{\rm DENIS} &   \multicolumn{4}{c}{\rm GLIMPSE}   &  \multicolumn{4}{c}{\rm MSX}& \multicolumn{4}{c}{\rm WISE} & \multicolumn{3}{c}{\rm NOMAD} & \multicolumn{5}{c}{\rm GAIA-DR2} \\ 
\hline 
 {\rm ID}& {\rm 2MASS-ID} &  {\it J} & {\it H} & {\it $K_{\rm S}$}  &
 {\it I} & {\it J} & {\it $K_{\rm S}$} & { [3.6]} & { [4.5]} & { [5.8]} & { [8.0]} &
 {\it A} & {\it D} & {\it C}& {\it D} &{\it W1} &{\it W2} & {\it W3} &{\it W4} & {\it B} &{\it V} & {\it R} & {\it G} & VR &{ plx} & { pmRa} & { pmDec} &  \\ 
\hline 
  & & {\rm (mag)}   &    {\rm (mag)}    & {\rm (mag)}     & {\rm (mag)} &{\rm (mag)}  & {\rm (mag)}  & 
 {\rm (mag)}  & {\rm (mag)} &{\rm (mag)} &{\rm (mag)}  & {\rm (mag)}&{\rm (mag)}&{\rm (mag)}&{\rm (mag)}&{\rm (mag)}&{\rm (mag)}& 
 {\rm (mag)}& {\rm (mag)}& {\rm (mag)}&{\rm (mag)}&{\rm (mag)}&{\rm (mag)}& (\kms)& (mas) & (mas/yr) & (mas/yr) \\ 
\hline                       
                           L40       &   J16192903-5042257  & 10.41 &  8.39 &  7.43 & 16.70 & 10.35 &  7.39 &  6.83 &  7.02 &  6.64 &  6.64 & \nodata & \nodata & \nodata & \nodata &  6.92 &  6.93 &  7.46 &  4.29 & \nodata& \nodata& \nodata&     17.50   &      $..$    &     0.13$\pm$     0.27 &   $-$3.86$\pm$   0.42 &   $-$3.82$\pm$   0.29 &  \\
                           L41       &   J16165636-5054263  & 10.91 &  8.59 &  7.46 & \nodata & \nodata & \nodata &  6.89 &  6.83 &  6.54 &  6.47 &  6.30 & \nodata & \nodata & \nodata &  6.79 &  6.79 &  6.24 &  4.54 & \nodata& \nodata& \nodata&     18.80   &      $..$    &     0.26$\pm$     0.36 &   $-$2.29$\pm$   0.69 &   $-$2.32$\pm$   0.45 &  \\
                           L42       &   J16215688-5103560  &  9.70 &  8.17 &  7.57 & 13.43 &  9.67 &  7.57 &  7.17 &  7.39 &  7.15 &  7.11 & \nodata & \nodata & \nodata & \nodata &  6.98 &  7.21 & \nodata & \nodata & \nodata& \nodata&  16.40 &     14.97   &      $..$    &    $-$0.08$\pm$     0.17 &   $-$3.98$\pm$   0.23 &   $-$7.05$\pm$   0.16 &  \\
                           L43       &   J16180517-5054355  & 10.24 &  8.43 &  7.63 & 15.22 & 10.27 &  7.59 &  7.11 &  7.29 &  6.99 &  6.96 & \nodata & \nodata & \nodata & \nodata &  7.05 &  7.13 &  6.62 &  5.24 & \nodata& \nodata& \nodata&     16.52   &      $..$    &     0.15$\pm$     0.21 &   $-$5.50$\pm$   0.34 &   $-$4.52$\pm$   0.22 &  \\
                           L44       &   J16172904-5100234  & 10.42 &  8.50 &  7.63 & 15.89 & 10.34 &  7.56 &  7.08 &  7.14 &  6.92 &  6.94 & \nodata & \nodata & \nodata & \nodata &  7.05 &  7.09 &  6.77 &  5.24 & \nodata&  17.57 &  20.63 &     17.06   &      $..$    &     0.17$\pm$     0.28 &   $-$2.89$\pm$   0.47 &   $-$3.75$\pm$   0.26 &  \\
                           L45       &   J16164039-5107327  &  8.35 &  7.80 &  7.68 & \nodata & \nodata & \nodata &  7.64 &  7.72 &  7.66 &  7.54 & \nodata & \nodata & \nodata & \nodata &  7.55 &  7.66 &  7.31 &  6.63 &  11.32 &  10.41 &   9.84 &     10.08   &    $-$33.47$\pm$     0.43  &     1.10$\pm$     0.07 &   $-$0.54$\pm$   0.09 &   $-$4.54$\pm$   0.06 &  \\
                           L46       &   J16172605-5102560  & 10.13 &  8.47 &  7.71 & 14.80 & 10.09 &  7.63 &  7.24 &  7.42 &  7.14 &  7.14 & \nodata & \nodata & \nodata & \nodata &  7.23 &  7.36 &  7.07 & \nodata &  17.24 &  17.06 &  21.00 &     16.13   &      $..$    &    $-$0.20$\pm$     0.18 &   $-$2.27$\pm$   0.33 &   $-$3.60$\pm$   0.20 &  \\
                           L47       &   J16164417-5104082  &  9.98 &  8.45 &  7.74 & \nodata & \nodata & \nodata &  7.31 &  7.52 &  7.16 &  7.20 & \nodata & \nodata & \nodata & \nodata &  7.35 &  7.49 &  7.33 &  5.47 &  20.79 & \nodata&  17.82 &     15.65   &      $..$    &    $-$0.08$\pm$     0.18 &   $-$1.01$\pm$   0.34 &   $-$1.01$\pm$   0.19 &  \\
                           L48       &   J16202739-5054420  & 11.66 &  9.12 &  7.94 & \nodata & 11.43 &  7.78 &  7.03 &  6.94 &  6.58 &  6.57 & \nodata & \nodata & \nodata & \nodata &  6.99 &  6.70 &  5.14 &  0.58 & \nodata& \nodata& \nodata&     19.98   &      $..$    &     1.83$\pm$     2.32 &   $-$1.87$\pm$   2.13 &   $-$0.43$\pm$   2.11 &  \\
                           L49       &   J16184348-5056111  & 10.43 &  8.72 &  7.95 & 14.84 & 10.33 &  7.91 &  7.49 &  7.61 &  7.36 &  7.30 & \nodata & \nodata & \nodata & \nodata &  7.46 &  7.48 &  6.45 &  5.10 &  16.98 &  16.24 &  17.24 &     16.25   &      $..$    &    \nodata$\pm$     0.15 &   $-$4.51$\pm$   0.23 &   $-$2.38$\pm$   0.16 &  \\
                           L50       &   J16181035-5101573  & 10.71 &  8.88 &  8.10 & 14.62 & 10.73 &  8.03 &  7.46 &  7.61 &  7.37 &  7.33 & \nodata & \nodata & \nodata & \nodata &  7.50 &  7.50 &  7.35 &  5.04 & \nodata& \nodata& \nodata&     17.27   &      $..$    &     0.37$\pm$     0.24 &   $-$4.53$\pm$   0.41 &   $-$3.98$\pm$   0.28 &  \\
                           L51       &   J16175105-5053143  & 11.43 &  9.20 &  8.32 & 17.24 & 11.34 &  8.33 &  7.69 &  7.78 &  7.55 &  7.47 & \nodata & \nodata & \nodata & \nodata &  7.78 &  7.78 &  7.09 & \nodata & \nodata& \nodata& \nodata&     18.37   &      $..$    &    $-$0.13$\pm$     0.32 &   $-$4.92$\pm$   0.47 &   $-$3.23$\pm$   0.36 &  \\
                           L52       &   J16193774-5058136  & 12.45 &  9.73 &  8.40 & \nodata & 12.42 &  8.35 &  7.45 &  7.37 &  6.96 &  6.95 & \nodata & \nodata & \nodata & \nodata &  7.47 &  7.10 &  5.41 &  2.50 & \nodata& \nodata& \nodata&    \nodata  &      $..$    &    \nodata&    $..$   &    $..$   &  \\
                           L53       &   J16175823-5100594  & 12.99 & 10.19 &  8.96 & \nodata & 12.86 &  8.84 &  8.02 &  8.08 &  7.79 &  7.77 & \nodata & \nodata & \nodata & \nodata &  8.11 &  8.04 &  7.65 &  4.88 & \nodata& \nodata& \nodata&    \nodata  &      $..$    &    \nodata&    $..$   &    $..$   &  \\
                           L54       &   J16164846-5108393  & 11.43 &  9.68 &  9.01 & \nodata & \nodata & \nodata &  8.49 &  8.66 &  8.46 &  8.41 & \nodata & \nodata & \nodata & \nodata &  8.54 &  8.57 &  7.80 &  5.69 & \nodata& \nodata&  21.00 &     17.36   &      $..$    &    $-$0.96$\pm$     0.35 &   $-$4.54$\pm$   0.43 &   $-$4.14$\pm$   0.30 &  \\
                           L55       &   J16174477-5057451  & 12.27 & 10.20 &  9.35 & \nodata & 12.28 &  9.26 &  8.62 &  8.81 &  8.53 &  8.60 & \nodata & \nodata & \nodata & \nodata &  8.77 &  8.80 &  7.05 &  4.86 & \nodata& \nodata& \nodata&     19.02   &      $..$    &     0.31$\pm$     0.67 &  $-$10.13$\pm$   1.04 &   $-$0.60$\pm$   0.73 &  \\
                           L56       &   J16201621-5058522  & 12.41 & 10.36 &  9.49 & 18.28 & 12.47 &  9.45 &  8.86 &  8.93 &  8.66 &  8.69 & \nodata & \nodata & \nodata & \nodata &  8.97 &  8.98 &  5.49 &  1.15 & \nodata& \nodata& \nodata&     19.23   &      $..$    &     0.26$\pm$     0.45 &   $-$4.65$\pm$   0.80 &   $-$6.00$\pm$   0.55 &  \\
                           L57       &   J16174628-5058235  & 13.54 & 10.85 &  9.68 & \nodata & 13.45 &  9.60 &  8.74 &  8.80 &  8.49 &  8.59 & \nodata & \nodata & \nodata & \nodata &  8.77 &  8.65 &  6.35 &  4.23 & \nodata& \nodata& \nodata&    \nodata  &      $..$    &    \nodata&    $..$   &    $..$   &  \\
                           L58  $^a$ &   J16170180-5047303  & 10.79 & 10.19 &  9.93 & \nodata & \nodata & \nodata &  9.82 &  9.81 &  9.86 & \nodata & \nodata & \nodata & \nodata & \nodata &  9.15 &  8.78 & \nodata & \nodata &  15.12 &  14.01 &  12.97 &     13.41   &    $-$41.33$\pm$     1.29  &     0.44$\pm$     0.02 &   $-$4.98$\pm$   0.04 &   $-$5.50$\pm$   0.03 &  \\
                           L59       &   J16171437-5047188  & 15.90 & 11.77 &  9.99 & \nodata & \nodata &  9.99 &  8.81 &  8.73 &  8.32 &  8.40 & \nodata & \nodata & \nodata & \nodata &  8.88 &  8.47 & \nodata &  2.38 & \nodata& \nodata& \nodata&    \nodata  &      $..$    &    \nodata&    $..$   &    $..$   &  \\
                           L60       &   J16200749-5047123  & 14.15 & 11.35 & 10.05 & \nodata & 14.04 &  9.94 &  9.07 &  9.10 &  8.73 &  8.78 & \nodata & \nodata & \nodata & \nodata &  9.09 &  8.91 &  5.59 &  2.04 & \nodata& \nodata& \nodata&    \nodata  &      $..$    &    \nodata&    $..$   &    $..$   &  \\
                           L61       &   J16185783-5051538  & 12.99 & 10.95 & 10.09 & \nodata & 12.89 & 10.03 &  9.45 &  9.48 &  9.17 &  9.21 & \nodata & \nodata & \nodata & \nodata &  9.48 &  9.43 &  6.47 &  3.67 & \nodata& \nodata& \nodata&     19.85   &      $..$    &     0.97$\pm$     0.70 &   $-$5.00$\pm$   1.39 &   $-$3.56$\pm$   1.00 &  \\
                           L62       &   J16200945-5049256  & 11.33 & 10.44 & 10.11 & 13.66 & 11.37 & 10.01 &  9.82 &  9.79 &  9.69 &  9.66 & \nodata & \nodata & \nodata & \nodata &  9.87 &  9.86 &  7.39 & \nodata &  18.45 &  14.94 &  15.07 &     15.08   &      $..$    &     0.21$\pm$     0.08 &   $-$2.95$\pm$   0.12 &   $-$4.34$\pm$   0.08 &  \\
                           L63       &   J16202826-5054493  & 11.71 & 10.57 & 10.16 & 14.19 & 11.70 & 10.06 &  9.91 &  9.97 &  9.86 & 10.13 & \nodata & \nodata & \nodata & \nodata & \nodata & \nodata & \nodata & \nodata &  16.63 &  16.62 &  15.40 &     15.75   &      $..$    &     0.18$\pm$     0.17 &   $-$2.17$\pm$   0.25 &   $-$2.57$\pm$   0.17 &  \\
                           L64       &   J16173473-5055106  & 11.89 & 10.70 & 10.26 & 14.87 & 11.88 & 10.11 &  9.92 &  9.91 &  9.79 &  9.69 & \nodata & \nodata & \nodata & \nodata &  9.95 &  9.93 &  8.91 & \nodata &  19.28 & \nodata&  16.33 &     16.51   &      $..$    &     0.30$\pm$     0.11 &   $-$4.52$\pm$   0.20 &   $-$3.35$\pm$   0.12 &  \\
                           L65       &   J16202094-5054467  & 14.35 & 11.57 & 10.29 & \nodata & 14.19 & 10.22 &  9.41 &  9.45 &  9.12 &  9.27 & \nodata & \nodata & \nodata & \nodata &  9.15 &  9.14 & \nodata & \nodata & \nodata& \nodata& \nodata&    \nodata  &      $..$    &    \nodata&    $..$   &    $..$   &  \\
                           L66  $^a$ &   J16201866-5055114  & 14.52 &\nodata& 10.50 & \nodata & \nodata & \nodata & \nodata & \nodata & \nodata & \nodata & \nodata & \nodata & \nodata & \nodata & \nodata & \nodata & \nodata & \nodata & \nodata& \nodata& \nodata&    \nodata  &      $..$    &    \nodata&    $..$   &    $..$   &  \\
                           L67       &   J16161247-5116217  & 14.27 & 11.64 & 10.51 & \nodata & \nodata & \nodata &  9.65 &  9.60 &  9.27 & \nodata & \nodata & \nodata & \nodata & \nodata &  9.53 &  9.22 &  5.13 &  1.97 & \nodata& \nodata& \nodata&    \nodata  &      $..$    &    \nodata&    $..$   &    $..$   &  \\
                           L68       &   J16202305-5052328  & 14.40 & 11.81 & 10.55 & \nodata & 14.37 & 10.47 &  9.62 &  9.62 &  9.25 & \nodata & \nodata & \nodata & \nodata & \nodata &  9.83 &  9.71 & \nodata & \nodata & \nodata& \nodata& \nodata&    \nodata  &      $..$    &    \nodata&    $..$   &    $..$   &  \\
                           L69       &   J16202632-5055100  & 15.48 & 12.30 & 10.77 & \nodata & 15.47 & 10.69 &  9.70 &  9.64 &  9.31 &  9.21 & \nodata & \nodata & \nodata & \nodata & \nodata & \nodata & \nodata & \nodata & \nodata& \nodata& \nodata&    \nodata  &      $..$    &    \nodata&    $..$   &    $..$   &  \\
                           L70  $^a$ &   J16171360-5048119  & 17.79 & 13.03 & 10.81 & \nodata & \nodata & 10.66 &  9.01 &  8.88 &  8.35 &  8.44 & \nodata & \nodata & \nodata & \nodata &  9.51 &  9.02 &  7.82 &  2.03 & \nodata& \nodata& \nodata&    \nodata  &      $..$    &    \nodata&    $..$   &    $..$   &  \\
                           L71  $^a$ &   J16202470-5054044  & 16.14 & 12.76 & 10.94 & \nodata & \nodata & 10.77 &  9.65 &  9.63 &  9.27 &  9.60 & \nodata & \nodata & \nodata & \nodata &  9.54 &  9.24 &  5.56 &  1.89 & \nodata& \nodata& \nodata&    \nodata  &      $..$    &    \nodata&    $..$   &    $..$   &  \\
                           L72       &   J16170784-5046454  & 12.36 & 11.37 & 10.98 & 14.65 & 12.24 & 10.96 & \nodata & \nodata & \nodata & \nodata & \nodata & \nodata & \nodata & \nodata & \nodata & \nodata & \nodata & \nodata &  20.05 &  17.17 &  15.98 &     16.12   &      $..$    &     0.13$\pm$     0.08 &   $-$0.56$\pm$   0.12 &   $-$1.09$\pm$   0.08 &  \\
                           L73  $^a$ &   J16202772-5054564  & 15.86 & 13.09 & 11.26 & 16.20 & 13.02 & 10.66 &  9.77 &  9.79 &  9.50 &  9.73 & \nodata & \nodata & \nodata & \nodata & \nodata & \nodata & \nodata & \nodata & \nodata& \nodata& \nodata&    \nodata  &      $..$    &    \nodata&    $..$   &    $..$   &  \\
                           L74       &   J16170437-5047164  & 13.26 & 12.45 & 11.78 & 14.54 & 13.38 & 11.86 & 10.95 & 10.90 & 10.57 & \nodata & \nodata & \nodata & \nodata & \nodata & \nodata & \nodata & \nodata & \nodata &  17.35 &  16.14 &  15.32 &     15.77   &      $..$    &     3.20$\pm$     0.05 &   10.55$\pm$   0.09 &  $-$35.08$\pm$   0.06 &  \\
                           L75  $^a$ &   J16170610-5047314  & 17.06 & 13.57 & 11.84 & \nodata & \nodata & 11.69 & 10.52 & 10.29 & 10.06 & \nodata & \nodata & \nodata & \nodata & \nodata & \nodata & \nodata & \nodata & \nodata & \nodata& \nodata& \nodata&    \nodata  &      $..$    &    \nodata&    $..$   &    $..$   &  \\
                           L76  $^a$ &   J16200887-5048510  & 18.15 & 14.31 & 12.19 & \nodata & \nodata & 11.95 & 10.40 & 10.15 &  9.86 & 10.06 & \nodata & \nodata & \nodata & \nodata & \nodata & \nodata & \nodata & \nodata & \nodata& \nodata& \nodata&    \nodata  &      $..$    &    \nodata&    $..$   &    $..$   &  \\
                            P1       &   J16143723-5126263  & 15.05 & 12.79 & 11.54 & \nodata & \nodata & \nodata & 10.49 & 10.04 &  9.84 &  9.53 & \nodata & \nodata & \nodata & \nodata & 10.59 & 10.04 &  8.86 &  5.84 & \nodata& \nodata& \nodata&    \nodata  &      $..$    &    \nodata&    $..$   &    $..$   &  \\
                            P2       &   J16161381-5136417  &  9.73 &  9.22 &  8.80 & \nodata & \nodata & \nodata &  8.27 &  8.00 &  7.78 &  7.44 & \nodata & \nodata & \nodata & \nodata &  8.21 &  7.83 &  7.71 &  4.93 &  13.78 &  12.97 &  11.35 &     12.60   &      $..$    &     0.14$\pm$     0.05 &   $-$3.47$\pm$   0.07 &   $-$3.86$\pm$   0.05 &  \\
                            P3       &   J16125464-5044510  &  7.08 &  6.17 &  5.79 &\nodata&\nodata&\nodata&  6.83 &  6.22 &  5.64 &  5.57 &\nodata&\nodata&\nodata&\nodata&  5.05 &  5.68 &  5.51 &  4.73 &  13.55 &  11.33 &  10.67 &     10.30   &    $-$39.82$\pm$     0.36  &     0.33$\pm$     0.06 &   $-$1.02$\pm$   0.08 &   $-$1.45$\pm$   0.06 &  \\
                            P4       &   J16151992-5026463  &  5.44 &  4.00 &  3.48 &\nodata&\nodata&\nodata&\nodata&\nodata&\nodata&\nodata&  2.76 &  2.05 &  1.97 &  1.36 &\nodata&  3.05 &  2.51 &  1.42 &  16.34 &  13.37 &  11.77 &     10.16   &    $-$49.68$\pm$     0.43  &     0.90$\pm$     0.12 &   $-$0.93$\pm$   0.18 &   $-$4.06$\pm$   0.12 &  \\
                            P5       &   J16164067-5014370  &  5.53 &  4.47 &  3.97 & \nodata & \nodata & \nodata &\nodata&  3.99 &  3.58 &  3.31 &  3.54 &  3.55 &  3.27 &\nodata&  3.85 &  3.58 &  3.47 &  2.62 &  13.63 &  12.78 &  11.23 &      9.66   &      $..$    &     1.36$\pm$     0.12 &    1.02$\pm$   0.17 &    0.98$\pm$   0.12 &  \\
                            P6       &   J16191454-5024243  &  6.60 &  4.07 &  2.83 & \nodata & \nodata & \nodata & \nodata & \nodata & \nodata & \nodata & \nodata & \nodata & \nodata & \nodata & \nodata & \nodata & \nodata & \nodata & \nodata& \nodata& \nodata&     14.39   &      $..$    &     0.05$\pm$     0.29 &   $-$5.56$\pm$   0.43 &   $-$4.01$\pm$   0.28 &  \\
                            P7       &   J16191948-5131264  &  5.57 &  4.20 &  3.84 &\nodata&\nodata&\nodata&\nodata&  4.13 &  3.65 &\nodata&  2.99 &  2.21 &  2.37 &  1.65 &  3.62 &  3.52 &  2.56 &  1.63 &  13.89 &  12.13 &  10.10 &      9.67   &    $-$59.33$\pm$     0.58  &     1.01$\pm$     0.10 &   $-$2.82$\pm$   0.17 &   $-$4.40$\pm$   0.16 &  \\
                            P8       &   J16202307-5104580  &  6.02 &  5.10 &  4.71 &\nodata&\nodata&\nodata&  4.69 &\nodata&  4.51 &  4.46 &  4.06 &\nodata&  3.33 &\nodata&  4.48 &  4.42 &  4.44 &  3.52 &  12.67 &  11.09 &   9.57 &      9.23   &    $-$42.74$\pm$     0.45  &     0.43$\pm$     0.04 &   $-$1.29$\pm$   0.07 &   $-$1.98$\pm$   0.04 &  \\

\hline
\end{tabular}
}
\begin{list}{}
\item {} 
\end{list}
\end{sidewaystable*}

\section{Spectra of Late-type Stars}

\begin{figure*}
\begin{center}
\resizebox{0.44\hsize}{!}{\includegraphics[angle=0]{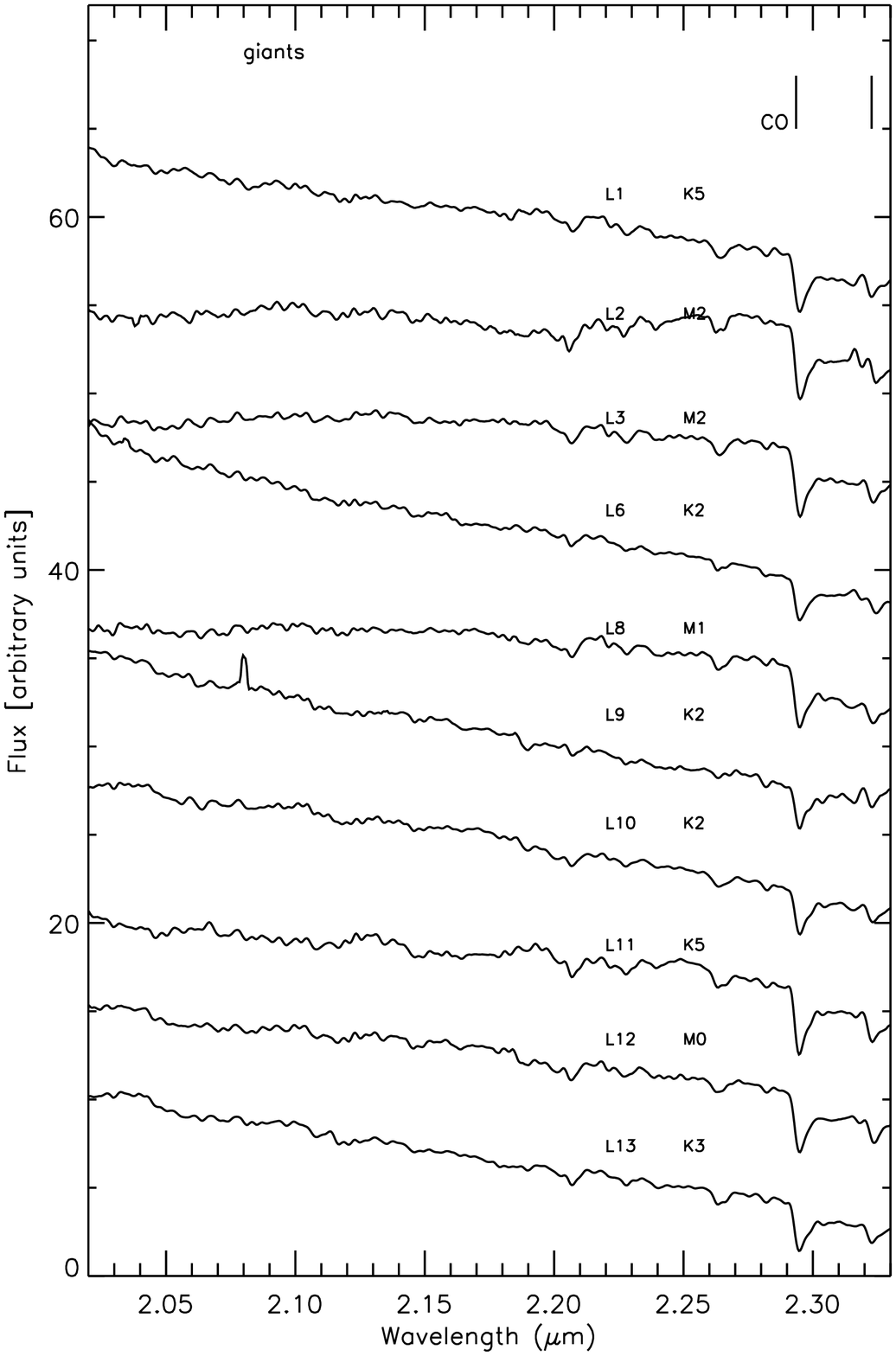}}
\resizebox{0.44\hsize}{!}{\includegraphics[angle=0]{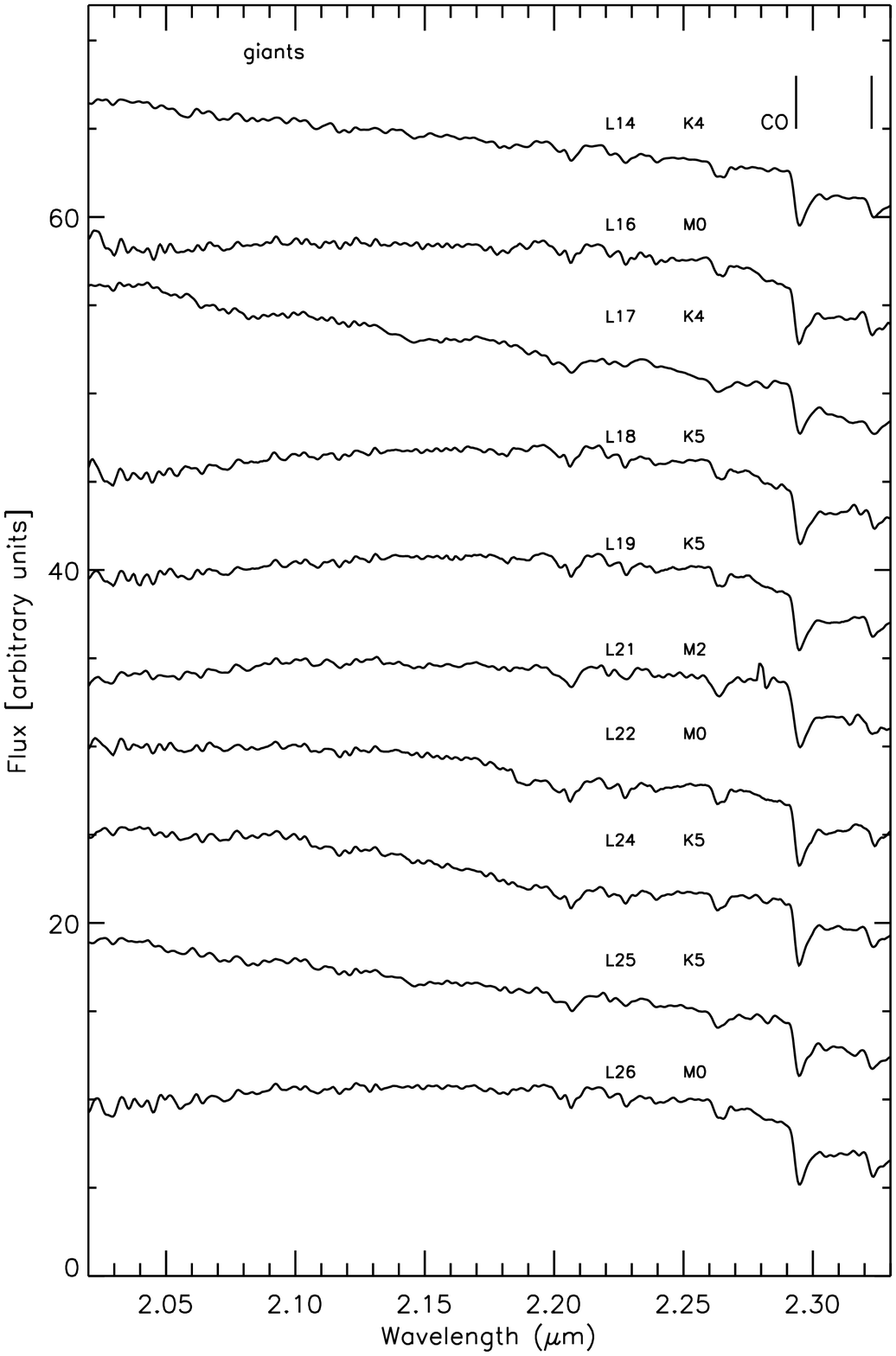}}
\end{center}
\begin{center}
\resizebox{0.44\hsize}{!}{\includegraphics[angle=0]{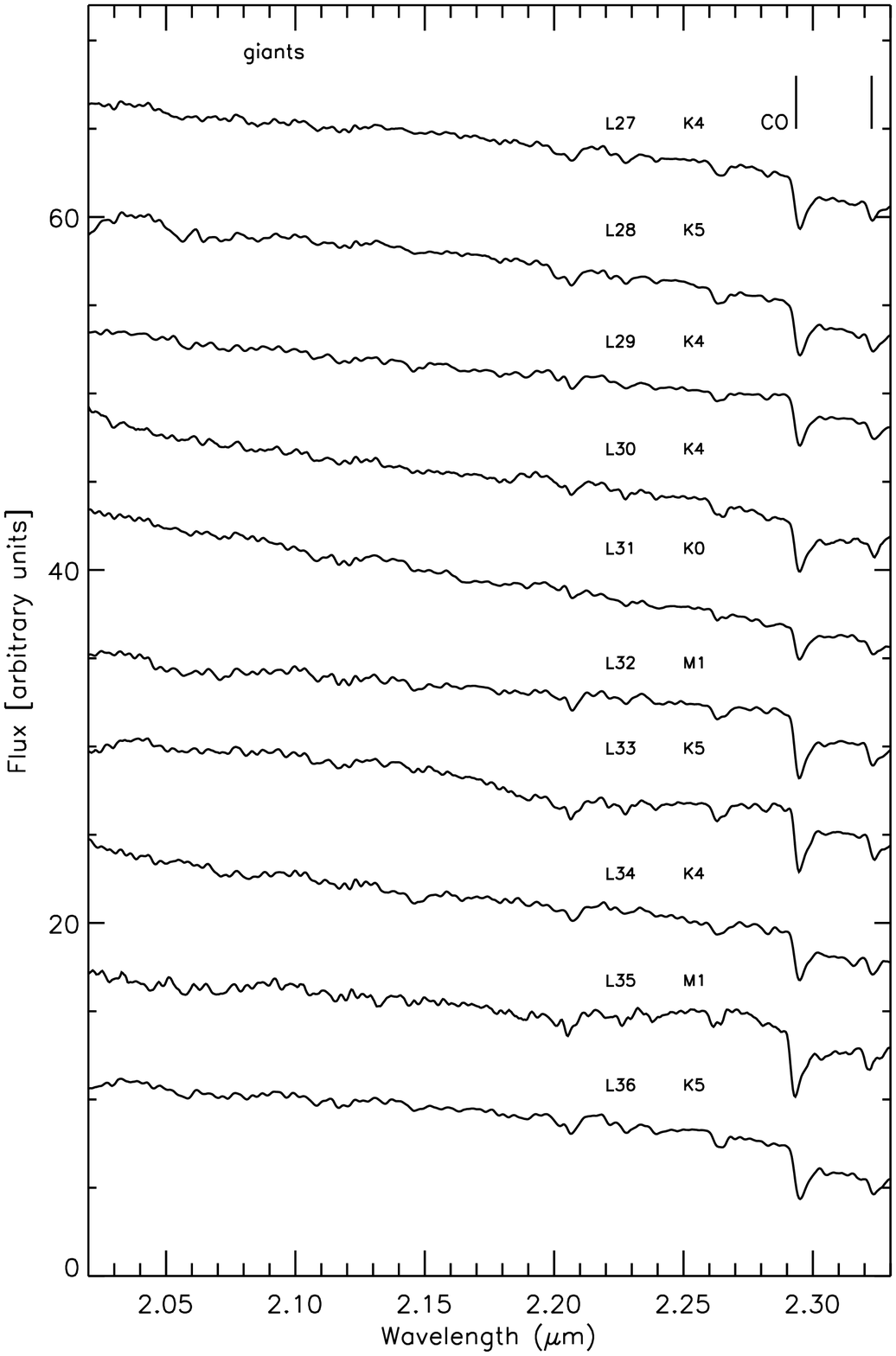}}
\resizebox{0.44\hsize}{!}{\includegraphics[angle=0]{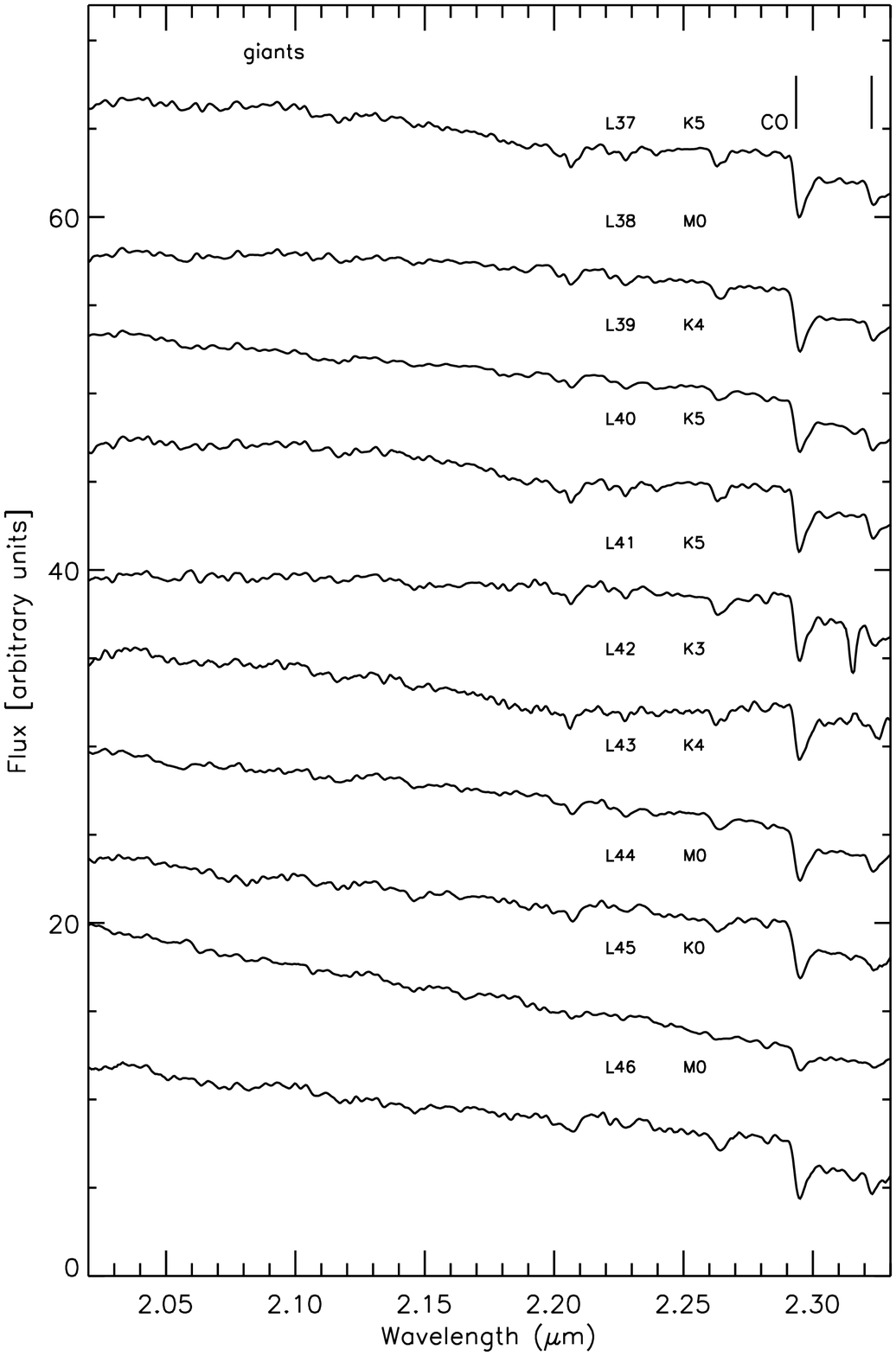}}
\end{center}
\caption{\label{giant.fig} 1. Dereddened $K$-band spectra (high-resolution mode) of 
giant stars in the direction of G332.809$-$0.132.
AGB stars are shown in the last panel.} 
\end{figure*}

We provide plots of the late-type stars spectroscopically detected in the direction of 
G332.809$-$0.132 and  CS 78 in Figs. \ref{giant.fig}1 and \ref{agb.fig}2.

\addtocounter{figure}{-1} 
\begin{figure*}
\begin{center}
\resizebox{0.44\hsize}{!}{\includegraphics[angle=0]{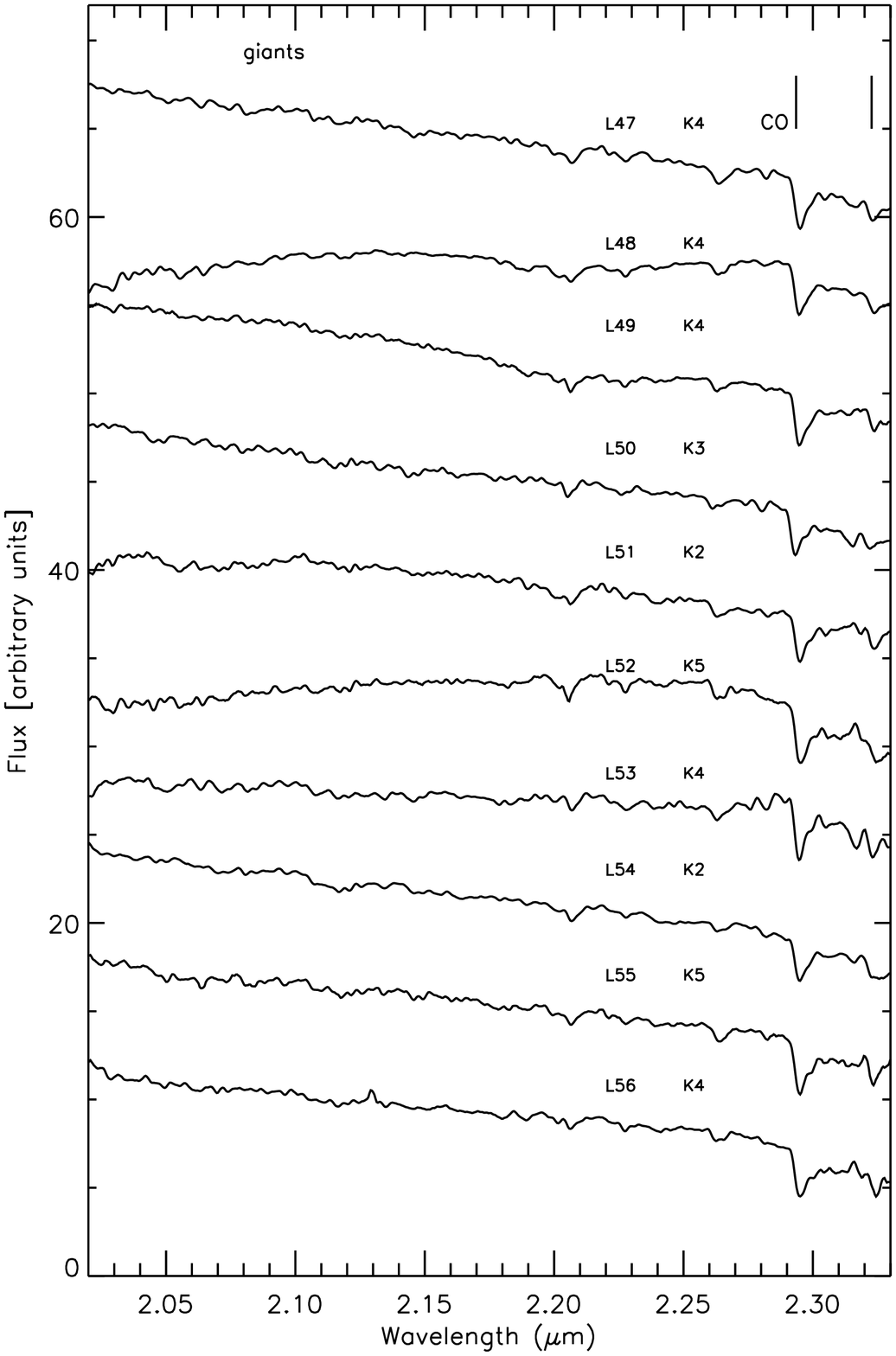}}
\resizebox{0.44\hsize}{!}{\includegraphics[angle=0]{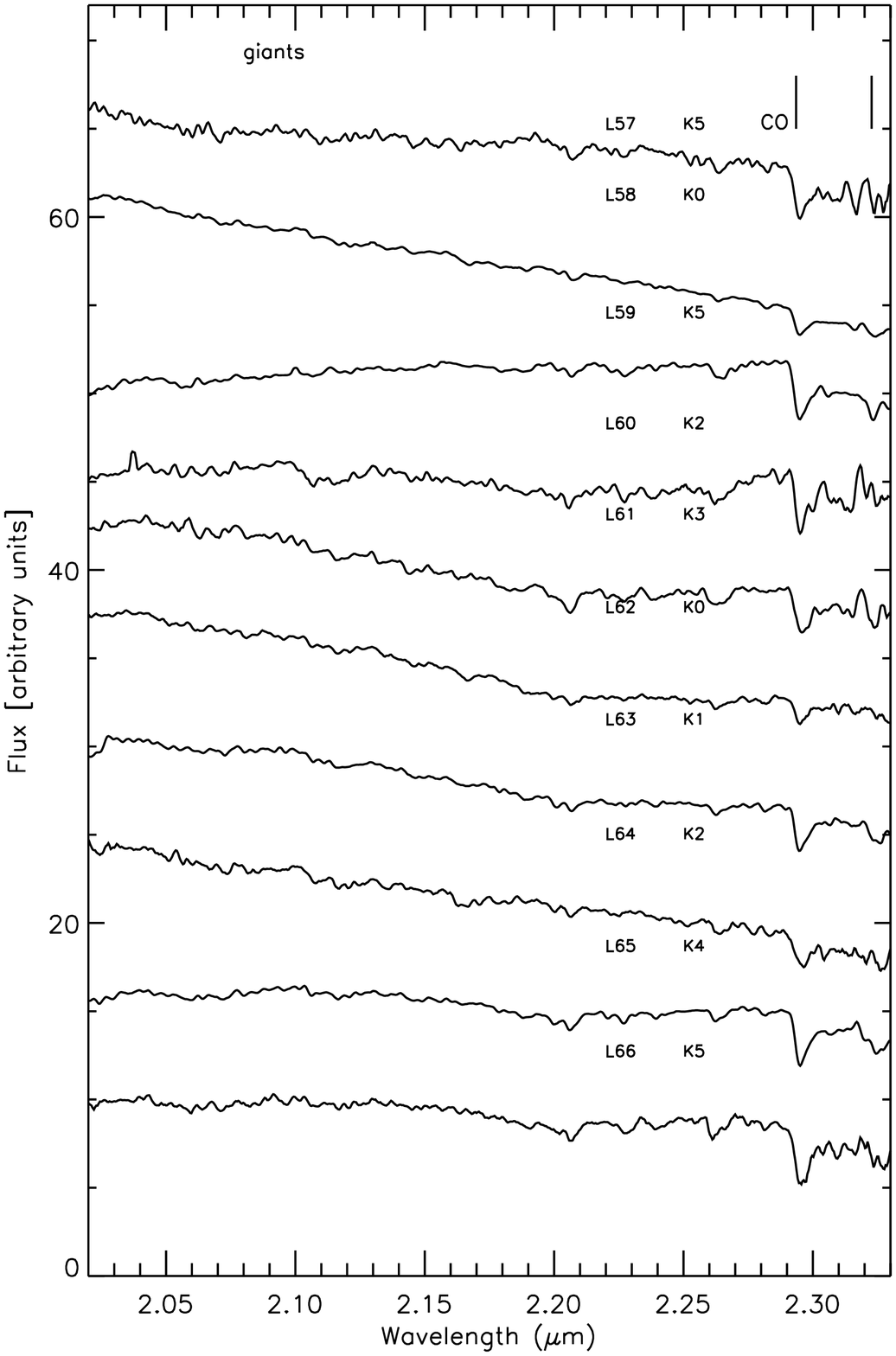}}
\end{center}
\begin{center}
\resizebox{0.44\hsize}{!}{\includegraphics[angle=0]{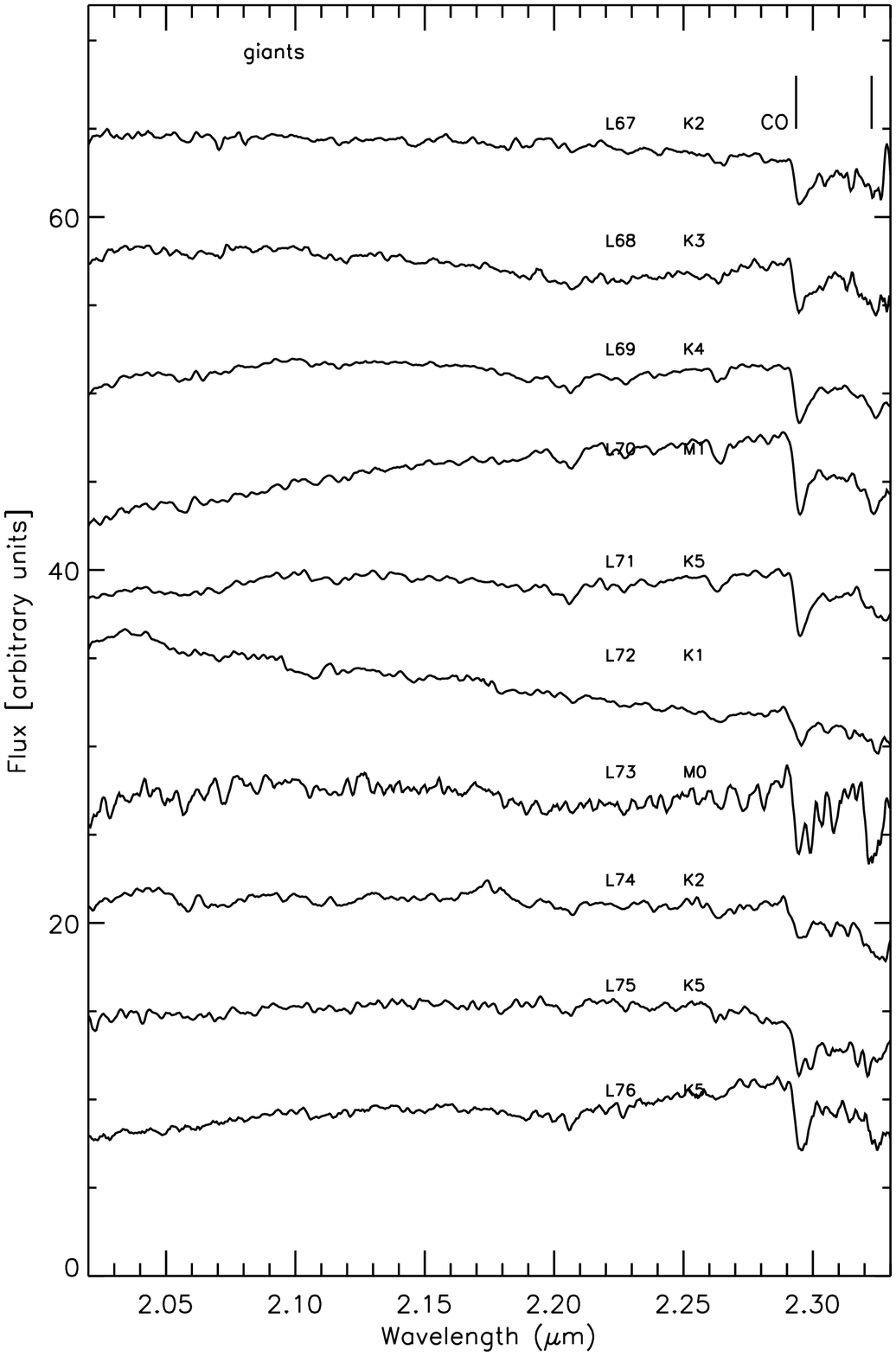}}
\resizebox{0.44\hsize}{!}{\includegraphics[angle=0]{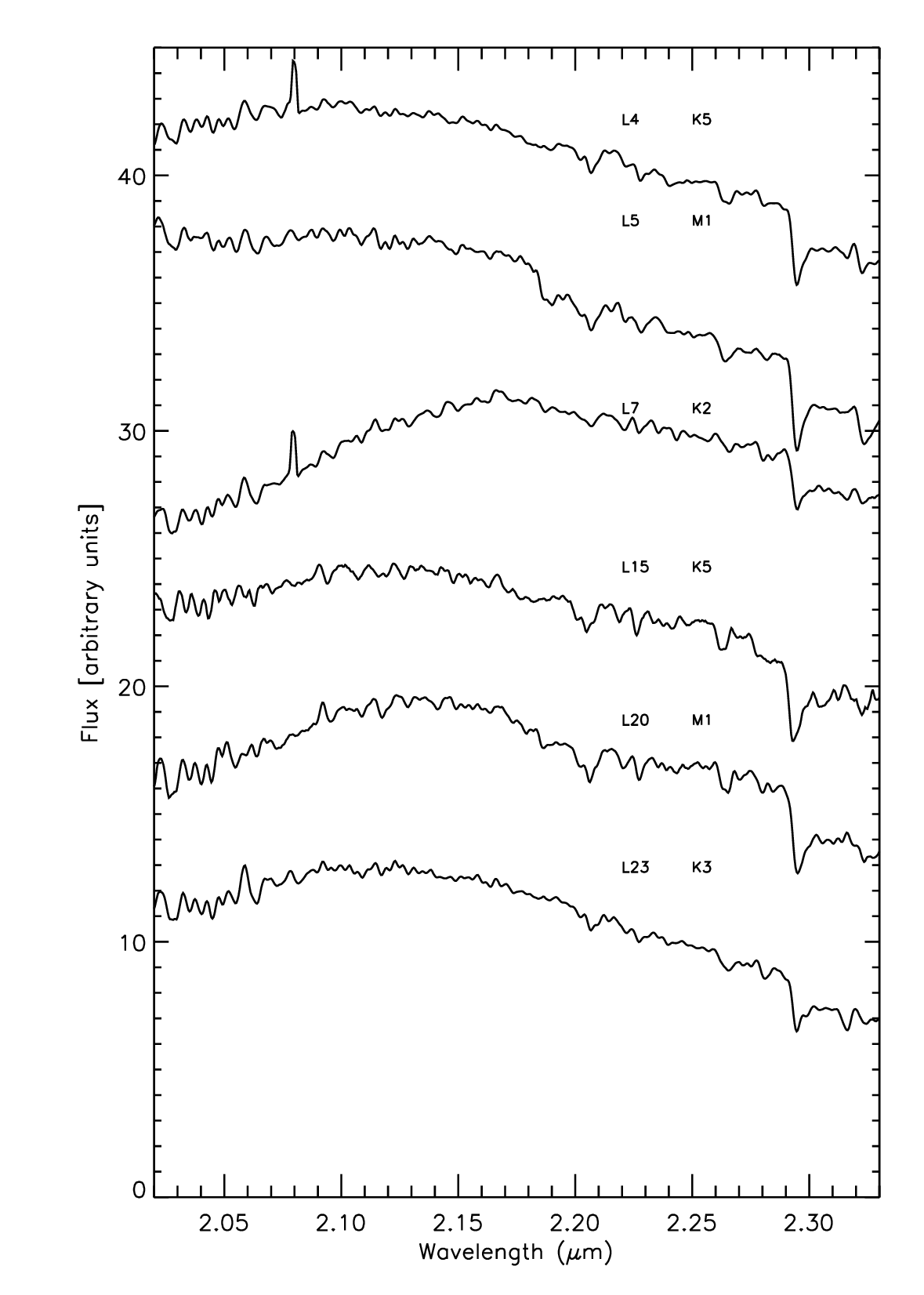}}
\end{center}
\caption{ 1. Continuation of Fig.\ \ref{giant.fig}.1.  } 
\end{figure*}

\begin{figure}[h]
\begin{center}
\resizebox{0.9\hsize}{!}{\includegraphics[angle=0]{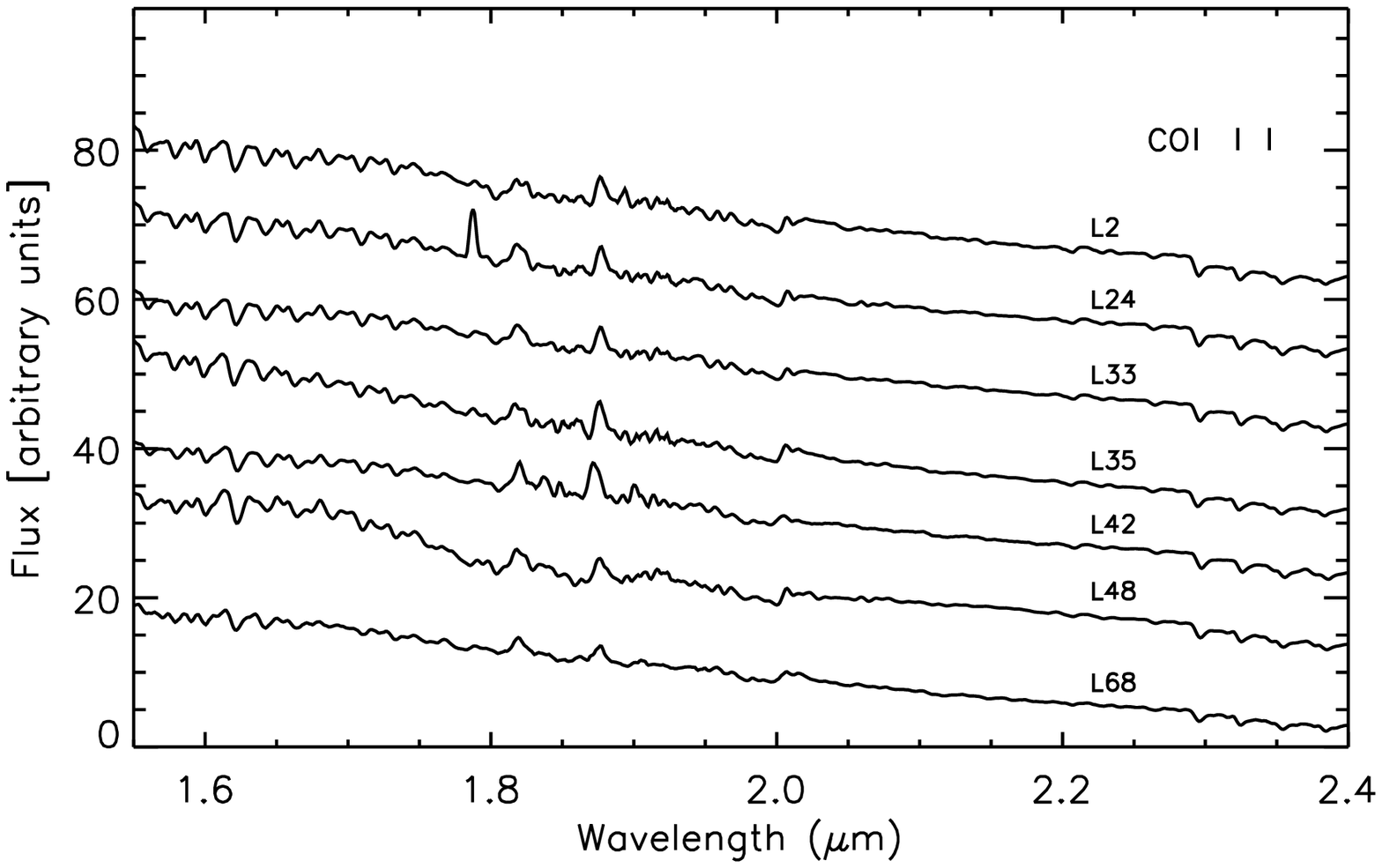}}
\end{center}
\caption{\label{agb.fig} 2. Dereddened $HK$ spectra of   
stars also observed in the low-resolution mode.} 
\end{figure}

\section{Identification Charts  }
\label{appendix.charts}

\begin{figure*}
\begin{center}
\resizebox{0.6\hsize}{!}{\includegraphics[angle=0]{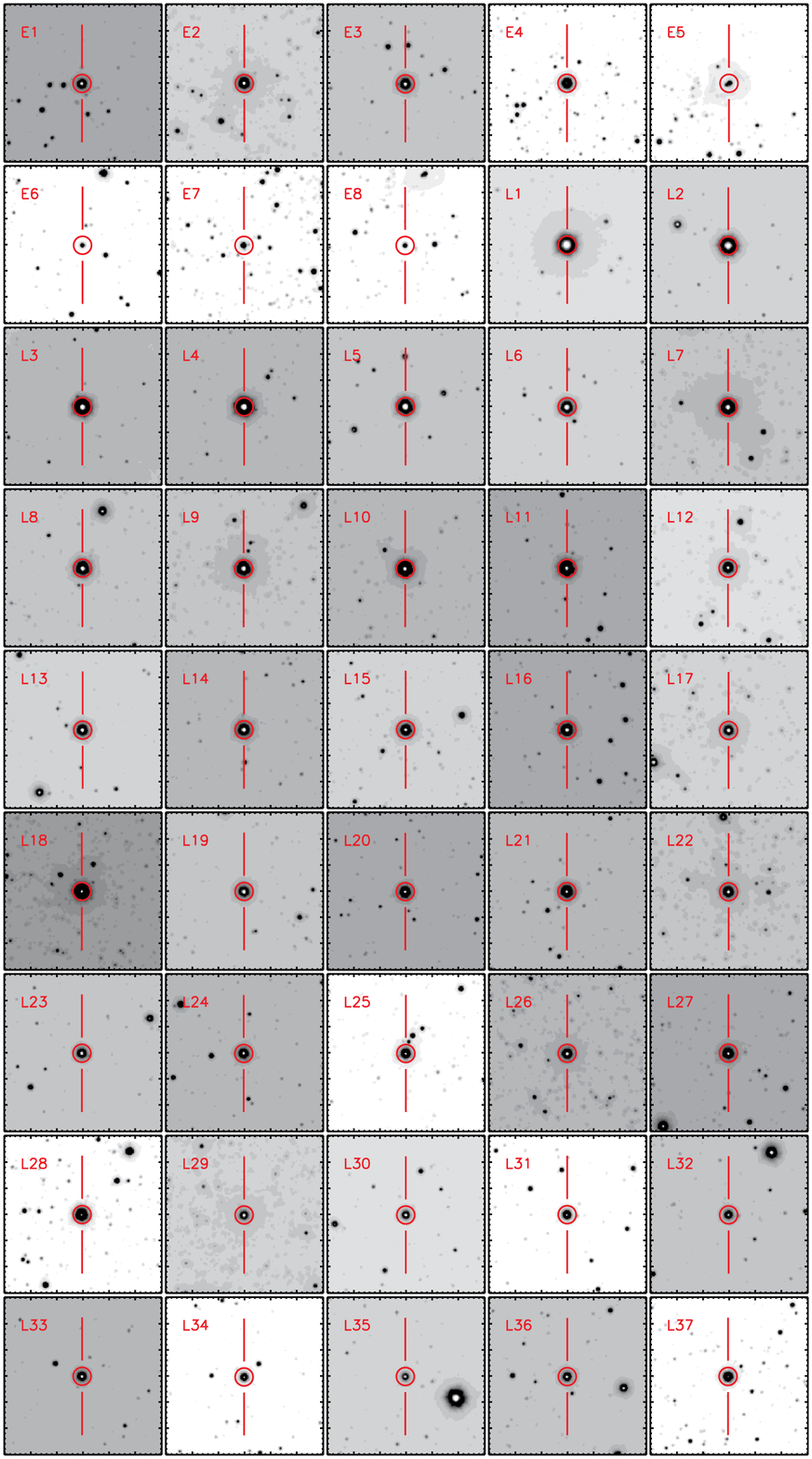}}
\caption{ \label{chart.fig} 1. VVV  \Ks-band images ( $\approx 1 ^\prime  \times \approx 1^\prime$ large) 
of the spectroscopically detected stars in the direction of 
G332.809$-$0.132 and  CS 78. North is up and east to the left. 
Targets are marked with circles. Identification numbers are as in Tables \ref{obs.early}
and {\bf \ref{table.latespectra}.}
} 
\end{center}
\end{figure*}

In Figure \ref{chart.fig}1 the  VVV  charts of the spectroscopically detected stars in the direction of 
G332.809$-$0.132 and  CS 78 are 
illustrated.

\addtocounter{figure}{-1}
\begin{figure*}
\begin{center}
\resizebox{0.6\hsize}{!}{\includegraphics[angle=0]{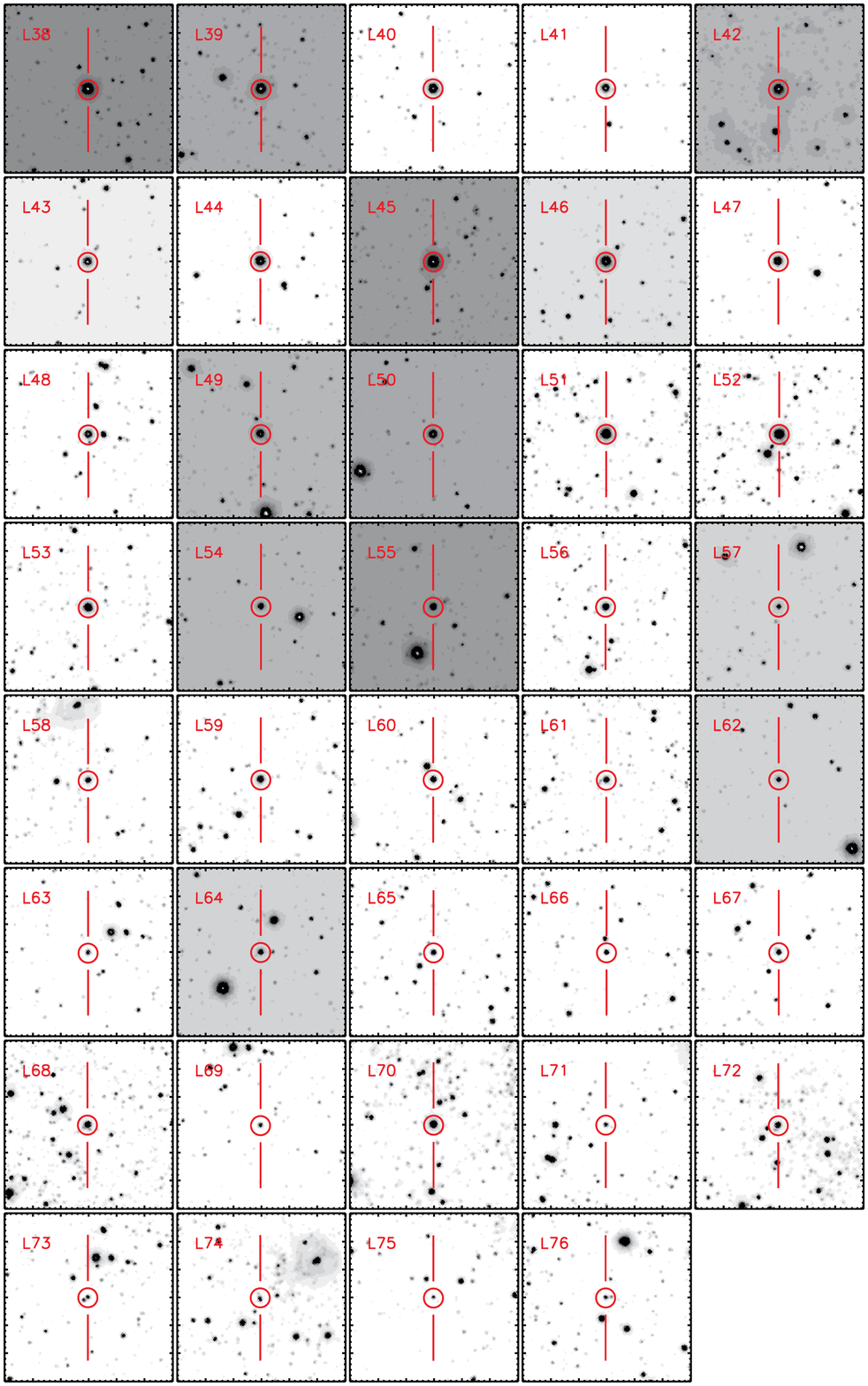}}
\caption{ 1. Continuation of Fig. \ref{chart.fig}1. } 
\end{center}
\end{figure*}

\section{SNRs and Stellar Clusters}

One of the first reports in the literature 
about the association of a  stellar cluster and an SNR was written by \citet{pauls77}.
SNR G127.1+00.5  coincides in position with the open cluster  NGC 559 
 \citep[note that the SNR was previously called G127.3+0.7][]{pauls77}.
For the SNR  a distance of 1.15$^{-0.25}_{+1.15}$ kpc has been reported \citep{acero16},
while the cluster is at a distance of 2.2 kpc \citep{kharchenko16}
and has an age of 630 Myr. The SNR's diameter is 45\arcmin. The cluster centroid is 
located within the remnant  at 50\% of the SNR's radius (11\arcmin).\\

One year after the  report of the  serendipitous discovery by \citet{pauls77},
a review of clusters projected onto SNRs was published  by \citet{kumar78}, who
searched a list of 117 SNRs for positional coincidences with
clusters and located two new coincidences. One was considered 
spurious based on distance measures, while  the 
association of the SNR G291.0-0.1 with the cluster Trumplet 18 was considered to be real.
However, nowadays, newer determinations place the cluster at 1.4 kpc \citep{kharchenko16}
and the SNR at 5 kpc \citep{acero16}.\\

Currently, a comprehensive  Catalogue of Galactic Supernova Remnants 
is maintained by Dr. D. Green with the latest update in 
2019\footnote{\url{http://www.mrao.cam.ac.uk/surveys/snrs/}}, while
\citet[][and references therein]{acero16} list  distances for 112 SNRs.

\citet{kharchenko16} provide a list of 3210  stellar clusters, which they detected
with ``a combination of uniform kinematic and near-infrared (NIR) photometric data 
gathered from the all-sky catalog PPMXL''. Parameters like distance and ages 
are uniformly determined.
We cross-matched the positions of SNRs from \citet{green19} 
with the positions of clusters by \citet{kharchenko16}.
We found a total of 90 cluster centroids located in projection within 48 SNRs.
Independently estimated distance is required to confirm
the physical association between the cluster and the SNR.
By shifting the coordinates in the longitudinal direction and repeating the matching procedures, 
we estimated that more than 80\% of the matches are random, and not physically meaningful. 
We limited  checking relative to  SNRs with a radius smaller than 25\arcmin\ the 28 matches. 
These 28 clusters have estimates of distance \citep{kharchenko16}, but
distance is known only for  13 SNRs \citep{acero16}.
{\it Three matches appear real;
BDSB-141 (10 Myr) is very likely associated with SNR G049.2-00.7, 
NGC 559 (630 Myr) with SNR G127.1+00.5 \citep[see also][]{pauls77}, 
and FSR-0891 (230 Myr) with SNR G189.1+03.0.}

\citet{bica19} collected Galactic candidate clusters from the literature
to generate a catalog of 10,978 star clusters 
(from Digital Sky Survey and incremented with the 2MASS, WISE, 
VVV, Spitzer and Herschel surveys).
We considered objects with diameters smaller than 1\degr, 
and found 37 possible matches with SNRs --
i.e., with SNR centroids enclosed within the cluster area.
When shifting their coordinates in longitudes, we found from 25 to 42 false matches.
By assuming  a mean of 32, we deduce that 86\% of the matches  are by chance.
Without distance (those are candidate clusters), one
would first look for the positional coincidences.
There are four entries with excellent matches
(separations of SNR centroids and cluster centroids  within 10\% of the cluster radius);
NGC 6834, NGC 3766, ASCC 64, and NGC 5281.
Note that these four entries are also included in the Kharchenko's catalog.
All but NGC 6834 are  random superpositions.
NGC 6834 is located at a likely distance of 3.1 kpc \citep{mathew14},
but  an estimate of the distance to SNR G065.7+01.2 is not available \citep{acero16}.\\

\end{appendix}

\begin{acknowledgements}
This publication makes use of data products from the Two Micron All Sky Survey, which is a joint project 
of the University of Massachusetts and the Infrared Processing and Analysis Center/California Institute of Technology, 
funded by the National Aeronautics and Space Administration and the National Science Foundation.
This work is based [in part] on observations made with the Spitzer Space Telescope, 
which is operated by the Jet Propulsion Laboratory, California Institute of Technology under a contract with NASA.
The DENIS project was supported, in France by the Institut National des Sciences de l'Univers, the Education Ministry 
and the Centre National de la Recherche Scientifique, in Germany by the State of Baden-Wuerttemberg, in Spain by the 
DGICYT, in Italy by the Consiglio Nazionale delle Ricerche, in Austria by the Fonds zur Foerderung der 
wissenschaftlichen Forschung and the Bundesministerium fuer Wissenschaft und Forschung. 
This research made use of data products from the
Midcourse Space Experiment, the processing of which was funded by the Ballistic Missile Defence Organization with additional
support from the NASA office of Space Science. 
This work makes use of the Naval Observatory Merged Astrometric Dataset (NOMAD).
This work has made use of data from the European Space Agency (ESA) mission {\it Gaia}
(\url{http://www.cosmos.esa.int/gaia}), processed by the {\it Gaia} Data Processing and Analysis
Consortium (DPAC, \url{http://www.cosmos.esa.int/web/gaia/dpac/consortium}). Funding for the DPAC
has been provided by national institutions, in particular the institutions participating in the 
{\it Gaia} Multilateral Agreement.
 This publication makes use of data products from
WISE, which is a joint project of the University of California, Los
Angeles, and the Jet Propulsion Laboratory/California Institute of Technology, 
funded by the National Aeronautics and Space Administration. 
This research has made use of the SIMBAD data base, operated at CDS, Strasbourg,
France. This work was partially supported by  the ESA fellowship,  
the ERC grant ERC Advanced Investigator Grant GLOSTAR (247078), 
the Fundamental Research Funds for the Central Universities in
China, and USTC grant KY2030000054. 
MM thanks  Maria Massi for useful discussions on free-free emitters,
and Mike Rich for discussions on explosive environments. 

%This paper is dedicated to Leonardo and to  Beethoven. 
\end{acknowledgements}

%\bibliographystyle{aa}
%\bibliography{biblio}

\end{document}